\documentclass[12pt,a4paper]{amsart}
\usepackage{amsfonts}
\usepackage{amsthm}
\usepackage{amsmath}
\usepackage{amscd}
\usepackage[latin2]{inputenc}
\usepackage{t1enc}
\usepackage[mathscr]{eucal}
\usepackage{indentfirst}
\usepackage{graphicx}
\usepackage{graphics}
\usepackage{pict2e}
\usepackage{epic}
\numberwithin{equation}{section}
\usepackage[margin=2.9cm]{geometry}
\usepackage{epstopdf} 
\usepackage{tikz}
\usepackage{subcaption}
\usepackage{epstopdf}
\usepackage{soul}
\usetikzlibrary{shapes,arrows,positioning,shadows,patterns,calc,decorations.markings}
\usepackage{amsaddr}
\usepackage{algorithm}
\usepackage[noend]{algpseudocode}
\usepackage{bbm}
\usepackage{caption}
\usepackage{subcaption}
\usepackage{bbm}
\usepackage{algorithm}
\usepackage[noend]{algpseudocode}
\usepackage{amsmath}
\usepackage{tikz}
\usepackage{amsbsy}
\usepackage{amssymb}
\usepackage{amsthm}
\usepackage{pgfplots}
\pgfplotsset{compat=1.3}
\usepackage{color}
\usepackage{xcolor,colortbl}
\definecolor{LightCyan}{rgb}{0.88,1,1}
\usepackage{bibunits}
\usepackage[textsize=tiny]{todonotes}
\usepackage{bm}
\usepackage[hyphens,spaces,obeyspaces]{url}
\usepackage{hyperref}

\allowdisplaybreaks

\usetikzlibrary{shapes,arrows}
\usetikzlibrary{arrows,positioning}
\usetikzlibrary{shapes,arrows}
\usetikzlibrary{arrows,positioning}
\usetikzlibrary{positioning,shadows,arrows}
\usetikzlibrary{arrows,positioning,decorations.markings,patterns,calc}

\usepackage{pgfplots}
\pgfplotsset{compat=1.3}

\theoremstyle{plain}

 \theoremstyle{definition}

\newtheorem{?}[Th]{Problem}

\makeatletter
\renewcommand{\email}[2][]{%
  \ifx\emails\@empty\relax\else{\g@addto@macro\emails{,\space}}\fi%
  \@ifnotempty{#1}{\g@addto@macro\emails{\textrm{(#1)}\space}}%
  \g@addto@macro\emails{#2}%
}
\makeatother

\begin{document}

\title[Network topology for inference of reaction models]{Exploiting network topology for large-scale inference of nonlinear reaction models}

\author{Nikhil Galagali and Youssef M.\ Marzouk}

\address{Massachusetts Institute of Technology\\
Cambridge MA, USA 02139} 

\email{nikhilg18@gmail.com}

\email{ymarz@mit.edu}

\keywords{reaction network | network inference | model selection | Bayesian inference | reversible-jump MCMC} 

\begin{abstract} 
The development of chemical reaction models aids understanding and prediction in areas ranging from biology to electrochemistry and combustion. A systematic approach to building reaction network models uses observational data not only to estimate unknown parameters, but also to learn model structure. Bayesian inference provides a natural approach to this data-driven construction of models. Yet traditional Bayesian model inference methodologies that numerically evaluate the evidence for each model 
are often
infeasible for nonlinear reaction network inference, as the number of plausible models can be combinatorially large. Alternative approaches based on model-space sampling can enable large-scale network inference, but their 
realization 
presents many challenges. In this paper, we present new computational methods that make large-scale nonlinear network inference tractable. First, we exploit the topology of networks describing potential interactions among chemical species to design improved ``between-model'' proposals for reversible-jump Markov chain Monte Carlo. Second, we introduce a sensitivity-based determination of move types which, when combined with network-aware proposals, yields significant additional gains in sampling performance. These algorithms are demonstrated on inference problems drawn from systems biology, with nonlinear differential equation models of species interactions.

\end{abstract}

\maketitle

\section{Introduction} 
Detailed chemical reaction networks are a critical component of simulation tools in a wide range of applications, including combustion, catalysis, electrochemistry, and biology. In addition to being used as predictive tools, reaction network models can encode a mechanistic understanding of the processes under study. The development of reaction networks typically entails three tasks: the selection of participating species, the identification of species interactions (referred to as reactions), and the calibration of unknown parameter values. To this end, we are often faced with the challenge of comparing a combinatorially large number of reaction networks, each containing parameterized reaction rate models derived from physical principles.
The ideas presented in this paper make large-scale Bayesian inference of network structure and parameters \textit{feasible} by exploiting essential aspects of network topology. 

A standard approach to building models is to postulate reaction networks and to compare them based on their ability to reproduce indirect system-level experimental data. Data-driven approaches to network learning involve defining a metric of fit, e.g., penalized least-squares, cross-validation, model evidence, etc., and selecting models that optimize this metric. As such, the development of models involves not only the identification of the best model structure, but also the estimation of underlying parameter values given available data. Bayesian inference provides a rigorous statistical framework for fusing data with prior knowledge to yield a full description of model and parameter uncertainties \cite{Gelman2004}. The application of Bayesian model inference to reaction networks, however, presents a significant computational challenge. Model discrimination in Bayesian analysis is based on computing model probabilities conditioned on available data, i.e., \textit{posterior} model probabilities. Formally, the posterior probability of a model $M_{m}$ is given by
\begin{align*}
p(M_{m} \vert {\mathcal{D}})=\frac{p(M_{m})p({\mathcal{D}} \vert M_{m})}{\sum_{m} p(M_{m})p({\mathcal{D}} \vert M_{m})},
\end{align*}
\noindent      
where
\begin{equation*}
p(\mathcal{D} \vert M_{m})=\int \cdots \int p(\mathcal{D} \vert {{k}_{m}}, M_m) p({{k}_{m}} \vert M_{m})d {{k}_{m}}
\end{equation*}
\noindent
is known as the model evidence, $p(M_{m})$ is the prior probability of model $M_{m}$, ${k_{m}}$ is the vector of parameters in model $M_{m}$, and ${\mathcal{D}}$ denotes the available data. A common approach to Bayesian model inference is to construct the candidate models $\{ M_m \}$ such that the prior and posterior on each ${{k}_{m}}$ are conjugate distributions, thereby making the calculation of individual model evidences analytically tractable. For instance, reaction network inference has been performed with network edge interactions represented by linear models or categorical distributions; both cases allow the specification of conjugate priors \cite{Friedman2000,Sachs2002}. %
Even in this simple setting where model evidences can be evaluated analytically and relatively cheaply, the combinatorial explosion of the number of networks, given species and their possible interactions, precludes direct enumeration of all models. \emph{Sampling} approaches have thus been used to infer such statistical models \cite{Ellis2008}. 

It is, however, widely believed that species interactions are more appropriately defined by the law of mass action, which gives the rate of an elementary chemical reaction (say $X+Y\rightarrow Z$) as the product of a reaction-specific rate constant $k$ with reactant concentrations $[X]$ and $[Y]$:
\begin{equation}
\text{Rate}=-k [X] [Y].
\label{massactionkineticsrate}
\end{equation}
%
Using the law of mass action to define reaction rates produces a system of ordinary differential equations (ODEs), such that the map from parameters (e.g., rate constants ${{k}_{n}}$) to observables (e.g., selected species concentrations) is typically nonlinear. ODE-based species interaction models have recently been incorporated into 
model inference frameworks \cite{Braman2013,Xu2010}. 
Rigorous computation of posterior model probabilities then requires numerical evaluation of a high-dimensional integral for each model.  A number of  Monte Carlo methods have been proposed for this purpose, but they are computationally taxing \cite{Chib2001,Gelman1998}. Alternative approaches such as Laplace approximations and, relatedly, the Bayesian information criterion are faster, but they involve potentially crude approximations of the posterior distribution \cite{Mackay2003,Schwarz1978}. And when the number of competing models becomes large, all of these methods become computationally {infeasible}.

Reaction network inference is particularly prone to this difficulty: instead of a few model hypotheses, one might start with a list of proposed reactions and form a collection of plausible models by considering all valid combinations of the proposed reactions. Across-model sampling \cite{Green2009} offers a useful solution in this setting. These sampling methods involve a Markov chain Monte Carlo (MCMC) algorithm that ``jumps'' between models to explore the joint posterior distribution over models and parameters. Posterior model probabilities are estimated by counting the number of times the sampler visits each model; direct evaluations of the evidence for each model are thus avoided. Efficient across-model sampling, however, is quite challenging: it requires the careful design of proposals for between-model moves. Many practical applications of across-model sampling have relied on pilot runs of within-model posterior sampling to characterize the relevant distributions
before employing an across-model sampler, although a few more automated methods do exist \cite{Green2009}. 
%
 Large-scale network inference with nonlinear forward models, in particular, has seen relatively limited attention \cite{Galagali2015}.
 A notable effort from Oates \textit{et al.}\ \cite{Oates2012} applies Bayesian model selection to ODE models systematically generated from potential species interactions, using reversible-jump MCMC (a general across-model sampling method) to simultaneously sample network topologies and their underlying rate parameters. Their approach employs standard methods for nested models: when adding reactions, the prior distribution is used to propose new reaction rates, and when swapping reactions, previous rate parameter values are simply retained. 
%
%
In general, however, the posterior distribution can deviate significantly from the prior, such that sampling from the prior yields poor across-model mixing. Further, with nonlinear forward models, the posterior distributions of common reaction parameters can differ across models, such that retaining parameter values during swap moves is not efficient.
%
%

In this paper, we present new methods for 
 Bayesian inference of chemical reaction networks. The essence of our approach is to exploit \emph{sensitivities} resulting from network structure. In particular, our approach performs an online analysis of network structure to construct better across-model moves, i.e., more effective ways of navigating through the discrete model space. This analysis accounts for sensitivities at the discrete level---the notion that, given a particular network structure, the presence of certain reactions will have no impact on the observables---and sensitivities at the continuous (parameter) level---the notion that observables are more sensitive to changes in certain combinations of rate parameters than in others. The resulting algorithms make fully Bayesian inference of reaction networks feasible in new problem settings, where existing approaches may fail to characterize the posterior distribution altogether.

\section{Reaction networks}
\begin{figure}[h]
{ \makeatletter
 \def\@captype{figure}
 \makeatother
 \begin{minipage}[h]{0.45\textwidth}
 \begin{subfigure}[b]{1.0\textwidth}
 \begin{center}
 \includegraphics[width=1.0\linewidth]{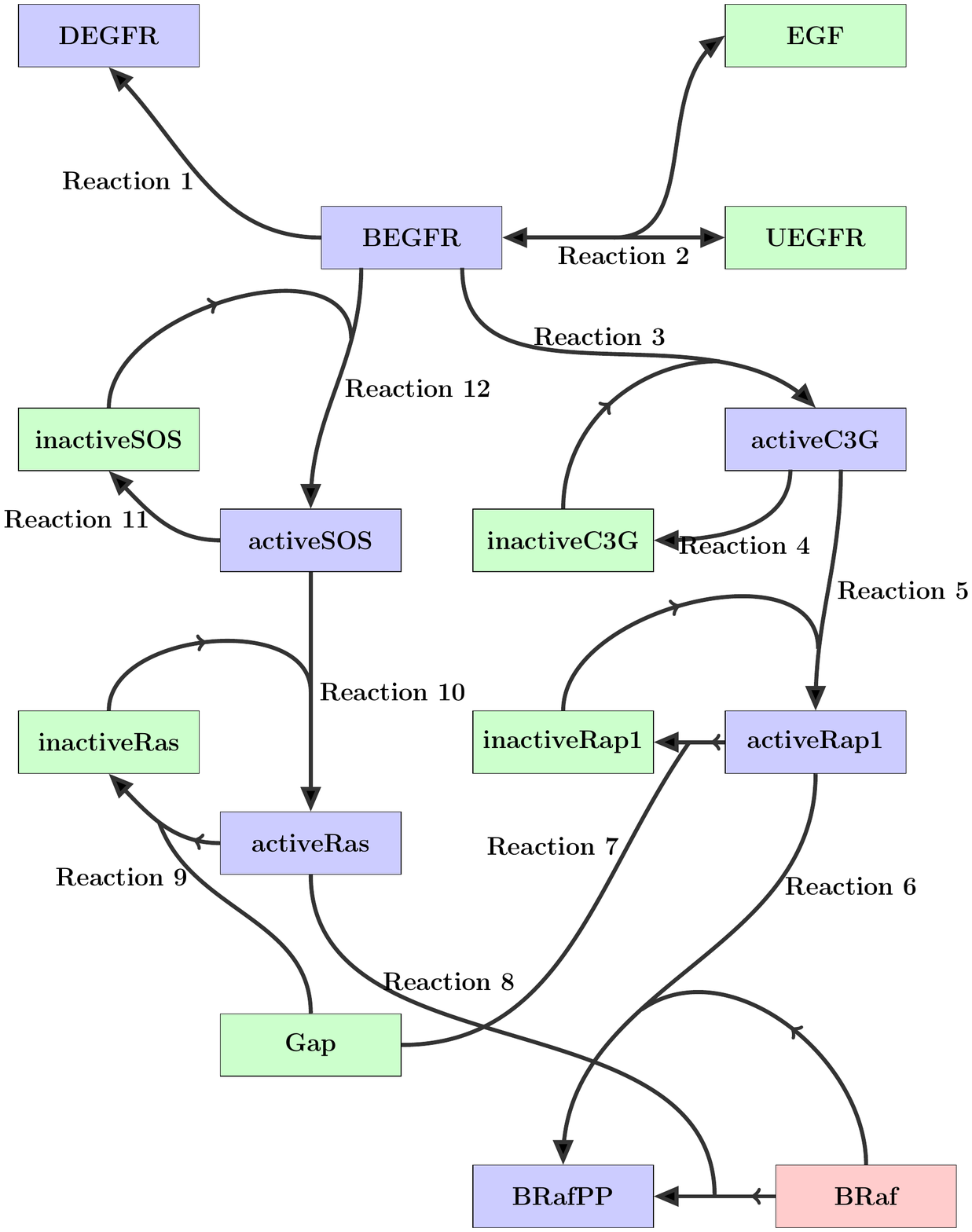}
 \end{center}  
  \caption{A reaction network}
  \label{fig:sub1}
 \end{subfigure} 
 \end{minipage}
 \begin{minipage}[h]{0.45\textwidth} 
\begin{subfigure}[b]{1.0\textwidth}
\begin{center}
\includegraphics[width=0.7\linewidth]{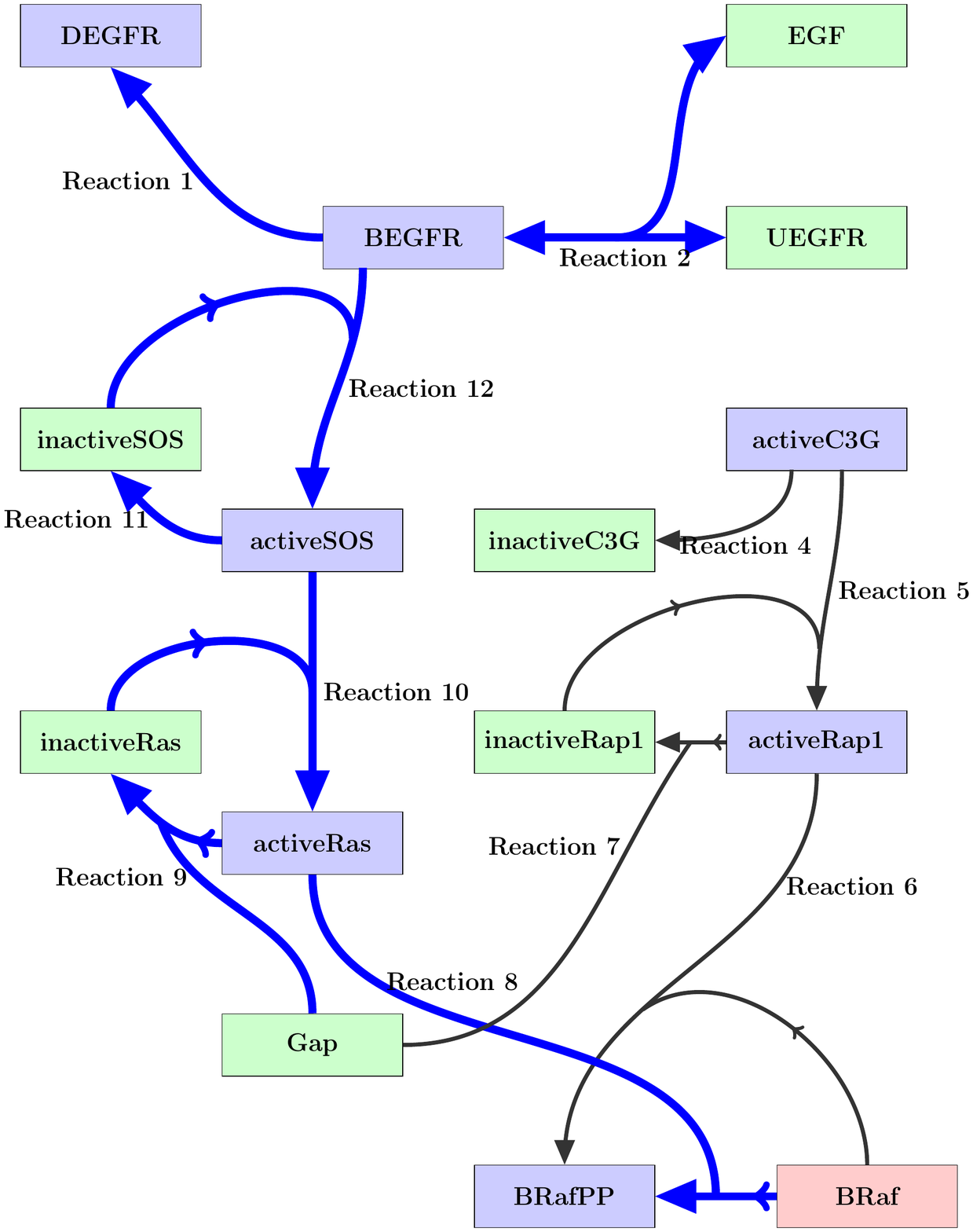}
\includegraphics[width=0.7\linewidth]{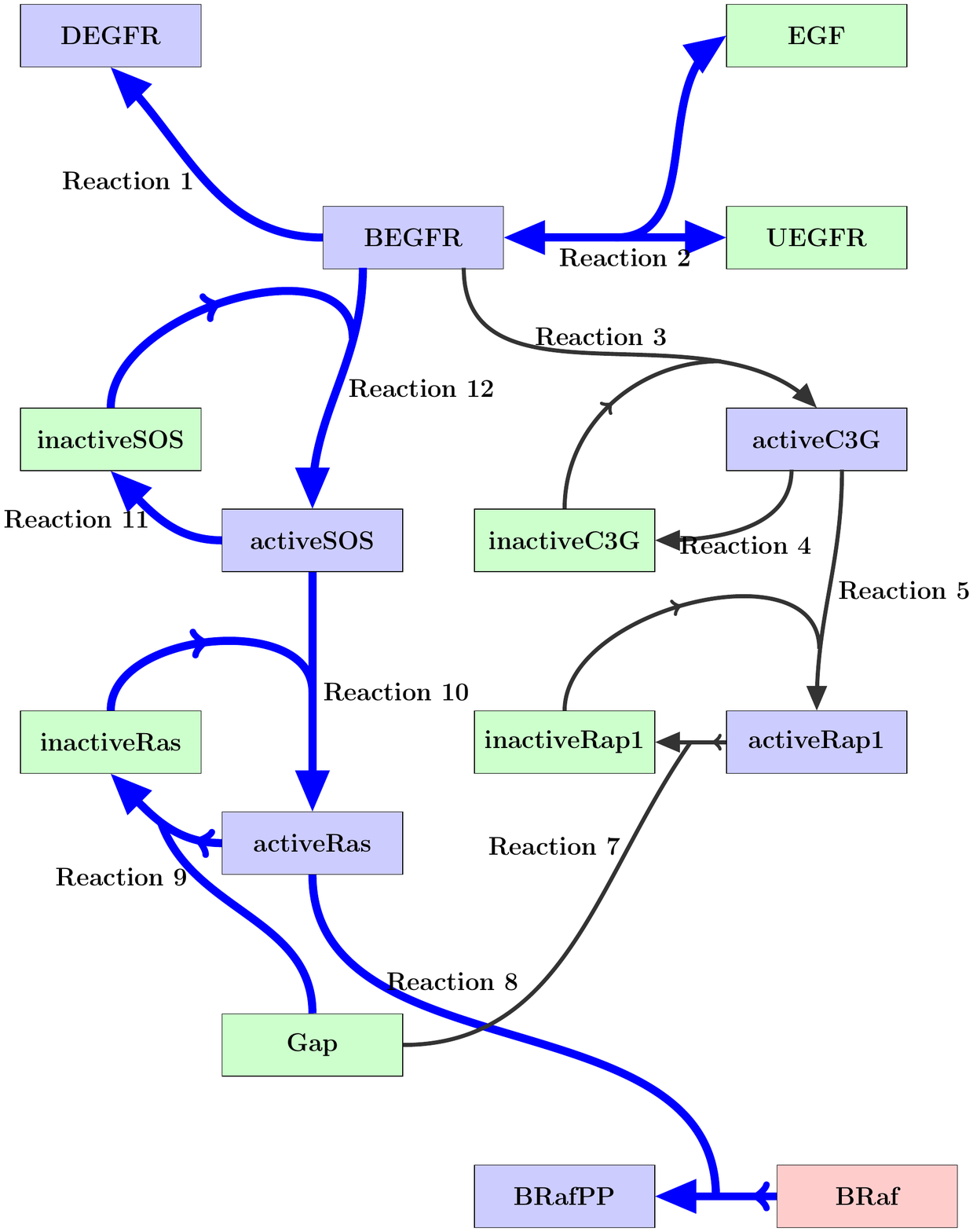}
\end{center}  
\caption{Two networks with identical effective network (in blue)}
\label{fig:sub2}
\end{subfigure}
\end{minipage}
\caption{Chemical reaction networks}
\label{fig:chemreactionnetwork}}
\end{figure}

\tikzstyle{SPECIESA}=[draw, fill=blue!20, text width=1em, text centered, minimum height=0.2em]
\tikzstyle{SPECIESB}=[draw, fill=blue!20, text width=1em, text centered, minimum height=0.2em]
\tikzstyle{SPECIESC}=[draw, fill=blue!20, text width=1em, text centered, minimum height=0.2em]
\tikzstyle{SPECIESD}=[draw, fill=blue!20, text width=1em, text centered, minimum height=0.2em]
\tikzstyle{SPECIESE}=[draw, fill=blue!20, text width=1em, text centered, minimum height=0.2em]
\tikzstyle{SPECIESF}=[draw, fill=blue!20, text width=1em, text centered, minimum height=0.2em]
\tikzstyle{SPECIESG}=[draw, fill=blue!20, text width=1em, text centered, minimum height=0.2em]

\tikzstyle{SPECIESINITIAL}=[draw, fill=green!20, text width=1em, text centered, minimum height=0.2em]
\tikzstyle{SPECIESINTERMEDIATE}=[draw, fill=blue!20, text width=1em, text centered, minimum height=0.2em]
\tikzstyle{SPECIESOBSERVABLE}=[draw, fill=red!20, text width=1em, text centered, minimum height=0.2em]

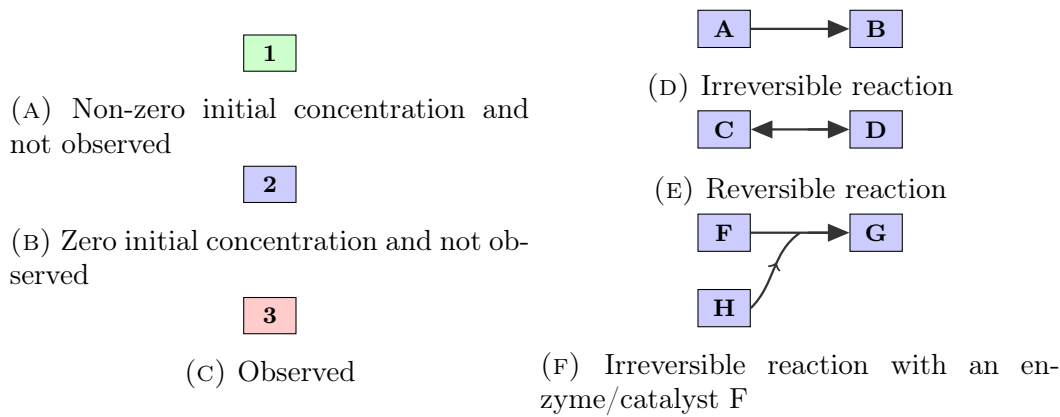
\begin{figure}[h]
\makeatletter
\def\@captype{figure}
\makeatother
\begin{minipage}{0.45\textwidth}
\begin{subfigure}[h]{1.0\textwidth}
\begin{center}
\begin{tikzpicture}
\begin{scope}
\node[SPECIESINITIAL](ySPECIESINITIAL) at (0,-2) {\bf \scriptsize 1};
\end{scope}
\end{tikzpicture}
\end{center}
\caption{Non-zero initial concentration and not observed}
\label{initialSpecies}
\end{subfigure}
\begin{subfigure}[h]{1.0\textwidth}
\begin{center}
\begin{tikzpicture}
\begin{scope}
\node[SPECIESINTERMEDIATE](ySPECIESINTERMEDIATE) at (3,-2) {\bf \scriptsize 2};
\end{scope}
\end{tikzpicture}
\end{center}
\caption{Zero initial concentration and not observed}
\label{intermediateSpecies}
\end{subfigure}
\begin{subfigure}[h]{1.0\textwidth}
\begin{center}
\begin{tikzpicture}
\begin{scope}
\node[SPECIESOBSERVABLE](ySPECIESOBSERVABLE) at (0,-3) {\bf \scriptsize 3};
\end{scope}
\end{tikzpicture}
\end{center}
\caption{Observed}
\label{observabedSpecies}
\end{subfigure}
\end{minipage}
\begin{minipage}{0.45\textwidth}
\begin{subfigure}[h]{1.0\textwidth}
\begin{center}
\begin{tikzpicture}
\begin{scope}
\node[SPECIESA](ySPECIESA) at (0,0) {\bf \scriptsize A};
\node[SPECIESB](ySPECIESB) at (2,0) {\bf \scriptsize B};
\draw [->,line width=0.3mm,draw=black!80,-triangle 45](ySPECIESA.east)to[out=0,in=180] node[above]{}(ySPECIESB.west);
\end{scope}
\end{tikzpicture}
\end{center}
\caption{Irreversible reaction}
\label{irreversibleReaction}
\end{subfigure}
\begin{subfigure}[h]{1.0\textwidth}
\begin{center}
\begin{tikzpicture}
\begin{scope}
\node[SPECIESE](ySPECIESC) at (0,-2) {\bf \scriptsize C};
\node[SPECIESF](ySPECIESD) at (2,-2) {\bf \scriptsize D};
\draw [->,line width=0.3mm,draw=black!80,-triangle 45](1.1,-2)to[out=0,in=180] node[above]{}(ySPECIESD.west);
\draw [->,line width=0.3mm,draw=black!80,-triangle 45](1.4,-2)to[out=180,in=0]node[below]{}(ySPECIESC.east);
\end{scope}
\end{tikzpicture}
\end{center}
\caption{Reversible reaction}
\label{reversibleReaction}
\end{subfigure}
\begin{subfigure}[h]{1.0\textwidth}
\begin{center}
\begin{tikzpicture}
\begin{scope}
\node[SPECIESE](ySPECIESF) at (0,-2) {\bf \scriptsize F};
\node[SPECIESF](ySPECIESG) at (2,-2) {\bf \scriptsize G};
\node[SPECIESG](ySPECIESH) at (0,-3) {\bf \scriptsize H};
\draw [->,line width=0.3mm,draw=black!80,-triangle 45](1.1,-2)to[out=0,in=180] node[above]{}(ySPECIESG.west);
\draw [-,line width=0.3mm,draw=black!80](1.4,-2)to[out=180,in=0]node[below]{}(ySPECIESF.east);
\draw [-,line width=0.3mm,draw=black!80,decoration={markings,mark=at position 0.5 with {\arrow{<}}},postaction={decorate}](1.0,-2)to[out=215,in=45]node[below]{}(ySPECIESH.east);
\end{scope}
\end{tikzpicture}
\end{center}
\caption{Irreversible reaction with an enzyme/catalyst F}
\label{irreversibleActivator}
\end{subfigure}
\end{minipage}
\caption{Taxonomy of reaction network elements: chemical species in the left column, reactions in the right column.}
\label{fig:reactionNetworkTaxonomy}
\end{figure}

\subsection{Reaction network elements}
A chemical reaction network generally consists of two different elements: chemical species $\emph{S}$, and interactions between species given by reactions $\emph{R}$. Consider a simple reaction network shown schematically in Figure \ref{fig:sub1}. The reaction network consists of 15 nodes representing the chemical species and 12 edges representing reactions. We can classify all species according to three categories (Figure \ref{fig:reactionNetworkTaxonomy}): species initially present (colored green) in the reaction system, species produced only during the operation (colored blue) of the system, and species that are either directly observed or directly linked to the observed data---referred to as \emph{observables} (colored red). At the same time, we designate three types of reactions (Figure \ref{fig:reactionNetworkTaxonomy}): irreversible reactions (single-headed arrow), reversible reactions (double-headed arrow), and irreversible or reversible reactions with enzymes (single- or double-headed arrows with pointers on branches connecting species that are consumed/produced). Enzymes are chemical species that are needed for the reaction to proceed, but are not consumed or produced during the course of the reaction. 
\subsection{Effective reaction network}
\label{sec:effectiveReactionNetwork}
A reaction network may contain reactions that are not active (because their reactants are not present) or reactions that are active and yet incapable of impacting the observables due to the \emph{structure} of the reaction network. Consider, for example, the two reaction networks shown in Figure \ref{fig:sub2}. We define the \emph{effective network} of a given reaction network to be the smallest subset of all reactions in the given network that produces an identical value of the observables as the given reaction network. Reactions beyond those in the effective network do not affect the observable concentrations for any rate parameter setting, and thus the given reaction network has the same marginal likelihood value as its effective reaction network. Different networks may have a common effective network; for example, both networks in Figure \ref{fig:sub2} have the same effective network (reactions 1, 2, 8, 9, 10, 11, 12). In reaction network 1, the non-production of species \texttt{activeC3G} renders reactions 4, 5, 6, and 7 inactive, and thus the value of the observable \texttt{BRaf} is independent of the rate constants of these reactions. In reaction network 2, although reactions 3, 4, 5, and 7 are active, they are linked to the observable \texttt{BRaf} through species \texttt{Gap} and \texttt{BEGFR}, which are enzymes. Recall that enzyme concentration is not affected by the reaction in which it participates. Thus, the observable is again independent of the rate constants of reactions 3, 4, 5, and 7. 
\subsection{Determining effective networks from proposed reactions}
In Section~\ref{sec:computation} we will describe how knowing the effective network corresponding to any given network can be used to define better across-model proposals and, in a further post-processing step, to reduce the variance of estimates of posterior model probabilities. In principle, one could learn the effective network for each candidate network \emph{before} performing inference.  If $N$ is the total number of proposed reactions, the set of candidate networks could have cardinality $2^{N}$ (although incorporating prior knowledge to eliminate highly implausible network structures might reduce this number in practice). Regardless, if the number of candidate networks is very high, it is more efficient to determine effective networks in an \emph{online} fashion, only for models visited by the sampler. Our procedure for identifying the effective network of a set of reactions and observables is given in Algorithm 1, described in the Appendix. This procedure involves first identifying all reactions that are active, given the species with nonzero initial concentrations, and then testing all active reactions to check whether they actually influence the observables. The algorithm is based on an analysis of the reaction network topology, and does not require any integration of the ODE system associated with the reaction network; thus it is computationally inexpensive.

\section{Reversible jump Markov chain Monte Carlo}
\label{sec:revJumpMCMCapproach}
Reversible jump MCMC (RJMCMC) provides a general framework for exploring a posterior distribution over a joint space of model choices and parameter values \cite{Green1995}. Consider the space of candidate models $\mathcal{M}=\{M_{m}\}$. Each model $M_{m}$ has an $n_{m}$-dimensional vector of unknown parameters $k_{m}\in \mathbb{R}^{n_{m}}_{+}$, where $n_{m}$ can take different values for different models. The RJMCMC algorithm simulates a Markov chain on the state space $\bigcup_{M_m\in \mathcal{M}} \{m \} \times \mathbb{R}^{n_m}$
whose invariant distribution is the joint model-and-parameter posterior distribution:
\begin{equation}
 p(M_{m},k_{m} \vert \mathcal{D})=p(k_{m} \vert M_{m},\mathcal{D})p(M_{m} \vert \mathcal{D}), 
 \end{equation}
where $p(k_{m} \vert M_{m},\mathcal{D})$ is the posterior distribution of the parameters. Each step of the algorithm consists of proposing a new pair of (model, parameter) values and accepting or rejecting the proposed values according to an acceptance probability that also depends on the current model and parameter values. At any point within the state space, many different proposals can be constructed. Generally, the proposals can be classified as \textit{between-model} and \textit{within-model} moves. A within-model move applies any standard Metropolis-Hastings proposal to the real-valued rate parameters, without changing the model. A between-model move involves proposing a different model (i.e., a different set of reactions) and a vector of parameter values associated with this model. Ensuring that the posterior distribution over the models and parameters is the invariant distribution of the Markov chain is accomplished by satisfying the detailed balance condition. (Further technical conditions on the proposal ensure that the Markov chain is irreducible and aperiodic, yielding an ergodic reversible jump algorithm \cite{Green1995}.) 
Detailed balance is enforced by constructing moves between any two models $M_m$ and $M_n$ according to a bijective map $f: (k_{m},u) \mapsto (k_{n},u')$, where $u$ and $u'$, known as dimension matching variables, are such that $\dim(k_{n})+\dim(u')=\dim(k_{m})+\dim(u)$. These variables have densities $q(u|k_{m})$ and $q(u'|k_{n})$, respectively. The choice of the distribution of $u$ is part of the proposal construction and, along with an appropriate map $f$, is key to an efficient RJMCMC simulation. In particular, these choices determine how parameter values in a proposed model are related to parameter values in the current model.  At each step of the simulation, given the current state $(M_m, k_{m})$, a move to a new model $M_n$ is first proposed according to some prescribed distribution $q(M_n \vert M_m)$. Next,  completing the proposal pair $(M_n, k_{n})$ given $(M_m, k_{m})$ involves generating  a sample of $u$ according to $q(u|k_{m})$ and accepting the proposed move with probability $\min\{1,\alpha\}$, where
\begin{equation}
\alpha=\frac{p(M_n, k_{n} \vert \mathcal{D}) q(M_m \vert M_n)q(u'|k_{n})}{p(M_m,k_{m} \vert \mathcal{D}) q(M_n \vert M_m)q(u|k_{m})}\left \vert \det \nabla f(k_{m},u) \right \vert,
\label{acceptanceRatio}
\end{equation}
and $(k_{n},u')=f(k_{m},u)$.
The reverse move from  $(k_{n},u')$ to $(k_{m},u)$ is given by $f^{-1}$ and has an acceptance probability $\min\{1,\alpha^{-1}\}$. 

\begin{figure} [h]
{
            \vspace{1cm}
            \centering
            \begin{subfigure}[b]{0.4\textwidth}
            \definecolor{mycolor1}{rgb}{0.00000,0.44700,0.74100}%
\begin{tikzpicture}[baseline,oneDSample/.style={minimum size=0.2cm,draw=magenta,fill=magenta,circle,inner sep=0pt},
        yscale=1, xscale = 1.1]

\begin{axis}[
width=0.7\textwidth,
height=0.7\textwidth,
scale only axis,
xlabel=$k_{1,1}$,
line width = 2pt,
axis lines = left,
ymin=0,
axis background/.style={fill=white},
xtick=\empty,
ytick=\empty
]
\addplot [color=red,solid,forget plot,very thick]
  table[row sep=crcr]{%
-1	0.0876082255551717\\
-0.97979797979798	0.0909453282738952\\
-0.95959595959596	0.0944703801364679\\
-0.939393939393939	0.0981972217601764\\
-0.919191919191919	0.102140932627116\\
-0.898989898989899	0.10631795861978\\
-0.878787878787879	0.110746253837813\\
-0.858585858585859	0.1154454382949\\
-0.838383838383838	0.120436973236213\\
-0.818181818181818	0.125744355952461\\
-0.797979797979798	0.131393336086754\\
-0.777777777777778	0.137412155520805\\
-0.757575757575758	0.14383181396695\\
-0.737373737373737	0.150686362352856\\
-0.717171717171717	0.158013225925353\\
-0.696969696969697	0.165853558661336\\
-0.676767676767677	0.174252629977565\\
-0.656565656565657	0.183260243768212\\
-0.636363636363636	0.19293118831978\\
-0.616161616161616	0.203325713454701\\
-0.595959595959596	0.214510028064101\\
-0.575757575757576	0.226556806640911\\
-0.555555555555556	0.239545687032088\\
-0.535353535353535	0.253563732758128\\
-0.515151515151515	0.26870582108019\\
-0.494949494949495	0.28507490149408\\
-0.474747474747475	0.302782047221595\\
-0.454545454545455	0.321946193047412\\
-0.434343434343434	0.342693414843963\\
-0.414141414141414	0.365155557684748\\
-0.393939393939394	0.389467959313301\\
-0.373737373737374	0.415765943763332\\
-0.353535353535353	0.444179678298601\\
-0.333333333333333	0.474826902042119\\
-0.313131313131313	0.507802960433945\\
-0.292929292929293	0.54316754107226\\
-0.272727272727273	0.580927544920644\\
-0.252525252525252	0.621015704666742\\
-0.232323232323232	0.66326496420937\\
-0.212121212121212	0.707379360698214\\
-0.191919191919192	0.752903296097788\\
-0.171717171717172	0.799192682326892\\
-0.151515151515151	0.845393381157991\\
-0.131313131313131	0.890434271419384\\
-0.111111111111111	0.933043454073114\\
-0.0909090909090909	0.971795531410564\\
-0.0707070707070707	1.00519451218056\\
-0.0505050505050505	1.03179018437663\\
-0.0303030303030303	1.05031653182754\\
-0.0101010101010101	1.0598314535848\\
0.0101010101010102	1.0598314535848\\
0.0303030303030303	1.05031653182754\\
0.0505050505050506	1.03179018437663\\
0.0707070707070707	1.00519451218056\\
0.0909090909090908	0.971795531410564\\
0.111111111111111	0.933043454073114\\
0.131313131313131	0.890434271419384\\
0.151515151515152	0.84539338115799\\
0.171717171717172	0.799192682326892\\
0.191919191919192	0.752903296097789\\
0.212121212121212	0.707379360698214\\
0.232323232323232	0.66326496420937\\
0.252525252525253	0.621015704666741\\
0.272727272727273	0.580927544920644\\
0.292929292929293	0.543167541072259\\
0.313131313131313	0.507802960433945\\
0.333333333333333	0.474826902042119\\
0.353535353535354	0.444179678298601\\
0.373737373737374	0.415765943763332\\
0.393939393939394	0.389467959313301\\
0.414141414141414	0.365155557684748\\
0.434343434343434	0.342693414843963\\
0.454545454545455	0.321946193047412\\
0.474747474747475	0.302782047221595\\
0.494949494949495	0.28507490149408\\
0.515151515151515	0.26870582108019\\
0.535353535353535	0.253563732758128\\
0.555555555555556	0.239545687032088\\
0.575757575757576	0.226556806640911\\
0.595959595959596	0.214510028064101\\
0.616161616161616	0.203325713454701\\
0.636363636363636	0.19293118831978\\
0.656565656565657	0.183260243768212\\
0.676767676767677	0.174252629977565\\
0.696969696969697	0.165853558661336\\
0.717171717171717	0.158013225925353\\
0.737373737373737	0.150686362352856\\
0.757575757575758	0.14383181396695\\
0.777777777777778	0.137412155520805\\
0.797979797979798	0.131393336086754\\
0.818181818181818	0.125744355952461\\
0.838383838383838	0.120436973236213\\
0.858585858585859	0.1154454382949\\
0.878787878787879	0.110746253837813\\
0.898989898989899	0.10631795861978\\
0.919191919191919	0.102140932627116\\
0.939393939393939	0.0981972217601764\\
0.95959595959596	0.0944703801364679\\
0.97979797979798	0.0909453282738951\\
1	0.0876082255551717\\
};
\end{axis}


\end{tikzpicture}%
            \end{subfigure}           
            \begin{subfigure}[b]{0.4\textwidth}
            \input{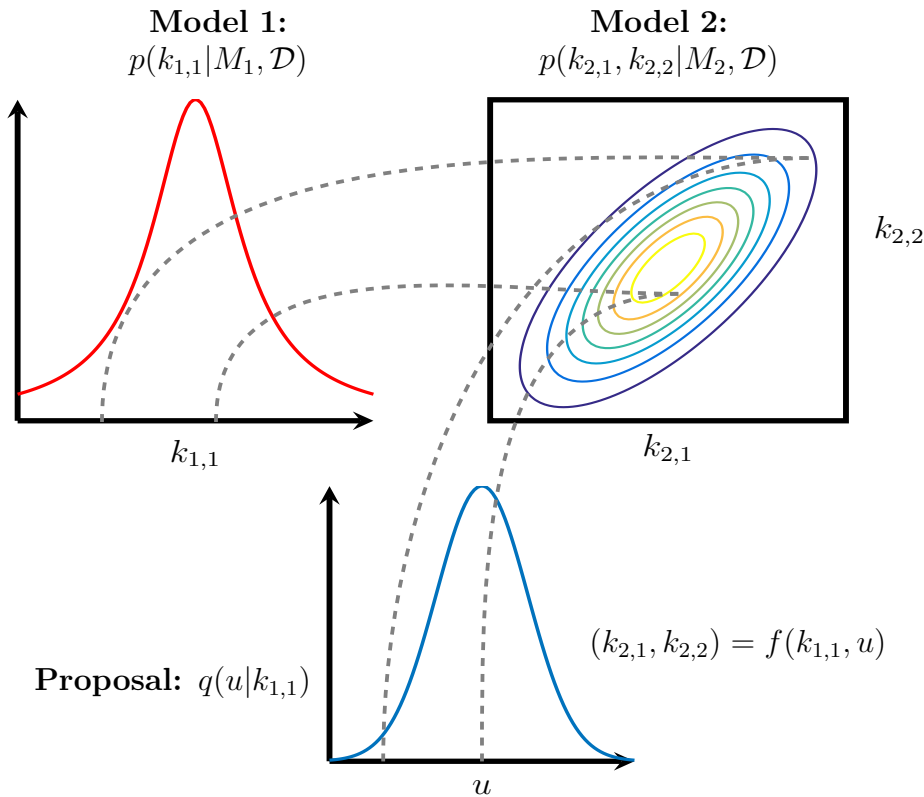}
            \end{subfigure}
           \begin{subfigure}[b]{0.8\textwidth}
           \begin{center}
%
%
\definecolor{mycolor1}{rgb}{0.00000,0.44700,0.74100}%
\begin{tikzpicture}[baseline,oneDSample/.style={minimum size=0.2cm,draw=blue,fill=blue,circle,inner sep=0pt},
        yscale=1, xscale = 1.1]

\begin{axis}[%
width=0.3\textwidth,
height=0.3\textwidth,
scale only axis,
xlabel=$u$,
axis lines= left,
ymin=0,
line width =2pt,
axis background/.style={fill=white},
xtick=\empty,
ytick=\empty
]
\addplot [color=mycolor1,solid,forget plot,very thick]
  table[row sep=crcr]{%
-1	0.00514092998763702\\
-0.97979797979798	0.00642009903340377\\
-0.95959595959596	0.00798127687034018\\
-0.939393939393939	0.00987719522093684\\
-0.919191919191919	0.0121681767630233\\
-0.898989898989899	0.014922720120256\\
-0.878787878787879	0.0182180164140247\\
-0.858585858585859	0.0221403659949995\\
-0.838383838383838	0.0267854603005296\\
-0.818181818181818	0.0322584906848667\\
-0.797979797979798	0.0386740438094208\\
-0.777777777777778	0.0461557420435137\\
-0.757575757575758	0.054835587600976\\
-0.737373737373737	0.0648529711033851\\
-0.717171717171717	0.0763533091689153\\
-0.696969696969697	0.0894862816829947\\
-0.676767676767677	0.104403647753532\\
-0.656565656565657	0.121256630043661\\
-0.636363636363636	0.140192870158084\\
-0.616161616161616	0.161352972863563\\
-0.595959595959596	0.184866673845496\\
-0.575757575757576	0.210848683997548\\
-0.555555555555556	0.239394282328685\\
-0.535353535353535	0.270574748741865\\
-0.515151515151515	0.30443274637019\\
-0.494949494949495	0.340977779945506\\
-0.474747474747475	0.380181870869257\\
-0.454545454545455	0.421975600300139\\
-0.434343434343434	0.466244677757922\\
-0.414141414141414	0.512827193657178\\
-0.393939393939394	0.561511709172504\\
-0.373737373737374	0.612036325448177\\
-0.353535353535353	0.664088856203459\\
-0.333333333333333	0.717308203343437\\
-0.313131313131313	0.77128700467323\\
-0.292929292929293	0.82557558696369\\
-0.272727272727273	0.8796872174791\\
-0.252525252525252	0.933104603991421\\
-0.232323232323232	0.985287548855602\\
-0.212121212121212	1.03568161867327\\
-0.191919191919192	1.08372764929946\\
-0.171717171717172	1.12887186833761\\
-0.151515151515151	1.1705763856312\\
-0.131313131313131	1.20832977823147\\
-0.111111111111111	1.241657481265\\
-0.0909090909090909	1.27013169106157\\
-0.0707070707070707	1.29338049243003\\
-0.0505050505050505	1.31109593822672\\
-0.0303030303030303	1.32304083599816\\
-0.0101010101010101	1.32905403266518\\
0.0101010101010102	1.32905403266518\\
0.0303030303030303	1.32304083599816\\
0.0505050505050506	1.31109593822672\\
0.0707070707070707	1.29338049243003\\
0.0909090909090908	1.27013169106157\\
0.111111111111111	1.241657481265\\
0.131313131313131	1.20832977823147\\
0.151515151515152	1.1705763856312\\
0.171717171717172	1.12887186833761\\
0.191919191919192	1.08372764929946\\
0.212121212121212	1.03568161867327\\
0.232323232323232	0.985287548855602\\
0.252525252525253	0.933104603991421\\
0.272727272727273	0.8796872174791\\
0.292929292929293	0.825575586963689\\
0.313131313131313	0.77128700467323\\
0.333333333333333	0.717308203343437\\
0.353535353535354	0.664088856203459\\
0.373737373737374	0.612036325448177\\
0.393939393939394	0.561511709172503\\
0.414141414141414	0.512827193657178\\
0.434343434343434	0.466244677757922\\
0.454545454545455	0.421975600300139\\
0.474747474747475	0.380181870869257\\
0.494949494949495	0.340977779945506\\
0.515151515151515	0.30443274637019\\
0.535353535353535	0.270574748741865\\
0.555555555555556	0.239394282328685\\
0.575757575757576	0.210848683997548\\
0.595959595959596	0.184866673845496\\
0.616161616161616	0.161352972863563\\
0.636363636363636	0.140192870158084\\
0.656565656565657	0.121256630043661\\
0.676767676767677	0.104403647753532\\
0.696969696969697	0.0894862816829947\\
0.717171717171717	0.0763533091689153\\
0.737373737373737	0.064852971103385\\
0.757575757575758	0.054835587600976\\
0.777777777777778	0.0461557420435137\\
0.797979797979798	0.0386740438094208\\
0.818181818181818	0.0322584906848667\\
0.838383838383838	0.0267854603005295\\
0.858585858585859	0.0221403659949995\\
0.878787878787879	0.0182180164140247\\
0.898989898989899	0.014922720120256\\
0.919191919191919	0.0121681767630233\\
0.939393939393939	0.00987719522093684\\
0.95959595959596	0.00798127687034018\\
0.97979797979798	0.00642009903340376\\
1	0.00514092998763702\\
};
\end{axis}


\end{tikzpicture}%
           \end{center}
           \end{subfigure}
           
           \begin{tikzpicture}[overlay]           
           \draw[-,line width=0.5mm,dashed,draw=black!50](-5,5.1)to[out=90,in=180]node[left]{}(4.3,8.6);
           \draw[-,line width=0.5mm,dashed,draw=black!50](-1.3,0.6)to[out=90,in=180]node[left]{}(4.3,8.6);          
           \draw[-,line width=0.5mm,dashed,draw=black!50](-3.5,5.1)to[out=90,in=180]node[left]{}(2.6,6.8);
           \draw[-,line width=0.5mm,dashed,draw=black!50](0.0,0.6)to[out=90,in=180]node[left]{}(2.6,6.8);           
           \end{tikzpicture}
           
           \begin{tikzpicture}[overlay]          
           \node[anchor=north,text width=4.0cm, text centered] at (-3.4,11.2) {\bf Model 1: $p(k_{1,1}| M_{1},\mathcal{D})$};
           \node[anchor=north,text width=5cm, text centered] at (2.4,11.2) {\bf Model 2: $p(k_{2,1},k_{2,2} | M_{2},\mathcal{D})$};          
           \node[anchor=north,text width=4.0cm, text centered] at (-4.0,2.5) {\bf Proposal: $q(u|k_{1,1})$};
           \end{tikzpicture}
         \begin{tikzpicture}[overlay]
         \node[anchor=north,text width=4.0cm, text centered] at (3.3,3.0) {\bf $(k_{2,1},k_{2,2})=f(k_{1,1},u)$};
         \end{tikzpicture}
           \caption{Efficient RJMCMC is achieved by ``aligning'' densities on $(k_{1,1},u)$ and $(k_{2,1},k_{2,2})$ via the choice of the map ${f}$ and the proposal $q(u \vert k_{1,1})$.}
           \label{RJMCMCdensities}  
}           
\end{figure} 

The discrete model-move proposal $q(M_n|M_m)$ is generally chosen so that every move adds or deletes one reaction. The selection of the jump function $f$ and the parameter proposals $q$ have the goal of improving the probability of accepting both the forward ($M_m \rightarrow M_n$) and reverse ($M_n \rightarrow M_m$) moves. In particular, higher between-model acceptance rates may be obtained by ``aligning'' the posterior and proposal parameter densities corresponding to the two models between which moves are proposed. As an example, consider moves between a one-dimensional model $M_{1}$ (with posterior parameter density $p(k_{1,1}|M_{1},\mathcal{D})$) and a two-dimensional model $M_{2}$ (with posterior parameter density $p(k_{2,1},k_{2,2}| M_{2},\mathcal{D})$) accomplished with proposal $q(u|k_{1,1})$ (Figure \ref{RJMCMCdensities}). By choosing the function $f$ and the shape of the proposal $q(u|k_{1,1})$ such that the regions of high density and low density in the two spaces $(k_{1,1},u)$ and $(k_{2,1},k_{2,2})$ \emph{map} to each other (formally, that $p(k_{1,1} \vert M_{1},\mathcal{D})q(u \vert k_{1,1})/\det \nabla f(k_{1,1},u)$ and $p(f(k_{1,1},u) \vert M_{2},\mathcal{D})$ have similar values), high between-model acceptance rates may be achieved. Intuitively, this construction attempts to choose $f$ and $q$ to make the acceptance rate close to one for all moves between the two spaces.
 
\section{Network analysis for improved sampling efficiency}
\label{sec:computation}
We now explain how we use the analysis of reaction network structure to design more effective across-model samplers. We propose three inter-related methods. First, we use the identification of effective networks to design better parameter proposals in across-model moves; we also explain why across-model samplers that do \textit{not} account for effective networks suffer in efficiency. Second, we use analysis of the observables' sensitivities to individual rate constants to further improve the design of between-model moves. Finally, we explain how analysis of network structure can be used to de-randomize posterior expectations already computed via RJMCMC, leading to further variance reduction.
 
\subsection{Centered Gaussian parameter proposals}
\label{sec:centering}
For nested models, as is the case in reaction network inference, a natural choice of jump function $f$ is simply the identity map. Thus, when proposing a move from a lower-dimensional model $M_m$ to a higher-dimensional model $M_n$, the rate constants of the newly added reactions are proposed according to $q(u \vert k_{m})$ and the values of the rate constants of reactions common to the two models are kept fixed. Suppose that model $M_m$ has $i$ reactions and that model $M_n$ has $a>i$ reactions, with the first $i$ reactions common; then an identity $f$ mapping $( {k}_{m,{1:i}}, {u} )$ to $( {k}_{n,{1:i}}, {k}_{n,{i+1:a}})$ is simply
\begin{equation}
{k}_{n,{1:i}} =  {k}_{m,{1:i}}, \  {k}_{n,{i+1:a}} = u
\end{equation}
and the acceptance probability is given by
\begin{equation}
\alpha(k_{m},k_{n})=\mbox{min}\left \{1, \frac{p(M_n, k_{n} \vert D) q(M_m \vert M_n)}{p(M_m, k_{m} \vert D) q(M_n \vert M_m)q(u|k_{m})}\right \}.
\label{identityAcceptanceRatio}
\end{equation}
The reverse move in this case is deterministic. 

Now let the proposal $q(u \vert k_{m})$ be a Gaussian distribution on $u$, with mean $\mu$ and covariance $\Sigma$. To improve the acceptance probability, we choose this Gaussian to approximate the conditional posterior density $p({k}_{n,{i+1:a}} \vert k_{m}, M_n, \mathcal{D})$. In particular, we choose $\mu$ to be the mode of the conditional posterior, and we set $-\Sigma$ to be the inverse Hessian of the logarithm of the conditional posterior density at $\mu$. This construction is equivalent to the centered second-order conditions in the framework of Brooks \textit{et al.}\ \cite{Brooks2003}.

\subsection{Network-aware parameter proposals}
\label{sec:networkaware}
The between-model moves described above, in which the parameter proposals adapt to conditional posterior densities, can produce efficient RJMCMC simulations in many settings \cite{Brooks2003,Godsill2001}. But their direct application to network inference, where many different networks can share the same effective network, presents a particular challenge. 


\tikzstyle{SPECIES1}=[draw, fill=blue!20, text width=0.2em, text centered, minimum height=0.1em]
\tikzstyle{SPECIES2}=[draw, fill=blue!20, text width=0.2em, text centered, minimum height=0.1em]
\tikzstyle{SPECIES3}=[draw, fill=green!20, text width=0.2em, text centered, minimum height=0.1em]    
\tikzstyle{SPECIES4}=[draw, fill=green!20, text width=0.2em, text centered, minimum height=0.1em]    
\tikzstyle{SPECIES5}=[draw, fill=green!20, text width=0.2em, text centered, minimum height=0.1em]
\tikzstyle{SPECIES6}=[draw, fill=blue!20, text width=0.2em, text centered, minimum height=0.1em]
\tikzstyle{SPECIES7}=[draw, fill=blue!20, text width=0.2em, text centered, minimum height=0.1em]
\tikzstyle{SPECIES8}=[draw, fill=red!20, text width=0.2em, text centered, minimum height=0.1em]
\tikzstyle{SPECIES9}=[draw, fill=blue!20, text width=0.2em, text centered, minimum height=0.1em]    
\tikzstyle{SPECIES10}=[draw, fill=blue!20, text width=0.2em, text centered, minimum height=0.1em]    
\tikzstyle{SPECIES11}=[draw, fill=green!20, text width=0.2em, text centered, minimum height=0.em]
\tikzstyle{SPECIES12}=[draw, fill=green!20, text width=0.2em, text centered, minimum height=0.1em]
\tikzstyle{SPECIES13}=[draw, fill=green!20, text width=0.2em, text centered, minimum height=0.1em]    
\tikzstyle{SPECIES14}=[draw, fill=blue!20, text width=0.2em, text centered, minimum height=0.1em]    
\tikzstyle{SPECIES15}=[draw, fill=green!20, text width=0.2em, text centered, minimum height=0.1em]
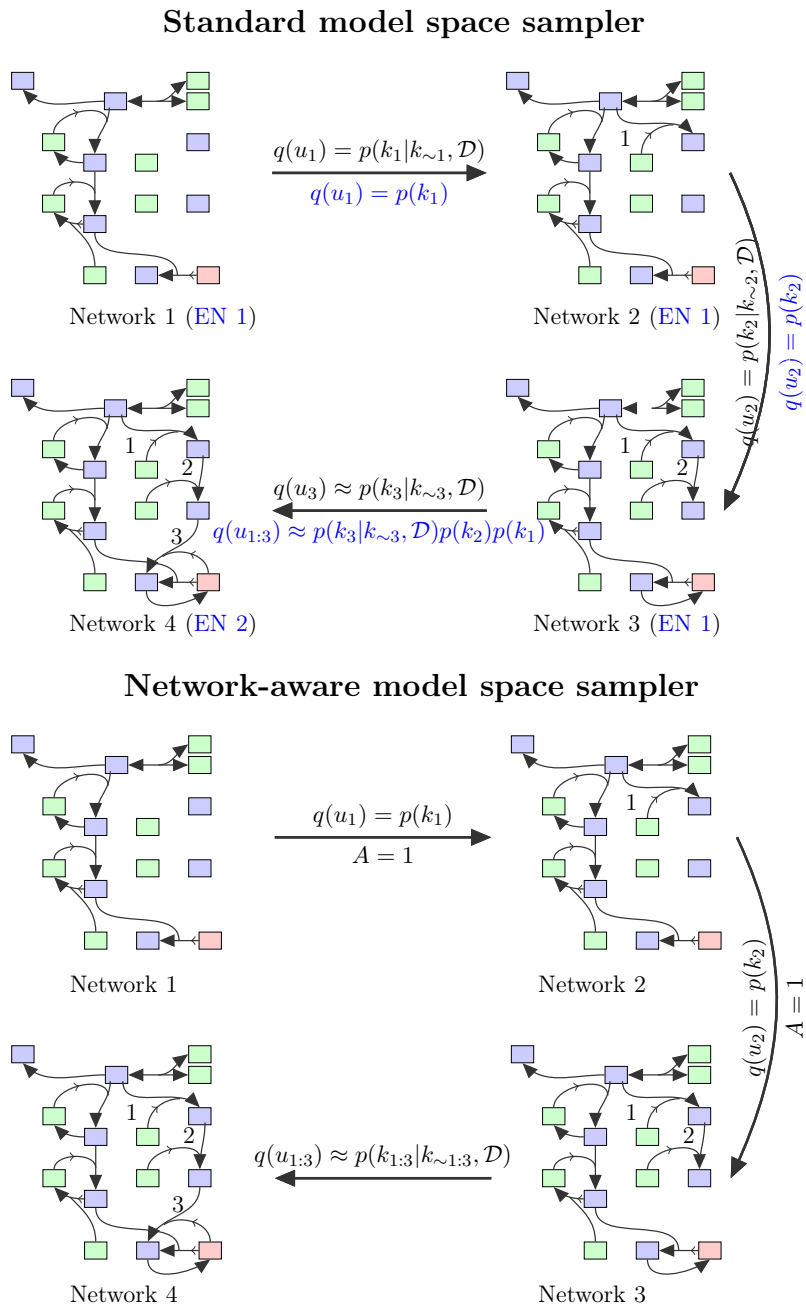
\begin{figure}[h]
\begin{center}
{\bf \hspace{2cm} Standard model space sampler}
\newline
\newline
\resizebox{0.7\textwidth}{!}{
\begin{tikzpicture}[]
\begin{scope}
\node[SPECIES1](ySPECIES1) at (0-1,2/6-1) {};
\node[SPECIES2](ySPECIES2) at (-9/6-1,4/6-1) {};
\node[SPECIES3](ySPECIES3) at (8/6-1,4/6-1) {};
\node[SPECIES4](ySPECIES4) at (8/6-1,2/6-1) {};
\node[SPECIES5](ySPECIES5) at (3/6-1,-4/6-1) {};
\node[SPECIES6](ySPECIES6) at (8/6-1,-2/6-1) {};
\node[SPECIES7](ySPECIES7) at (8/6-1,-8/6-1) {};
\node[SPECIES8](ySPECIES8) at (9/6-1, -15/6-1) {};
\node[SPECIES9](ySPECIES9) at (3/6-1, -15/6-1) {};
\node[SPECIES10](ySPECIES10) at (-2/6-1,-10/6-1) {};
\node[SPECIES11](ySPECIES11) at (-2/6-1,-15/6-1) {};
\node[SPECIES12](ySPECIES12) at (3/6-1,-8/6-1) {};
\node[SPECIES13](ySPECIES13) at (-6/6-1, -8/6-1) {};
\node[SPECIES14](ySPECIES14) at (-2/6-1,-4/6-1) {};
\node[SPECIES15](ySPECIES15) at (-6/6-1,-2/6-1) {};


\draw [->,line width=0.2mm,draw=black!80,-triangle 45](ySPECIES1.west)to[out=180,in=315]node[below]{}(ySPECIES2.south) ;

\draw [->,line width=0.2mm,draw=black!80,-triangle 45](4/6-1,2/6-1)to[out=0,in=215] node[above]{}(ySPECIES3.west);
\draw [->,line width=0.2mm,draw=black!80,-triangle 45](2.2/6-1,2/6-1)to[out=0,in=180]node[below]{}(ySPECIES4.west);
\draw [->,line width=0.2mm,draw=black!80,-triangle 45](2.2/6-1,2/6-1)to[out=180,in=0](ySPECIES1.east)node[below]{};

\draw[-,line width=0.2mm,draw=black!80](ySPECIES10.south)to[out=270,in=90]node[near start,right]{}(6/6-1,-15/6-1);
\draw[->,line width=0.2mm,draw=black!80,-triangle 45,decoration={markings,mark=at position 0.2 with {\arrow{>}}},postaction={decorate}](ySPECIES8.west)to[out=180,in=0]node[right]{}(ySPECIES9.east);

\draw[->,line width=0.2mm,draw=black!80,-triangle 45,decoration={markings,mark=at position 0.2 with {\arrow{>}}},postaction={decorate}](ySPECIES10.west)to[out=180,in=315]node[left]{}(ySPECIES13.south);
\draw[-,line width=0.2mm,draw=black!80](ySPECIES11.north)to[out=90,in=315]node[left]{}(-4.5/6-1,-9.8/6-1);

\draw [->, line width=0.2mm,draw=black!80,-triangle 45](ySPECIES14.south)to[out=270,in=90]node[right]{}(ySPECIES10.north);
\draw [-, line width=0.2mm,draw=black!80,decoration={markings,mark=at position 0.5 with {\arrow{>}}},postaction={decorate}](ySPECIES13.north)to[out=90,in=90]node[right]{}(-2/6-1,-7/6-1);

\draw [->, line width=0.2mm,draw=black!80,-triangle 45](ySPECIES14.west)to[out=180,in=315]node[left]{}(ySPECIES15.south);

\draw  [->,line width=0.2mm,draw=black!80,-triangle 45](-0.6/6-1,1.4/6-1)to[out=270,in=90] node[right]{}(ySPECIES14.north);
\draw  [-,line width=0.2mm,draw=black!80,decoration={markings,mark=at position 0.5 with {\arrow{>}}},postaction={decorate}](ySPECIES15.north)to[out=90,in=90] node[right]{}(-0.8/6-1,0-1);

\node[text width=3.5cm] at (1.0-1,-4.2) {\small Network 1 ({\color{blue} EN 1})};
{
\draw [->,line width=0.4mm,draw=black!80,-triangle 45](8/6+0.7-0.5,-5/6-1)to[out=0,in=180]node[above]{\small $q(u_{1})=p(k_{1}|k_{\sim 1},\mathcal{D})$}(5-0.5+0.5,-5/6-1);
}
{
\draw [->,line width=0.4mm,draw=black!80,-triangle 45](8/6+0.7-0.5,-5/6-1)to[out=0,in=180]node[below]{\color{blue} \small $q(u_{1})=p(k_{1})$}(5-0.5+0.5,-5/6-1);
}

{
\node[SPECIES1](ySPECIES1NE) at (0+7,2/6-1) {};
\node[SPECIES2](ySPECIES2NE) at (-9/6+7,4/6-1) {};
\node[SPECIES3](ySPECIES3NE) at (8/6+7,4/6-1) {};
\node[SPECIES4](ySPECIES4NE) at (8/6+7,2/6-1) {};
\node[SPECIES5](ySPECIES5NE) at (3/6+7,-4/6-1) {};
\node[SPECIES6](ySPECIES6NE) at (8/6+7,-2/6-1) {};
\node[SPECIES7](ySPECIES7NE) at (8/6+7,-8/6-1) {};
\node[SPECIES8](ySPECIES8NE) at (9/6+7, -15/6-1) {};
\node[SPECIES9](ySPECIES9NE) at (3/6+7, -15/6-1) {};
\node[SPECIES10](ySPECIES10NE) at (-2/6+7,-10/6-1) {};
\node[SPECIES11](ySPECIES11NE) at (-2/6+7,-15/6-1) {};
\node[SPECIES12](ySPECIES12NE) at (3/6+7,-8/6-1) {};
\node[SPECIES13](ySPECIES13NE) at (-6/6+7, -8/6-1) {};
\node[SPECIES14](ySPECIES14NE) at (-2/6+7,-4/6-1) {};
\node[SPECIES15](ySPECIES15NE) at (-6/6+7,-2/6-1) {};

\draw [->,line width=0.2mm,draw=black!80,-triangle 45](ySPECIES1NE.west)to[out=180,in=315]node[below]{}(ySPECIES2NE.south) ;

\draw [->,line width=0.2mm,draw=black!80,-triangle 45](4/6+7,2/6-1)to[out=0,in=215] node[above]{}(ySPECIES3NE.west);
\draw [->,line width=0.2mm,draw=black!80,-triangle 45](2.2/6+7,2/6-1)to[out=0,in=180]node[below]{}(ySPECIES4NE.west);
\draw [->,line width=0.2mm,draw=black!80,-triangle 45](2.2/6+7,2/6-1)to[out=180,in=0](ySPECIES1NE.east)node[below]{};

\draw[->,line width=0.2mm,draw=black!80,-triangle 45](0.6/6+7,1.4/6-1)to[out=270,in=135] node[near start,below]{\small 1}(ySPECIES6NE.north);
\draw[-,line width=0.2mm,draw=black!80,decoration={markings,mark=at position 0.5 with {\arrow{>}}},postaction={decorate}](ySPECIES5NE.north)to[out=90,in=180](6.1/6+7,-0.3/6-1) node[above]{}(6/6+7,2/6-1);

\draw[-,line width=0.2mm,draw=black!80](ySPECIES10NE.south)to[out=270,in=90]node[near start,right]{}(6/6+7,-15/6-1);
\draw[->,line width=0.2mm,draw=black!80,-triangle 45,decoration={markings,mark=at position 0.2 with {\arrow{>}}},postaction={decorate}](ySPECIES8NE.west)to[out=180,in=0]node[right]{}(ySPECIES9NE.east);

\draw[->,line width=0.2mm,draw=black!80,-triangle 45,decoration={markings,mark=at position 0.2 with {\arrow{>}}},postaction={decorate}](ySPECIES10NE.west)to[out=180,in=315]node[left]{}(ySPECIES13NE.south);
\draw[-,line width=0.2mm,draw=black!80](ySPECIES11NE.north)to[out=90,in=315]node[left]{}(-4.5/6+7,-9.8/6-1);

\draw [->, line width=0.2mm,draw=black!80,-triangle 45](ySPECIES14NE.south)to[out=270,in=90]node[right]{}(ySPECIES10NE.north);
\draw [-, line width=0.2mm,draw=black!80,decoration={markings,mark=at position 0.5 with {\arrow{>}}},postaction={decorate}](ySPECIES13NE.north)to[out=90,in=90]node[right]{}(-2/6+7,-7/6-1);

\draw [->, line width=0.2mm,draw=black!80,-triangle 45](ySPECIES14NE.west)to[out=180,in=315]node[left]{}(ySPECIES15NE.south);

\draw  [->,line width=0.2mm,draw=black!80,-triangle 45](-0.6/6+7,1.4/6-1)to[out=270,in=90] node[right]{}(ySPECIES14NE.north);
\draw  [-,line width=0.2mm,draw=black!80,decoration={markings,mark=at position 0.5 with {\arrow{>}}},postaction={decorate}](ySPECIES15NE.north)to[out=90,in=90] node[right]{}(-0.8/6+7,0-1);

\node[text width=3.5cm] at (1.0+6.5,-4.2) {\small Network 2 ({\color{blue} EN 1})};

}

{
\node[SPECIES1](ySPECIES1SE) at (0+7,2/6-6.0) {};
\node[SPECIES2](ySPECIES2SE) at (-9/6+7,4/6-6.0) {};
\node[SPECIES3](ySPECIES3SE) at (8/6+7,4/6-6.0) {};
\node[SPECIES4](ySPECIES4SE) at (8/6+7,2/6-6.0) {};
\node[SPECIES5](ySPECIES5SE) at (3/6+7,-4/6-6.0) {};
\node[SPECIES6](ySPECIES6SE) at (8/6+7,-2/6-6.0) {};
\node[SPECIES7](ySPECIES7SE) at (8/6+7,-8/6-6.0) {};
\node[SPECIES8](ySPECIES8SE) at (9/6+7, -15/6-6.0) {};
\node[SPECIES9](ySPECIES9SE) at (3/6+7, -15/6-6.0) {};
\node[SPECIES10](ySPECIES10SE) at (-2/6+7,-10/6-6.0) {};
\node[SPECIES11](ySPECIES11SE) at (-2/6+7,-15/6-6.0) {};
\node[SPECIES12](ySPECIES12SE) at (3/6+7,-8/6-6.0) {};
\node[SPECIES13](ySPECIES13SE) at (-6/6+7, -8/6-6.0) {};
\node[SPECIES14](ySPECIES14SE) at (-2/6+7,-4/6-6.0) {};
\node[SPECIES15](ySPECIES15SE) at (-6/6+7,-2/6-6.0) {};

\draw [->,line width=0.2mm,draw=black!80,-triangle 45](ySPECIES1SE.west)to[out=180,in=315]node[below]{}(ySPECIES2SE.south) ;

\draw [->,line width=0.2mm,draw=black!80,-triangle 45](4/6+7,2/6-6.0)to[out=0,in=215] node[above]{}(ySPECIES3SE.west);
\draw [->,line width=0.2mm,draw=black!80,-triangle 45](2.2/6+7.3,2/6-6.0)to[out=0,in=180]node[below]{}(ySPECIES4SE.west);
\draw [->,line width=0.2mm,draw=black!80,-triangle 45](2.2/6+7,2/6-6.0)to[out=180,in=0](ySPECIES1SE.east)node[below]{};

\draw [->,line width=0.2mm,draw=black!80,-triangle 45](0.6/6+7,1.4/6-6.0)to[out=270,in=135] node[near start,below]{\small 1}(ySPECIES6SE.north);
\draw [-,line width=0.2mm,draw=black!80,decoration={markings,mark=at position 0.5 with {\arrow{>}}},postaction={decorate}](ySPECIES5SE.north)to[out=90,in=180](6.1/6+7,-0.3/6-6.0) node[above]{}(6/6+7,2/6-6.0);


\draw[->,line width=0.2mm,draw=black!80,-triangle 45](8.5/6+7,-2.6/6-6.0)to[out=270,in=90]node[near start,left]{\small 2}(ySPECIES7SE.north);
\draw[-,line width=0.2mm,draw=black!80,decoration={markings,mark=at position 0.5 with {\arrow{>}}},postaction={decorate}](ySPECIES12SE.north)to[out=90,in=90]node[right]{}(8.05/6+7,-6.15/6-6.0);


\draw[->,line width=0.2mm,draw=black!80,-triangle 45](ySPECIES9SE.south)to[out=270,in=215]node[below]{}(ySPECIES8SE.south);

\draw[-,line width=0.2mm,draw=black!80](ySPECIES10SE.south)to[out=270,in=90]node[near start,right]{}(6/6+7,-15/6-6.0);
\draw[->,line width=0.2mm,draw=black!80,-triangle 45,decoration={markings,mark=at position 0.2 with {\arrow{>}}},postaction={decorate}](ySPECIES8SE.west)to[out=180,in=0]node[right]{}(ySPECIES9SE.east);


\draw[->,line width=0.2mm,draw=black!80,-triangle 45,decoration={markings,mark=at position 0.2 with {\arrow{>}}},postaction={decorate}](ySPECIES10SE.west)to[out=180,in=315]node[left]{}(ySPECIES13SE.south);
\draw[-,line width=0.2mm,draw=black!80](ySPECIES11SE.north)to[out=90,in=315]node[left]{}(-4.5/6+7,-9.8/6-6.0);

\draw [->, line width=0.2mm,draw=black!80,-triangle 45](ySPECIES14SE.south)to[out=270,in=90]node[right]{}(ySPECIES10SE.north);
\draw [-, line width=0.2mm,draw=black!80,decoration={markings,mark=at position 0.5 with {\arrow{>}}},postaction={decorate}](ySPECIES13SE.north)to[out=90,in=90]node[right]{}(-2/6+7,-7/6-6.0);

\draw [->, line width=0.2mm,draw=black!80,-triangle 45](ySPECIES14SE.west)to[out=180,in=315]node[left]{}(ySPECIES15SE.south);

\draw  [->,line width=0.2mm,draw=black!80,-triangle 45](-0.6/6+7,1.4/6-6.0)to[out=270,in=90] node[right]{}(ySPECIES14SE.north);
\draw  [-,line width=0.2mm,draw=black!80,decoration={markings,mark=at position 0.5 with {\arrow{>}}},postaction={decorate}](ySPECIES15SE.north)to[out=90,in=90] node[right]{}(-0.8/6+7,0-6.0);

\node[text width=3.5cm] at (1.0+6.5,-9.2) {\small Network 3 ({\color{blue} EN 1})};
\draw [->,line width=0.4mm,draw=black!80,-triangle 45](8/6+6.5+1.1,-5/6-1)to[out=295,in=65]node[left]{\small \rotatebox{90} {$q(u_{2})=p(k_{2}|k_{\sim 2},\mathcal{D})$}}(8/6+6.5+1.0,-5/6-6.5);
\draw [->,line width=0.4mm,draw=black!80,-triangle 45](8/6+6.5+1.1,-5/6-1)to[out=295,in=65]node[right]{\color{blue} \small \rotatebox{90} {$q(u_{2})=p(k_{2})$}}(8/6+6.5+1.0,-5/6-6.5);
\draw [->,line width=0.4mm,draw=black!80,-triangle 45](8/6+6.5+1.1,-5/6-1)to[out=295,in=65]node[right]{\small \rotatebox{90} {\color{white} A=1}}(8/6+6.5+1.0,-5/6-6.5);
}

{
\node[SPECIES1](ySPECIES1SW) at (0-1,2/6-6.0) {};
\node[SPECIES2](ySPECIES2SW) at (-9/6-1,4/6-6.0) {};
\node[SPECIES3](ySPECIES3SW) at (8/6-1,4/6-6.0) {};
\node[SPECIES4](ySPECIES4SW) at (8/6-1,2/6-6.0) {};
\node[SPECIES5](ySPECIES5SW) at (3/6-1,-4/6-6.0) {};
\node[SPECIES6](ySPECIES6SW) at (8/6-1,-2/6-6.0) {};
\node[SPECIES7](ySPECIES7SW) at (8/6-1,-8/6-6.0) {};
\node[SPECIES8](ySPECIES8SW) at (9/6-1, -15/6-6.0) {};
\node[SPECIES9](ySPECIES9SW) at (3/6-1, -15/6-6.0) {};
\node[SPECIES10](ySPECIES10SW) at (-2/6-1,-10/6-6.0) {};
\node[SPECIES11](ySPECIES11SW) at (-2/6-1,-15/6-6.0) {};
\node[SPECIES12](ySPECIES12SW) at (3/6-1,-8/6-6.0) {};
\node[SPECIES13](ySPECIES13SW) at (-6/6-1, -8/6-6.0) {};
\node[SPECIES14](ySPECIES14SW) at (-2/6-1,-4/6-6.0) {};
\node[SPECIES15](ySPECIES15SW) at (-6/6-1,-2/6-6.0) {};

\draw [->,line width=0.2mm,draw=black!80,-triangle 45](ySPECIES1SW.west)to[out=180,in=315]node[below]{}(ySPECIES2SW.south) ;

\draw [->,line width=0.2mm,draw=black!80,-triangle 45](4/6-1,2/6-6.0)to[out=0,in=215] node[above]{}(ySPECIES3SW.west);
\draw [->,line width=0.2mm,draw=black!80,-triangle 45](2.2/6-1,2/6-6.0)to[out=0,in=180]node[below]{}(ySPECIES4SW.west);
\draw [->,line width=0.2mm,draw=black!80,-triangle 45](2.2/6-1,2/6-6.0)to[out=180,in=0](ySPECIES1SW.east)node[below]{};

\draw [->,line width=0.2mm,draw=black!80,-triangle 45](0.6/6-1,1.4/6-6.0)to[out=270,in=135] node[near start, below]{\small 1}(ySPECIES6SW.north);
\draw [-,line width=0.2mm,draw=black!80,decoration={markings,mark=at position 0.5 with {\arrow{>}}},postaction={decorate}](ySPECIES5SW.north)to[out=90,in=180](6.1/6-1,-0.3/6-6.0) node[above]{}(6/6-1,2/6-6.0);


\draw[->,line width=0.2mm,draw=black!80,-triangle 45](8.5/6-1,-2.6/6-6.0)to[out=270,in=90]node[near start,left]{\small 2}(ySPECIES7SW.north);
\draw[-,line width=0.2mm,draw=black!80,decoration={markings,mark=at position 0.5 with {\arrow{>}}},postaction={decorate}](ySPECIES12SW.north)to[out=90,in=90]node[right]{}(8.05/6-1,-6.15/6-6.0);

\draw[->,line width=0.2mm,draw=black!80,-triangle 45](ySPECIES7SW.south)to[out=270,in=65]node[near start,left]{\small 3}(ySPECIES9SW.north);
\draw[-,line width=0.2mm,draw=black!80,decoration={markings,mark=at position 0.5 with {\arrow{>}}},postaction={decorate}](ySPECIES8SW.north)to[out=90,in=35]node[right]{}(4.55/6-1,-12.5/6-6.0);

\draw[->,line width=0.2mm,draw=black!80,-triangle 45](ySPECIES9SW.south)to[out=270,in=215]node[below]{}(ySPECIES8SW.south);

\draw[-,line width=0.2mm,draw=black!80](ySPECIES10SW.south)to[out=270,in=90]node[near start,right]{}(6/6-1,-15/6-6.0);
\draw[->,line width=0.2mm,draw=black!80,-triangle 45,decoration={markings,mark=at position 0.2 with {\arrow{>}}},postaction={decorate}](ySPECIES8SW.west)to[out=180,in=0]node[right]{}(ySPECIES9SW.east);


\draw[->,line width=0.2mm,draw=black!80,-triangle 45,decoration={markings,mark=at position 0.2 with {\arrow{>}}},postaction={decorate}](ySPECIES10SW.west)to[out=180,in=315]node[left]{}(ySPECIES13SW.south);
\draw[-,line width=0.2mm,draw=black!80](ySPECIES11SW.north)to[out=90,in=315]node[left]{}(-4.5/6-1,-9.8/6-6.0);

\draw [->, line width=0.2mm,draw=black!80,-triangle 45](ySPECIES14SW.south)to[out=270,in=90]node[right]{}(ySPECIES10SW.north);
\draw [-, line width=0.2mm,draw=black!80,decoration={markings,mark=at position 0.5 with {\arrow{>}}},postaction={decorate}](ySPECIES13SW.north)to[out=90,in=90]node[right]{}(-2/6-1,-7/6-6.0);

\draw [->, line width=0.2mm,draw=black!80,-triangle 45](ySPECIES14SW.west)to[out=180,in=315]node[left]{}(ySPECIES15SW.south);

\draw  [->,line width=0.2mm,draw=black!80,-triangle 45](-0.6/6-1,1.4/6-6.0)to[out=270,in=90] node[right]{}(ySPECIES14SW.north);
\draw  [-,line width=0.2mm,draw=black!80,decoration={markings,mark=at position 0.5 with {\arrow{>}}},postaction={decorate}](ySPECIES15SW.north)to[out=90,in=90] node[right]{}(-0.8/6-1,0-6.0);

\node[text width=3.5cm] at (1.0-1,-9.2) {\small Network 4 ({\color{blue} EN 2})};
\draw [->,line width=0.4mm,draw=black!80,-triangle 45](5-0.5+0.5,-5/6-6.5)to[out=180,in=0]node[above]{{\bf \small $q(u_{3}) \approx p(k_{3} \vert k_{\sim 3},\mathcal{D})$}}(8/6+0.7-0.5,-5/6-6.5);
}

{
\draw [->,line width=0.4mm,draw=black!80,-triangle 45](5-0.5+0.5,-5/6-6.5)to[out=180,in=0]node[below]{{\color{blue} \bf \small $q(u_{1:3}) \approx p(k_{3} \vert k_{\sim 3},\mathcal{D})p(k_{2})p(k_{1})$}}(8/6+0.7-0.5,-5/6-6.5);}

\end{scope}
\end{tikzpicture}}
\end{center}

\label{movesinmodelspace}

\tikzstyle{SPECIES1}=[draw, fill=blue!20, text width=0.2em, text centered, minimum height=0.1em]
\tikzstyle{SPECIES2}=[draw, fill=blue!20, text width=0.2em, text centered, minimum height=0.1em]
\tikzstyle{SPECIES3}=[draw, fill=green!20, text width=0.2em, text centered, minimum height=0.1em]    
\tikzstyle{SPECIES4}=[draw, fill=green!20, text width=0.2em, text centered, minimum height=0.1em]    
\tikzstyle{SPECIES5}=[draw, fill=green!20, text width=0.2em, text centered, minimum height=0.1em]
\tikzstyle{SPECIES6}=[draw, fill=blue!20, text width=0.2em, text centered, minimum height=0.1em]
\tikzstyle{SPECIES7}=[draw, fill=blue!20, text width=0.2em, text centered, minimum height=0.1em]
\tikzstyle{SPECIES8}=[draw, fill=red!20, text width=0.2em, text centered, minimum height=0.1em]
\tikzstyle{SPECIES9}=[draw, fill=blue!20, text width=0.2em, text centered, minimum height=0.1em]    
\tikzstyle{SPECIES10}=[draw, fill=blue!20, text width=0.2em, text centered, minimum height=0.1em]    
\tikzstyle{SPECIES11}=[draw, fill=green!20, text width=0.2em, text centered, minimum height=0.em]
\tikzstyle{SPECIES12}=[draw, fill=green!20, text width=0.2em, text centered, minimum height=0.1em]
\tikzstyle{SPECIES13}=[draw, fill=green!20, text width=0.2em, text centered, minimum height=0.1em]    
\tikzstyle{SPECIES14}=[draw, fill=blue!20, text width=0.2em, text centered, minimum height=0.1em]    
\tikzstyle{SPECIES15}=[draw, fill=green!20, text width=0.2em, text centered, minimum height=0.1em]

\begin{center}
{\bf \hspace{2cm}Network-aware model space sampler}
\newline
\newline
\resizebox{0.7\textwidth}{!}{
\begin{tikzpicture}[]
\begin{scope}
\node[SPECIES1](ySPECIES1) at (0-1,2/6-1) {};
\node[SPECIES2](ySPECIES2) at (-9/6-1,4/6-1) {};
\node[SPECIES3](ySPECIES3) at (8/6-1,4/6-1) {};
\node[SPECIES4](ySPECIES4) at (8/6-1,2/6-1) {};
\node[SPECIES5](ySPECIES5) at (3/6-1,-4/6-1) {};
\node[SPECIES6](ySPECIES6) at (8/6-1,-2/6-1) {};
\node[SPECIES7](ySPECIES7) at (8/6-1,-8/6-1) {};
\node[SPECIES8](ySPECIES8) at (9/6-1, -15/6-1) {};
\node[SPECIES9](ySPECIES9) at (3/6-1, -15/6-1) {};
\node[SPECIES10](ySPECIES10) at (-2/6-1,-10/6-1) {};
\node[SPECIES11](ySPECIES11) at (-2/6-1,-15/6-1) {};
\node[SPECIES12](ySPECIES12) at (3/6-1,-8/6-1) {};
\node[SPECIES13](ySPECIES13) at (-6/6-1, -8/6-1) {};
\node[SPECIES14](ySPECIES14) at (-2/6-1,-4/6-1) {};
\node[SPECIES15](ySPECIES15) at (-6/6-1,-2/6-1) {};





\draw [->,line width=0.2mm,draw=black!80,-triangle 45](ySPECIES1.west)to[out=180,in=315]node[below]{}(ySPECIES2.south) ;

\draw [->,line width=0.2mm,draw=black!80,-triangle 45](4/6-1,2/6-1)to[out=0,in=215] node[above]{}(ySPECIES3.west);
\draw [->,line width=0.2mm,draw=black!80,-triangle 45](2.2/6-1,2/6-1)to[out=0,in=180]node[below]{}(ySPECIES4.west);
\draw [->,line width=0.2mm,draw=black!80,-triangle 45](2.2/6-1,2/6-1)to[out=180,in=0](ySPECIES1.east)node[below]{};






\draw[-,line width=0.2mm,draw=black!80](ySPECIES10.south)to[out=270,in=90]node[near start,right]{}(6/6-1,-15/6-1);
\draw[->,line width=0.2mm,draw=black!80,-triangle 45,decoration={markings,mark=at position 0.2 with {\arrow{>}}},postaction={decorate}](ySPECIES8.west)to[out=180,in=0]node[right]{}(ySPECIES9.east);


\draw[->,line width=0.2mm,draw=black!80,-triangle 45,decoration={markings,mark=at position 0.2 with {\arrow{>}}},postaction={decorate}](ySPECIES10.west)to[out=180,in=315]node[left]{}(ySPECIES13.south);
\draw[-,line width=0.2mm,draw=black!80](ySPECIES11.north)to[out=90,in=315]node[left]{}(-4.5/6-1,-9.8/6-1);

\draw [->, line width=0.2mm,draw=black!80,-triangle 45](ySPECIES14.south)to[out=270,in=90]node[right]{}(ySPECIES10.north);
\draw [-, line width=0.2mm,draw=black!80,decoration={markings,mark=at position 0.5 with {\arrow{>}}},postaction={decorate}](ySPECIES13.north)to[out=90,in=90]node[right]{}(-2/6-1,-7/6-1);

\draw [->, line width=0.2mm,draw=black!80,-triangle 45](ySPECIES14.west)to[out=180,in=315]node[left]{}(ySPECIES15.south);

\draw  [->,line width=0.2mm,draw=black!80,-triangle 45](-0.6/6-1,1.4/6-1)to[out=270,in=90] node[right]{}(ySPECIES14.north);
\draw  [-,line width=0.2mm,draw=black!80,decoration={markings,mark=at position 0.5 with {\arrow{>}}},postaction={decorate}](ySPECIES15.north)to[out=90,in=90] node[right]{}(-0.8/6-1,0-1);

\node[text width=3.5cm] at (1.0-1,-4.2) {\small Network 1};

\node[SPECIES1](ySPECIES1NE) at (0+7,2/6-1) {};
\node[SPECIES2](ySPECIES2NE) at (-9/6+7,4/6-1) {};
\node[SPECIES3](ySPECIES3NE) at (8/6+7,4/6-1) {};
\node[SPECIES4](ySPECIES4NE) at (8/6+7,2/6-1) {};
\node[SPECIES5](ySPECIES5NE) at (3/6+7,-4/6-1) {};
\node[SPECIES6](ySPECIES6NE) at (8/6+7,-2/6-1) {};
\node[SPECIES7](ySPECIES7NE) at (8/6+7,-8/6-1) {};
\node[SPECIES8](ySPECIES8NE) at (9/6+7, -15/6-1) {};
\node[SPECIES9](ySPECIES9NE) at (3/6+7, -15/6-1) {};
\node[SPECIES10](ySPECIES10NE) at (-2/6+7,-10/6-1) {};
\node[SPECIES11](ySPECIES11NE) at (-2/6+7,-15/6-1) {};
\node[SPECIES12](ySPECIES12NE) at (3/6+7,-8/6-1) {};
\node[SPECIES13](ySPECIES13NE) at (-6/6+7, -8/6-1) {};
\node[SPECIES14](ySPECIES14NE) at (-2/6+7,-4/6-1) {};
\node[SPECIES15](ySPECIES15NE) at (-6/6+7,-2/6-1) {};

\draw [->,line width=0.2mm,draw=black!80,-triangle 45](ySPECIES1NE.west)to[out=180,in=315]node[below]{}(ySPECIES2NE.south) ;

\draw [->,line width=0.2mm,draw=black!80,-triangle 45](4/6+7,2/6-1)to[out=0,in=215] node[above]{}(ySPECIES3NE.west);
\draw [->,line width=0.2mm,draw=black!80,-triangle 45](2.2/6+7,2/6-1)to[out=0,in=180]node[below]{}(ySPECIES4NE.west);
\draw [->,line width=0.2mm,draw=black!80,-triangle 45](2.2/6+7,2/6-1)to[out=180,in=0](ySPECIES1NE.east)node[below]{};

\draw [->,line width=0.2mm,draw=black!80,-triangle 45](0.6/6+7,1.4/6-1)to[out=270,in=135] node[near start,below]{\small 1}(ySPECIES6NE.north);
\draw [-,line width=0.2mm,draw=black!80,decoration={markings,mark=at position 0.5 with {\arrow{>}}},postaction={decorate}](ySPECIES5NE.north)to[out=90,in=180](6.1/6+7,-0.3/6-1) node[above]{}(6/6+7,2/6-1);





\draw[-,line width=0.2mm,draw=black!80](ySPECIES10NE.south)to[out=270,in=90]node[near start,right]{}(6/6+7,-15/6-1);
\draw[->,line width=0.2mm,draw=black!80,-triangle 45,decoration={markings,mark=at position 0.2 with {\arrow{>}}},postaction={decorate}](ySPECIES8NE.west)to[out=180,in=0]node[right]{}(ySPECIES9NE.east);


\draw[->,line width=0.2mm,draw=black!80,-triangle 45,decoration={markings,mark=at position 0.2 with {\arrow{>}}},postaction={decorate}](ySPECIES10NE.west)to[out=180,in=315]node[left]{}(ySPECIES13NE.south);
\draw[-,line width=0.2mm,draw=black!80](ySPECIES11NE.north)to[out=90,in=315]node[left]{}(-4.5/6+7,-9.8/6-1);

\draw [->, line width=0.2mm,draw=black!80,-triangle 45](ySPECIES14NE.south)to[out=270,in=90]node[right]{}(ySPECIES10NE.north);
\draw [-, line width=0.2mm,draw=black!80,decoration={markings,mark=at position 0.5 with {\arrow{>}}},postaction={decorate}](ySPECIES13NE.north)to[out=90,in=90]node[right]{}(-2/6+7,-7/6-1);

\draw [->, line width=0.2mm,draw=black!80,-triangle 45](ySPECIES14NE.west)to[out=180,in=315]node[left]{}(ySPECIES15NE.south);

\draw  [->,line width=0.2mm,draw=black!80,-triangle 45](-0.6/6+7,1.4/6-1)to[out=270,in=90] node[right]{}(ySPECIES14NE.north);
\draw  [-,line width=0.2mm,draw=black!80,decoration={markings,mark=at position 0.5 with {\arrow{>}}},postaction={decorate}](ySPECIES15NE.north)to[out=90,in=90] node[right]{}(-0.8/6+7,0-1);

\node[text width=3.5cm] at (1.0+6.5,-4.2) {\small Network 2};
\draw [->,line width=0.4mm,draw=black!80,-triangle 45](8/6+0.7-0.5,-5/6-1)to[out=0,in=180]node[above]{\small $q(u_{1})=p(k_{1})$}(5-0.5+0.5,-5/6-1);
\draw [->,line width=0.4mm,draw=black!80,-triangle 45](8/6+0.7-0.5,-5/6-1)to[out=0,in=180]node[below]{\small $A=1$}(5-0.5+0.5,-5/6-1);

\node[SPECIES1](ySPECIES1SE) at (0+7,2/6-6.0) {};
\node[SPECIES2](ySPECIES2SE) at (-9/6+7,4/6-6.0) {};
\node[SPECIES3](ySPECIES3SE) at (8/6+7,4/6-6.0) {};
\node[SPECIES4](ySPECIES4SE) at (8/6+7,2/6-6.0) {};
\node[SPECIES5](ySPECIES5SE) at (3/6+7,-4/6-6.0) {};
\node[SPECIES6](ySPECIES6SE) at (8/6+7,-2/6-6.0) {};
\node[SPECIES7](ySPECIES7SE) at (8/6+7,-8/6-6.0) {};
\node[SPECIES8](ySPECIES8SE) at (9/6+7, -15/6-6.0) {};
\node[SPECIES9](ySPECIES9SE) at (3/6+7, -15/6-6.0) {};
\node[SPECIES10](ySPECIES10SE) at (-2/6+7,-10/6-6.0) {};
\node[SPECIES11](ySPECIES11SE) at (-2/6+7,-15/6-6.0) {};
\node[SPECIES12](ySPECIES12SE) at (3/6+7,-8/6-6.0) {};
\node[SPECIES13](ySPECIES13SE) at (-6/6+7, -8/6-6.0) {};
\node[SPECIES14](ySPECIES14SE) at (-2/6+7,-4/6-6.0) {};
\node[SPECIES15](ySPECIES15SE) at (-6/6+7,-2/6-6.0) {};

\draw [->,line width=0.2mm,draw=black!80,-triangle 45](ySPECIES1SE.west)to[out=180,in=315]node[below]{}(ySPECIES2SE.south) ;

\draw [->,line width=0.2mm,draw=black!80,-triangle 45](4/6+7,2/6-6.0)to[out=0,in=215] node[above]{}(ySPECIES3SE.west);
\draw [->,line width=0.2mm,draw=black!80,-triangle 45](2.2/6+7,2/6-6.0)to[out=0,in=180]node[below]{}(ySPECIES4SE.west);
\draw [->,line width=0.2mm,draw=black!80,-triangle 45](2.2/6+7,2/6-6.0)to[out=180,in=0](ySPECIES1SE.east)node[below]{};

\draw[->,line width=0.2mm,draw=black!80,-triangle 45](0.6/6+7,1.4/6-6.0)to[out=270,in=135] node[near start,below]{\small 1}(ySPECIES6SE.north);
\draw[-,line width=0.2mm,draw=black!80,decoration={markings,mark=at position 0.5 with {\arrow{>}}},postaction={decorate}](ySPECIES5SE.north)to[out=90,in=180](6.1/6+7,-0.3/6-6.0) node[above]{}(6/6+7,2/6-6.0);


\draw[->,line width=0.2mm,draw=black!80,-triangle 45](8.5/6+7,-2.6/6-6.0)to[out=270,in=90]node[near start,left]{\small 2}(ySPECIES7SE.north);
\draw[-,line width=0.2mm,draw=black!80,decoration={markings,mark=at position 0.5 with {\arrow{>}}},postaction={decorate}](ySPECIES12SE.north)to[out=90,in=90]node[right]{}(8.05/6+7,-6.15/6-6.0);


\draw[->,line width=0.2mm,draw=black!80,-triangle 45](ySPECIES9SE.south)to[out=270,in=215]node[below]{}(ySPECIES8SE.south);

\draw[-,line width=0.2mm,draw=black!80](ySPECIES10SE.south)to[out=270,in=90]node[near start,right]{}(6/6+7,-15/6-6.0);
\draw[->,line width=0.2mm,draw=black!80,-triangle 45,decoration={markings,mark=at position 0.2 with {\arrow{>}}},postaction={decorate}](ySPECIES8SE.west)to[out=180,in=0]node[right]{}(ySPECIES9SE.east);


\draw[->,line width=0.2mm,draw=black!80,-triangle 45,decoration={markings,mark=at position 0.2 with {\arrow{>}}},postaction={decorate}](ySPECIES10SE.west)to[out=180,in=315]node[left]{}(ySPECIES13SE.south);
\draw[-,line width=0.2mm,draw=black!80](ySPECIES11SE.north)to[out=90,in=315]node[left]{}(-4.5/6+7,-9.8/6-6.0);

\draw [->, line width=0.2mm,draw=black!80,-triangle 45](ySPECIES14SE.south)to[out=270,in=90]node[right]{}(ySPECIES10SE.north);
\draw [-, line width=0.2mm,draw=black!80,decoration={markings,mark=at position 0.5 with {\arrow{>}}},postaction={decorate}](ySPECIES13SE.north)to[out=90,in=90]node[right]{}(-2/6+7,-7/6-6.0);

\draw [->, line width=0.2mm,draw=black!80,-triangle 45](ySPECIES14SE.west)to[out=180,in=315]node[left]{}(ySPECIES15SE.south);

\draw  [->,line width=0.2mm,draw=black!80,-triangle 45](-0.6/6+7,1.4/6-6.0)to[out=270,in=90] node[right]{}(ySPECIES14SE.north);
\draw  [-,line width=0.2mm,draw=black!80,decoration={markings,mark=at position 0.5 with {\arrow{>}}},postaction={decorate}](ySPECIES15SE.north)to[out=90,in=90] node[right]{}(-0.8/6+7,0-6.0);

\node[text width=3.5cm] at (1.0+6.5,-9.2) {\small Network 3};
\draw [->,line width=0.4mm,draw=black!80,-triangle 45](8/6+6.5+1.1,-5/6-1)to[out=295,in=65]node[left]{\small \rotatebox{90} {$q(u_{2})=p(k_{2})$}}(8/6+6.5+1.0,-5/6-6.5);
\draw [->,line width=0.4mm,draw=black!80,-triangle 45](8/6+6.5+1.1,-5/6-1)to[out=295,in=65]node[right]{\small \rotatebox{90} {$A=1$}}(8/6+6.5+1.0,-5/6-6.5);

\node[SPECIES1](ySPECIES1SW) at (0-1,2/6-6.0) {};
\node[SPECIES2](ySPECIES2SW) at (-9/6-1,4/6-6.0) {};
\node[SPECIES3](ySPECIES3SW) at (8/6-1,4/6-6.0) {};
\node[SPECIES4](ySPECIES4SW) at (8/6-1,2/6-6.0) {};
\node[SPECIES5](ySPECIES5SW) at (3/6-1,-4/6-6.0) {};
\node[SPECIES6](ySPECIES6SW) at (8/6-1,-2/6-6.0) {};
\node[SPECIES7](ySPECIES7SW) at (8/6-1,-8/6-6.0) {};
\node[SPECIES8](ySPECIES8SW) at (9/6-1, -15/6-6.0) {};
\node[SPECIES9](ySPECIES9SW) at (3/6-1, -15/6-6.0) {};
\node[SPECIES10](ySPECIES10SW) at (-2/6-1,-10/6-6.0) {};
\node[SPECIES11](ySPECIES11SW) at (-2/6-1,-15/6-6.0) {};
\node[SPECIES12](ySPECIES12SW) at (3/6-1,-8/6-6.0) {};
\node[SPECIES13](ySPECIES13SW) at (-6/6-1, -8/6-6.0) {};
\node[SPECIES14](ySPECIES14SW) at (-2/6-1,-4/6-6.0) {};
\node[SPECIES15](ySPECIES15SW) at (-6/6-1,-2/6-6.0) {};

\draw [->,line width=0.2mm,draw=black!80,-triangle 45](ySPECIES1SW.west)to[out=180,in=315]node[below]{}(ySPECIES2SW.south) ;

\draw [->,line width=0.2mm,draw=black!80,-triangle 45](4/6-1,2/6-6.0)to[out=0,in=215] node[above]{}(ySPECIES3SW.west);
\draw [->,line width=0.2mm,draw=black!80,-triangle 45](2.2/6-1,2/6-6.0)to[out=0,in=180]node[below]{}(ySPECIES4SW.west);
\draw [->,line width=0.2mm,draw=black!80,-triangle 45](2.2/6-1,2/6-6.0)to[out=180,in=0](ySPECIES1SW.east)node[below]{};

\draw [->,line width=0.2mm,draw=black!80,-triangle 45](0.6/6-1,1.4/6-6.0)to[out=270,in=135] node[near start, below]{\small 1}(ySPECIES6SW.north);
\draw [-,line width=0.2mm,draw=black!80,decoration={markings,mark=at position 0.5 with {\arrow{>}}},postaction={decorate}](ySPECIES5SW.north)to[out=90,in=180](6.1/6-1,-0.3/6-6.0) node[above]{}(6/6-1,2/6-6.0);


\draw[->,line width=0.2mm,draw=black!80,-triangle 45](8.5/6-1,-2.6/6-6.0)to[out=270,in=90]node[near start,left]{\small 2}(ySPECIES7SW.north);
\draw[-,line width=0.2mm,draw=black!80,decoration={markings,mark=at position 0.5 with {\arrow{>}}},postaction={decorate}](ySPECIES12SW.north)to[out=90,in=90]node[right]{}(8.05/6-1,-6.15/6-6.0);

\draw[->,line width=0.2mm,draw=black!80,-triangle 45](ySPECIES7SW.south)to[out=270,in=65]node[near start,left]{\small 3}(ySPECIES9SW.north);
\draw[-,line width=0.2mm,draw=black!80,decoration={markings,mark=at position 0.5 with {\arrow{>}}},postaction={decorate}](ySPECIES8SW.north)to[out=90,in=35]node[right]{}(4.55/6-1,-12.5/6-6.0);

\draw[->,line width=0.2mm,draw=black!80,-triangle 45](ySPECIES9SW.south)to[out=270,in=215]node[below]{}(ySPECIES8SW.south);

\draw[-,line width=0.2mm,draw=black!80](ySPECIES10SW.south)to[out=270,in=90]node[near start,right]{}(6/6-1,-15/6-6.0);
\draw[->,line width=0.2mm,draw=black!80,-triangle 45,decoration={markings,mark=at position 0.2 with {\arrow{>}}},postaction={decorate}](ySPECIES8SW.west)to[out=180,in=0]node[right]{}(ySPECIES9SW.east);


\draw[->,line width=0.2mm,draw=black!80,-triangle 45,decoration={markings,mark=at position 0.2 with {\arrow{>}}},postaction={decorate}](ySPECIES10SW.west)to[out=180,in=315]node[left]{}(ySPECIES13SW.south);
\draw[-,line width=0.2mm,draw=black!80](ySPECIES11SW.north)to[out=90,in=315]node[left]{}(-4.5/6-1,-9.8/6-6.0);

\draw [->, line width=0.2mm,draw=black!80,-triangle 45](ySPECIES14SW.south)to[out=270,in=90]node[right]{}(ySPECIES10SW.north);
\draw [-, line width=0.2mm,draw=black!80,decoration={markings,mark=at position 0.5 with {\arrow{>}}},postaction={decorate}](ySPECIES13SW.north)to[out=90,in=90]node[right]{}(-2/6-1,-7/6-6.0);

\draw [->, line width=0.2mm,draw=black!80,-triangle 45](ySPECIES14SW.west)to[out=180,in=315]node[left]{}(ySPECIES15SW.south);

\draw  [->,line width=0.2mm,draw=black!80,-triangle 45](-0.6/6-1,1.4/6-6.0)to[out=270,in=90] node[right]{}(ySPECIES14SW.north);
\draw  [-,line width=0.2mm,draw=black!80,decoration={markings,mark=at position 0.5 with {\arrow{>}}},postaction={decorate}](ySPECIES15SW.north)to[out=90,in=90] node[right]{}(-0.8/6-1,0-6.0);

\node[text width=3.5cm] at (1.0-1,-9.2) {\small Network 4};

{
\draw [->,line width=0.4mm,draw=black!80,-triangle 45](5-0.5+0.5,-5/6-6.5)to[out=180,in=0]node[above]{}(8/6+0.7-0.5,-5/6-6.5);}

{
\draw [->,line width=0.4mm,draw=black!80,-triangle 45](5-0.5+0.5,-5/6-6.5)to[out=180,in=0]node[above]{{\bf \small $q(u_{1:3}) \approx p(k_{1:3} \vert k_{\sim 1:3},\mathcal{D})$}}(8/6+0.7-0.5,-5/6-6.5);}
\end{scope}
\end{tikzpicture}}
\end{center}
\caption{Moving from Network 1 to Network 2 or from Network 2 to Network 3  with network-unaware (NuA) and network-aware (NA) approaches leads to the proposal adapting to the prior. For the move between Networks 3 and 4, our NA approach employs a proposal that approximates the joint conditional posterior $p(k_{1},k_{2},k_{3} \vert k_{\sim 1:3},\mathcal{D})$. In contrast the NuA approach retains the samples from the first two steps and constructs a proposal for the last move according to a proposal that only approximates the conditional posterior $p(k_{3} \vert k_{\sim 3},\mathcal{D})$.}
\label{movesinmodelspace}
\end{figure}

For example, consider a model-space sampler as shown in the top diagram of Figure~\ref{movesinmodelspace}. Using the centering technique of Section~\ref{sec:centering}, the proposal adapts to the conditional posterior of the rate constant of the newly added reaction. If the sampler moves between two models with the same effective network, the proposal then simply adapts to the prior distribution, since the newly added reaction has no impact on the observed nodes. (The effective network name and the effective proposal are shown in blue.)
Assuming that the prior distribution of each rate constant (or its logarithm, see Section~\ref{sec:results}) is independent and Gaussian, the proposal matches the prior distribution for all rate constant values and the acceptance rate is identically one for all moves. The difficulty comes when, after one or more moves within the same effective network, the sampler now attempts to move to a new effective network. Now the \emph{effective proposal} is the product of the prior distributions of all rate constants that did not produce a change in the effective network, and single Gaussian on the final rate constant---corresponding to the reaction whose addition finally allowed the effective network to change. (To illustrate, see the move in Figure~\ref{movesinmodelspace} from Network 3 to Network 4, or equivalently, from EN1 to EN2.) The resulting proposal is \emph{not} well adapted to EN2, and has a low probability of acceptance! 

We instead propose a network-aware (NA) approach in which, because we have determined the effective networks, we design parameter proposals that adapt to the \textit{difference between the effective networks} of any two networks. When the proposed move is between two networks with the same EN, we once again recover the prior as the proposal distribution. When the proposed move is between two networks with different ENs, however, we construct a proposal that approximates the conditional posterior distribution of the rate constants of \emph{all} reactions accounting for the difference between the two effective networks. (See bottom diagram of Figure~\ref{movesinmodelspace} for an illustration.) At each step of the algorithm we thus obtain better alignment between the posterior and proposal densities and therefore superior sampling efficiency.  Details of our NA sampler, along with pseudocode for all the steps involved, are given in the supplementary material. 

%
\subsection{Sensitivity-based network-aware proposals} 
\label{sec:sensitivitynetworkaware}
Between-model moves with deterministic reverse moves, as described in Section~\ref{sec:centering}, are a natural choice for nested models. In some cases, however, MCMC mixing may be improved by adopting non-deterministic reverse moves.
In chemical reaction networks, it is sometimes observed that the \textit{maximum a posteriori} value of the rate constant of a reaction $R^\ast$ common to two networks differs substantially between the two networks. Intuitively, the rate constant value shifts in order to compensate for the presence or absence of other reactions. In this case, keeping the rate constant of $R^\ast$ fixed when proposing moves between the two networks leads to very poor acceptance rates. 

\begin{figure}
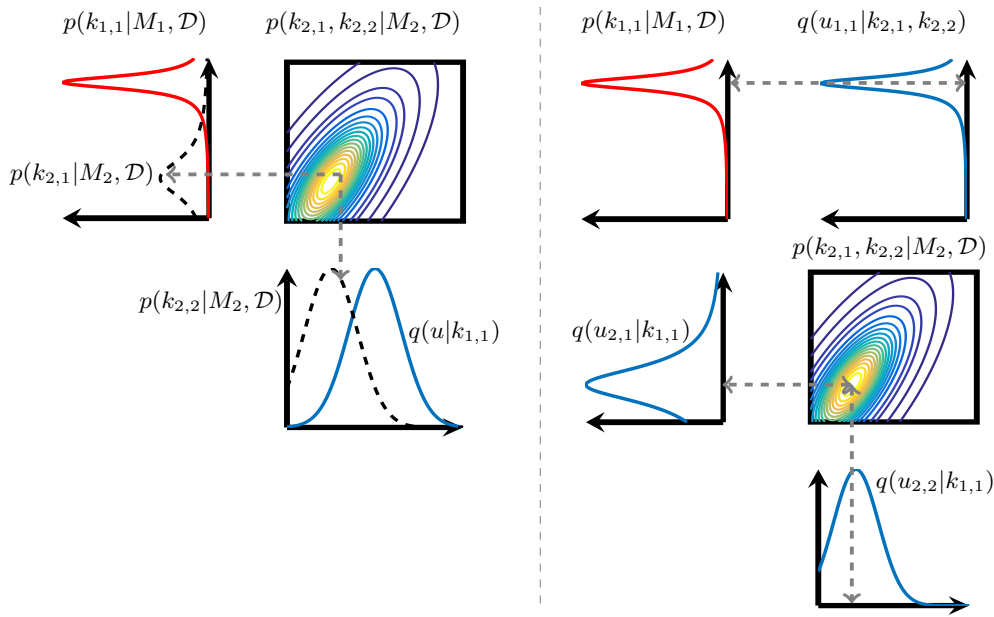

\begin{subfigure}[h]{0.44\textwidth}
\vspace*{15pt} 
\begin{minipage}[t]{0.45\textwidth}
\begin{center}
%
%
\definecolor{mycolor1}{rgb}{0.00000,0.44700,0.74100}%
\begin{tikzpicture}[rotate=90,oneDSample/.style={minimum size=0.2cm,draw=magenta,fill=magenta,circle,inner sep=0pt},
        yscale=1, xscale = 1.1]

\begin{axis}[
width=0.64\textwidth,
height=0.64\textwidth,
scale only axis,
line width = 2pt,
axis lines = left,
ymin=0,
axis background/.style={fill=white},
xtick=\empty,
ytick=\empty
]
\addplot [color=red,solid,forget plot,very thick]
  table[row sep=crcr]{%
-1	0.0109762029718549\\
-0.97979797979798	0.0112408622795783\\
-0.95959595959596	0.0115151986508714\\
-0.939393939393939	0.0117996890783316\\
-0.919191919191919	0.0120948402715634\\
-0.898989898989899	0.0124011909038369\\
-0.878787878787879	0.0127193140591549\\
-0.858585858585859	0.0130498199003925\\
-0.838383838383838	0.0133933585815979\\
-0.818181818181818	0.0137506234302887\\
-0.797979797979798	0.0141223544286893\\
-0.777777777777778	0.0145093420263855\\
-0.757575757575758	0.0149124313208926\\
-0.737373737373737	0.0153325266472079\\
-0.717171717171717	0.0157705966226403\\
-0.696969696969697	0.0162276796991783\\
-0.676767676767677	0.0167048902824919\\
-0.656565656565657	0.0172034254845038\\
-0.636363636363636	0.0177245725854757\\
-0.616161616161616	0.0182697172919304\\
-0.595959595959596	0.0188403528887023\\
-0.575757575757576	0.0194380903972494\\
-0.555555555555556	0.0200646698683946\\
-0.535353535353535	0.0207219729562834\\
-0.515151515151515	0.0214120369420068\\
-0.494949494949495	0.0221370704006083\\
-0.474747474747475	0.0228994707347294\\
-0.454545454545455	0.0237018438327623\\
-0.434343434343434	0.0245470261500423\\
-0.414141414141414	0.0254381095595057\\
-0.393939393939394	0.0263784693747925\\
-0.373737373737374	0.0273717960157517\\
-0.353535353535353	0.0284221308658255\\
-0.333333333333333	0.0295339069655064\\
-0.313131313131313	0.0307119952991931\\
-0.292929292929293	0.0319617575683321\\
-0.272727272727273	0.0332891065066886\\
-0.252525252525252	0.0347005749901266\\
-0.232323232323232	0.0362033954311366\\
-0.212121212121212	0.0378055912372285\\
-0.191919191919192	0.0395160824644686\\
-0.171717171717172	0.0413448082283596\\
-0.151515151515151	0.0433028689636662\\
-0.131313131313131	0.0454026922778416\\
-0.111111111111111	0.0476582269517321\\
-0.0909090909090909	0.0500851706479046\\
-0.0707070707070707	0.0527012381453001\\
-0.0505050505050505	0.0555264784993741\\
-0.0303030303030303	0.0585836515217421\\
-0.0101010101010101	0.0618986765041831\\
0.0101010101010102	0.065501169339842\\
0.0303030303030303	0.0694250883344979\\
0.0505050505050506	0.0737095143411065\\
0.0707070707070707	0.0783995977806985\\
0.0909090909090908	0.0835477141610383\\
0.111111111111111	0.0892148815947649\\
0.131313131313131	0.0954725095476125\\
0.151515151515152	0.102404568996794\\
0.171717171717172	0.11011030227958\\
0.191919191919192	0.118707628875893\\
0.212121212121212	0.128337455036708\\
0.232323232323232	0.139169166011836\\
0.252525252525253	0.151407677480579\\
0.272727272727273	0.165302558919479\\
0.292929292929293	0.181159932320268\\
0.313131313131313	0.199358118377362\\
0.333333333333333	0.220368382742624\\
0.353535353535354	0.244782675126507\\
0.373737373737374	0.273351020282777\\
0.393939393939394	0.307032299427944\\
0.414141414141414	0.347063654965773\\
0.434343434343434	0.395055741989025\\
0.454545454545455	0.453123485038102\\
0.474747474747475	0.524064369979394\\
0.494949494949495	0.611596783863426\\
0.515151515151515	0.720664170590745\\
0.535353535353535	0.857782566534872\\
0.555555555555556	1.03132403123548\\
0.575757575757576	1.25140601463591\\
0.595959595959596	1.52854247647591\\
0.616161616161616	1.86923618603195\\
0.636363636363636	2.26561742519051\\
0.656565656565657	2.67790145449557\\
0.676767676767677	3.02009215342433\\
0.696969696969697	3.18017858765273\\
0.717171717171717	3.0919278438923\\
0.737373737373737	2.79297689748194\\
0.757575757575758	2.39061700726999\\
0.777777777777778	1.98331544468362\\
0.797979797979798	1.62402664991532\\
0.818181818181818	1.32812055959444\\
0.838383838383838	1.09196891651639\\
0.858585858585859	0.905589316251765\\
0.878787878787879	0.758510866639273\\
0.898989898989899	0.641792880988959\\
0.919191919191919	0.54838375716072\\
0.939393939393939	0.472905137863776\\
0.95959595959596	0.411305892483498\\
0.97979797979798	0.360540297525405\\
1	0.318309886183791\\
};
{\addplot[color=black,dashed,forget plot,very thick]
  table[row sep=crcr]{%
-1	0.280861664279815\\
-0.97979797979798	0.298223442439697\\
-0.95959595959596	0.317011779925821\\
-0.939393939393939	0.337351066684761\\
-0.919191919191919	0.3593718786127\\
-0.898989898989899	0.383208954668965\\
-0.878787878787879	0.408998032786299\\
-0.858585858585859	0.436871158801217\\
-0.838383838383838	0.466949997678138\\
-0.818181818181818	0.499336597600329\\
-0.797979797979798	0.534101006851524\\
-0.777777777777778	0.571265157626744\\
-0.757575757575758	0.610782566758161\\
-0.737373737373737	0.652513740358771\\
-0.717171717171717	0.696197805872175\\
-0.696969696969697	0.741421929390021\\
-0.676767676767677	0.787591563312015\\
-0.656565656565657	0.833906443988631\\
-0.636363636363636	0.879349228955221\\
-0.616161616161616	0.922695110462172\\
-0.595959595959596	0.962550711012814\\
-0.575757575757576	0.997427966777714\\
-0.555555555555556	1.02585281621566\\
-0.535353535353535	1.04649971861507\\
-0.515151515151515	1.05833339931045\\
-0.494949494949495	1.0607323235331\\
-0.474747474747475	1.05356795635252\\
-0.454545454545455	1.03722162194539\\
-0.434343434343434	1.01253495288119\\
-0.414141414141414	0.980705574895949\\
-0.393939393939394	0.943151095739565\\
-0.373737373737374	0.901368105193096\\
-0.353535353535353	0.856808830900818\\
-0.333333333333333	0.810789332732297\\
-0.313131313131313	0.764433538352255\\
-0.292929292929293	0.718649936534392\\
-0.272727272727273	0.674133539584107\\
-0.252525252525252	0.63138453954302\\
-0.232323232323232	0.590735926850423\\
-0.212121212121212	0.552384148604294\\
-0.191919191919192	0.516418860890451\\
-0.171717171717172	0.482849530188821\\
-0.151515151515151	0.451627898098864\\
-0.131313131313131	0.42266614808304\\
-0.111111111111111	0.39585108670758\\
-0.0909090909090909	0.371054876957983\\
-0.0707070707070707	0.348142927734661\\
-0.0505050505050505	0.326979519674881\\
-0.0303030303030303	0.307431679229718\\
-0.0101010101010101	0.289371729115121\\
0.0101010101010102	0.27267886000402\\
0.0303030303030303	0.257239993608679\\
0.0505050505050506	0.242950144158148\\
0.0707070707070707	0.229712434000648\\
0.0909090909090908	0.217437878593745\\
0.111111111111111	0.206045024887217\\
0.131313131313131	0.195459503365718\\
0.151515151515152	0.185613536244323\\
0.171717171717172	0.176445431165084\\
0.191919191919192	0.167899080132572\\
0.212121212121212	0.159923476480553\\
0.232323232323232	0.152472257702604\\
0.252525252525253	0.145503278487487\\
0.272727272727273	0.138978215882507\\
0.292929292929293	0.132862206870509\\
0.313131313131313	0.127123517569984\\
0.333333333333333	0.121733242591535\\
0.353535353535354	0.11666503268949\\
0.373737373737374	0.111894848648692\\
0.393939393939394	0.107400739281024\\
0.414141414141414	0.103162641429466\\
0.434343434343434	0.0991621999574289\\
0.454545454545455	0.0953826058153508\\
0.474747474747475	0.0918084504093644\\
0.494949494949495	0.0884255946375683\\
0.515151515151515	0.0852210511007854\\
0.535353535353535	0.0821828781320368\\
0.555555555555556	0.0793000844193778\\
0.575757575757576	0.0765625431185814\\
0.595959595959596	0.0739609144646184\\
0.616161616161616	0.0714865759937643\\
0.636363636363636	0.0691315595816178\\
0.656565656565657	0.0668884945867667\\
0.676767676767677	0.0647505564658227\\
0.696969696969697	0.0627114202937081\\
0.717171717171717	0.0607652186840571\\
0.737373737373737	0.0589065036590481\\
0.757575757575758	0.0571302120665424\\
0.777777777777778	0.0554316341856537\\
0.797979797979798	0.0538063852003698\\
0.818181818181818	0.0522503792550945\\
0.838383838383838	0.0507598058364364\\
0.858585858585859	0.0493311082526562\\
0.878787878787879	0.0479609640062743\\
0.898989898989899	0.0466462668767674\\
0.919191919191919	0.045384110549346\\
0.939393939393939	0.0441717736427771\\
0.95959595959596	0.0430067060043267\\
0.97979797979798	0.0418865161533614\\
1	0.0408089597671527\\
};}
\end{axis}
\end{tikzpicture}%
\end{center}
\end{minipage}
\begin{minipage}[t]{0.45\textwidth}
\begin{center}
\input{multivariateGaussianShifted1.tex}
\end{center}
\vspace*{10pt}
\begin{center}
%
%
\definecolor{mycolor1}{rgb}{0.00000,0.44700,0.74100}%
\begin{tikzpicture}[baseline,oneDSample/.style={minimum size=0.2cm,draw=blue,fill=blue,circle,inner sep=0pt},
        yscale=1, xscale = 1.1]

\begin{axis}[%
width=0.7\textwidth,
height=0.7\textwidth,
scale only axis,
y label style={at={(axis description cs:1.15,.5)},rotate=270,anchor=south},
axis lines= left,
ymin=0,
line width =2pt,
axis background/.style={fill=white},
xtick=\empty,
ytick=\empty
]
\addplot[color=mycolor1,solid,forget plot,very thick]
  table[row sep=crcr]{%
-1	0.00514092998763702\\
-0.97979797979798	0.00642009903340377\\
-0.95959595959596	0.00798127687034018\\
-0.939393939393939	0.00987719522093684\\
-0.919191919191919	0.0121681767630233\\
-0.898989898989899	0.014922720120256\\
-0.878787878787879	0.0182180164140247\\
-0.858585858585859	0.0221403659949995\\
-0.838383838383838	0.0267854603005296\\
-0.818181818181818	0.0322584906848667\\
-0.797979797979798	0.0386740438094208\\
-0.777777777777778	0.0461557420435137\\
-0.757575757575758	0.054835587600976\\
-0.737373737373737	0.0648529711033851\\
-0.717171717171717	0.0763533091689153\\
-0.696969696969697	0.0894862816829947\\
-0.676767676767677	0.104403647753532\\
-0.656565656565657	0.121256630043661\\
-0.636363636363636	0.140192870158084\\
-0.616161616161616	0.161352972863563\\
-0.595959595959596	0.184866673845496\\
-0.575757575757576	0.210848683997548\\
-0.555555555555556	0.239394282328685\\
-0.535353535353535	0.270574748741865\\
-0.515151515151515	0.30443274637019\\
-0.494949494949495	0.340977779945506\\
-0.474747474747475	0.380181870869257\\
-0.454545454545455	0.421975600300139\\
-0.434343434343434	0.466244677757922\\
-0.414141414141414	0.512827193657178\\
-0.393939393939394	0.561511709172504\\
-0.373737373737374	0.612036325448177\\
-0.353535353535353	0.664088856203459\\
-0.333333333333333	0.717308203343437\\
-0.313131313131313	0.77128700467323\\
-0.292929292929293	0.82557558696369\\
-0.272727272727273	0.8796872174791\\
-0.252525252525252	0.933104603991421\\
-0.232323232323232	0.985287548855602\\
-0.212121212121212	1.03568161867327\\
-0.191919191919192	1.08372764929946\\
-0.171717171717172	1.12887186833761\\
-0.151515151515151	1.1705763856312\\
-0.131313131313131	1.20832977823147\\
-0.111111111111111	1.241657481265\\
-0.0909090909090909	1.27013169106157\\
-0.0707070707070707	1.29338049243003\\
-0.0505050505050505	1.31109593822672\\
-0.0303030303030303	1.32304083599816\\
-0.0101010101010101	1.32905403266518\\
0.0101010101010102	1.32905403266518\\
0.0303030303030303	1.32304083599816\\
0.0505050505050506	1.31109593822672\\
0.0707070707070707	1.29338049243003\\
0.0909090909090908	1.27013169106157\\
0.111111111111111	1.241657481265\\
0.131313131313131	1.20832977823147\\
0.151515151515152	1.1705763856312\\
0.171717171717172	1.12887186833761\\
0.191919191919192	1.08372764929946\\
0.212121212121212	1.03568161867327\\
0.232323232323232	0.985287548855602\\
0.252525252525253	0.933104603991421\\
0.272727272727273	0.8796872174791\\
0.292929292929293	0.825575586963689\\
0.313131313131313	0.77128700467323\\
0.333333333333333	0.717308203343437\\
0.353535353535354	0.664088856203459\\
0.373737373737374	0.612036325448177\\
0.393939393939394	0.561511709172503\\
0.414141414141414	0.512827193657178\\
0.434343434343434	0.466244677757922\\
0.454545454545455	0.421975600300139\\
0.474747474747475	0.380181870869257\\
0.494949494949495	0.340977779945506\\
0.515151515151515	0.30443274637019\\
0.535353535353535	0.270574748741865\\
0.555555555555556	0.239394282328685\\
0.575757575757576	0.210848683997548\\
0.595959595959596	0.184866673845496\\
0.616161616161616	0.161352972863563\\
0.636363636363636	0.140192870158084\\
0.656565656565657	0.121256630043661\\
0.676767676767677	0.104403647753532\\
0.696969696969697	0.0894862816829947\\
0.717171717171717	0.0763533091689153\\
0.737373737373737	0.064852971103385\\
0.757575757575758	0.054835587600976\\
0.777777777777778	0.0461557420435137\\
0.797979797979798	0.0386740438094208\\
0.818181818181818	0.0322584906848667\\
0.838383838383838	0.0267854603005295\\
0.858585858585859	0.0221403659949995\\
0.878787878787879	0.0182180164140247\\
0.898989898989899	0.014922720120256\\
0.919191919191919	0.0121681767630233\\
0.939393939393939	0.00987719522093684\\
0.95959595959596	0.00798127687034018\\
0.97979797979798	0.00642009903340376\\
1	0.00514092998763702\\
};
{\addplot [color=black,dashed,forget plot,very thick]
  table[row sep=crcr]{%
-1	0.331590462642496\\
-0.97979797979798	0.370134615241341\\
-0.95959595959596	0.411289844212404\\
-0.939393939393939	0.454953363958235\\
-0.919191919191919	0.500975383990722\\
-0.898989898989899	0.549156975593069\\
-0.878787878787879	0.599248883749375\\
-0.858585858585859	0.65095141349983\\
-0.838383838383838	0.703915497032365\\
-0.818181818181818	0.757745018771203\\
-0.797979797979798	0.812000441121054\\
-0.777777777777778	0.866203734345221\\
-0.757575757575758	0.919844571567051\\
-0.737373737373737	0.972387705615475\\
-0.717171717171717	1.02328140011459\\
-0.696969696969697	1.07196674469803\\
-0.676767676767677	1.11788764540414\\
-0.656565656565657	1.16050124801635\\
-0.636363636363636	1.19928852604208\\
-0.616161616161616	1.23376474761423\\
-0.595959595959596	1.26348952796446\\
-0.575757575757576	1.28807617696523\\
-0.555555555555556	1.30720006482618\\
-0.535353535353535	1.32060575313111\\
-0.515151515151515	1.32811267230449\\
-0.494949494949495	1.32961916912272\\
-0.474747474747475	1.32510479744693\\
-0.454545454545455	1.3146307800164\\
-0.434343434343434	1.29833862672258\\
-0.414141414141414	1.27644695295563\\
-0.393939393939394	1.2492465980213\\
-0.373737373737374	1.21709419599786\\
-0.353535353535353	1.18040439766851\\
-0.333333333333333	1.13964098055376\\
-0.313131313131313	1.09530711318874\\
-0.292929292929293	1.0479350586939\\
-0.272727272727273	0.998075610903666\\
-0.252525252525252	0.946287553862521\\
-0.232323232323232	0.893127422860809\\
-0.212121212121212	0.839139823280798\\
-0.191919191919192	0.784848533644616\\
-0.171717171717172	0.730748582974799\\
-0.151515151515151	0.677299451670746\\
-0.131313131313131	0.624919501448109\\
-0.111111111111111	0.573981695367632\\
-0.0909090909090909	0.524810625396146\\
-0.0707070707070707	0.477680823931664\\
-0.0505050505050505	0.432816298692498\\
-0.0303030303030303	0.390391198442482\\
-0.0101010101010101	0.35053149101441\\
0.0101010101010102	0.313317515491416\\
0.0303030303030303	0.278787257381724\\
0.0505050505050506	0.24694018904899\\
0.0707070707070707	0.21774151714747\\
0.0909090909090908	0.191126683749377\\
0.111111111111111	0.167005977463881\\
0.131313131313131	0.145269124243218\\
0.151515151515152	0.125789743802133\\
0.171717171717172	0.108429575685181\\
0.191919191919192	0.0930423980824404\\
0.212121212121212	0.0794775816734762\\
0.232323232323232	0.0675832393322401\\
0.252525252525253	0.0572089498374323\\
0.272727272727273	0.0482080493254473\\
0.292929292929293	0.0404394977576075\\
0.313131313131313	0.0337693389460401\\
0.333333333333333	0.0280717816137279\\
0.353535353535354	0.023229935583696\\
0.373737373737374	0.0191362416206187\\
0.393939393939394	0.0156926358770778\\
0.414141414141414	0.0128104905688122\\
0.434343434343434	0.01041037169251\\
0.454545454545455	0.00842165259367977\\
0.474747474747475	0.006782019276936\\
0.494949494949495	0.00543689979848966\\
0.515151515151515	0.00433884613882971\\
0.535353535353535	0.00344689284005235\\
0.555555555555556	0.00272591259016744\\
0.575757575757576	0.00214598499308955\\
0.595959595959596	0.00168179108954534\\
0.616161616161616	0.00131204286907698\\
0.636363636363636	0.00101895408492685\\
0.656565656565657	0.000787756174117749\\
0.676767676767677	0.000606260995268282\\
0.696969696969697	0.000464470410449896\\
0.717171717171717	0.000354231425998118\\
0.737373737373737	0.00026893463341847\\
0.757575757575758	0.000203253013176632\\
0.777777777777778	0.000152917737024689\\
0.797979797979798	0.000114527384761479\\
0.818181818181818	8.53869373171276e-05\\
0.838383838383838	6.33729816244044e-05\\
0.858585858585859	4.68217300087542e-05\\
0.878787878787879	3.44366886044056e-05\\
0.898989898989899	2.52130811785188e-05\\
0.919191919191919	1.83764269372351e-05\\
0.939393939393939	1.33329679080554e-05\\
0.95959595959596	9.62993169790184e-06\\
0.97979797979798	6.92389050567992e-06\\
1	4.95573171578099e-06\\
};}
\end{axis}
\end{tikzpicture}%
\end{center}
\vspace*{10pt}
\begin{center}
%
%
\definecolor{mycolor1}{rgb}{0.0,0.0,0.0}%
\begin{tikzpicture}[baseline,oneDSample/.style={minimum size=0.2cm,draw=white,fill=white,circle,inner sep=0pt},
        yscale=1, xscale = 1.1]

\begin{axis}[%
width=0.58\textwidth,
height=0.58\textwidth,
color=white,
scale only axis,
axis lines= left,
ymin=0,
line width =2pt,
axis background/.style={fill=white},
xtick=\empty,
ytick=\empty
]
\addplot [color=white,solid,forget plot,very thick]
  table[row sep=crcr]{%
-1	0.331590462642496\\
-0.97979797979798	0.370134615241341\\
-0.95959595959596	0.411289844212404\\
-0.939393939393939	0.454953363958235\\
-0.919191919191919	0.500975383990722\\
-0.898989898989899	0.549156975593069\\
-0.878787878787879	0.599248883749375\\
-0.858585858585859	0.65095141349983\\
-0.838383838383838	0.703915497032365\\
-0.818181818181818	0.757745018771203\\
-0.797979797979798	0.812000441121054\\
-0.777777777777778	0.866203734345221\\
-0.757575757575758	0.919844571567051\\
-0.737373737373737	0.972387705615475\\
-0.717171717171717	1.02328140011459\\
-0.696969696969697	1.07196674469803\\
-0.676767676767677	1.11788764540414\\
-0.656565656565657	1.16050124801635\\
-0.636363636363636	1.19928852604208\\
-0.616161616161616	1.23376474761423\\
-0.595959595959596	1.26348952796446\\
-0.575757575757576	1.28807617696523\\
-0.555555555555556	1.30720006482618\\
-0.535353535353535	1.32060575313111\\
-0.515151515151515	1.32811267230449\\
-0.494949494949495	1.32961916912272\\
-0.474747474747475	1.32510479744693\\
-0.454545454545455	1.3146307800164\\
-0.434343434343434	1.29833862672258\\
-0.414141414141414	1.27644695295563\\
-0.393939393939394	1.2492465980213\\
-0.373737373737374	1.21709419599786\\
-0.353535353535353	1.18040439766851\\
-0.333333333333333	1.13964098055376\\
-0.313131313131313	1.09530711318874\\
-0.292929292929293	1.0479350586939\\
-0.272727272727273	0.998075610903666\\
-0.252525252525252	0.946287553862521\\
-0.232323232323232	0.893127422860809\\
-0.212121212121212	0.839139823280798\\
-0.191919191919192	0.784848533644616\\
-0.171717171717172	0.730748582974799\\
-0.151515151515151	0.677299451670746\\
-0.131313131313131	0.624919501448109\\
-0.111111111111111	0.573981695367632\\
-0.0909090909090909	0.524810625396146\\
-0.0707070707070707	0.477680823931664\\
-0.0505050505050505	0.432816298692498\\
-0.0303030303030303	0.390391198442482\\
-0.0101010101010101	0.35053149101441\\
0.0101010101010102	0.313317515491416\\
0.0303030303030303	0.278787257381724\\
0.0505050505050506	0.24694018904899\\
0.0707070707070707	0.21774151714747\\
0.0909090909090908	0.191126683749377\\
0.111111111111111	0.167005977463881\\
0.131313131313131	0.145269124243218\\
0.151515151515152	0.125789743802133\\
0.171717171717172	0.108429575685181\\
0.191919191919192	0.0930423980824404\\
0.212121212121212	0.0794775816734762\\
0.232323232323232	0.0675832393322401\\
0.252525252525253	0.0572089498374323\\
0.272727272727273	0.0482080493254473\\
0.292929292929293	0.0404394977576075\\
0.313131313131313	0.0337693389460401\\
0.333333333333333	0.0280717816137279\\
0.353535353535354	0.023229935583696\\
0.373737373737374	0.0191362416206187\\
0.393939393939394	0.0156926358770778\\
0.414141414141414	0.0128104905688122\\
0.434343434343434	0.01041037169251\\
0.454545454545455	0.00842165259367977\\
0.474747474747475	0.006782019276936\\
0.494949494949495	0.00543689979848966\\
0.515151515151515	0.00433884613882971\\
0.535353535353535	0.00344689284005235\\
0.555555555555556	0.00272591259016744\\
0.575757575757576	0.00214598499308955\\
0.595959595959596	0.00168179108954534\\
0.616161616161616	0.00131204286907698\\
0.636363636363636	0.00101895408492685\\
0.656565656565657	0.000787756174117749\\
0.676767676767677	0.000606260995268282\\
0.696969696969697	0.000464470410449896\\
0.717171717171717	0.000354231425998118\\
0.737373737373737	0.00026893463341847\\
0.757575757575758	0.000203253013176632\\
0.777777777777778	0.000152917737024689\\
0.797979797979798	0.000114527384761479\\
0.818181818181818	8.53869373171276e-05\\
0.838383838383838	6.33729816244044e-05\\
0.858585858585859	4.68217300087542e-05\\
0.878787878787879	3.44366886044056e-05\\
0.898989898989899	2.52130811785188e-05\\
0.919191919191919	1.83764269372351e-05\\
0.939393939393939	1.33329679080554e-05\\
0.95959595959596	9.62993169790184e-06\\
0.97979797979798	6.92389050567992e-06\\
1	4.95573171578099e-06\\
};
\end{axis}
\end{tikzpicture}%
\end{center}
\end{minipage}

\begin{tikzpicture}[overlay]
\node[anchor=north,circle,color=magenta] at (3.59,7.3) {};
\draw[->,line width=0.5mm,dashed,draw=black!50](4.2,5.8)to[out=180,in=0]node[left]{}(1.9,5.8);
\draw[->,line width=0.5mm,dashed,draw=black!50](4.2,5.8)to[out=270,in=90]node[left]{}(4.2,4.4);    
\end{tikzpicture}

\begin{tikzpicture}[overlay]
\node[anchor=north,text width=4.0cm, text centered] at (1.5,8.6) {\bf \scriptsize $p(k_{1,1} \vert M_{1},\mathcal{D})$};
\node[anchor=north,text width=5.0cm, text centered] at (4.5,8.6) {\bf \scriptsize $p(k_{2,1},k_{2,2} \vert M_{2},\mathcal{D})$};
\node[anchor=north,text width=4.0cm, text centered] at (5.7,4.5) {\bf \scriptsize $q(u \vert k_{1,1})$};
\node[anchor=north,text width=4.0cm, text centered] at (0.8,6.6) {\bf \scriptsize $p(k_{2,1} \vert M_{2},\mathcal{D})$};
\node[anchor=north,text width=4.0cm, text centered] at (2.5,4.9) {\bf \scriptsize $p(k_{2,2} \vert M_{2},\mathcal{D})$};
\end{tikzpicture}
\label{highSensitivityReaction}
\end{subfigure}
\begin{subfigure}[h]{0.44\textwidth}
\vspace*{15pt} 
\begin{minipage}[t]{0.45\textwidth}
\begin{center}
%
%
\definecolor{mycolor1}{rgb}{0.00000,0.44700,0.74100}%
\begin{tikzpicture}[rotate=90,oneDSample/.style={minimum size=0.2cm,draw=magenta,fill=magenta,circle,inner sep=0pt},
        yscale=1, xscale = 1.1]

\begin{axis}[
width=0.64\textwidth,
height=0.64\textwidth,
scale only axis,
line width = 2pt,
axis lines = left,
ymin=0,
axis background/.style={fill=white},
xtick=\empty,
ytick=\empty
]
\addplot [color=red,solid,forget plot,very thick]
  table[row sep=crcr]{%
-1	0.0109762029718549\\
-0.97979797979798	0.0112408622795783\\
-0.95959595959596	0.0115151986508714\\
-0.939393939393939	0.0117996890783316\\
-0.919191919191919	0.0120948402715634\\
-0.898989898989899	0.0124011909038369\\
-0.878787878787879	0.0127193140591549\\
-0.858585858585859	0.0130498199003925\\
-0.838383838383838	0.0133933585815979\\
-0.818181818181818	0.0137506234302887\\
-0.797979797979798	0.0141223544286893\\
-0.777777777777778	0.0145093420263855\\
-0.757575757575758	0.0149124313208926\\
-0.737373737373737	0.0153325266472079\\
-0.717171717171717	0.0157705966226403\\
-0.696969696969697	0.0162276796991783\\
-0.676767676767677	0.0167048902824919\\
-0.656565656565657	0.0172034254845038\\
-0.636363636363636	0.0177245725854757\\
-0.616161616161616	0.0182697172919304\\
-0.595959595959596	0.0188403528887023\\
-0.575757575757576	0.0194380903972494\\
-0.555555555555556	0.0200646698683946\\
-0.535353535353535	0.0207219729562834\\
-0.515151515151515	0.0214120369420068\\
-0.494949494949495	0.0221370704006083\\
-0.474747474747475	0.0228994707347294\\
-0.454545454545455	0.0237018438327623\\
-0.434343434343434	0.0245470261500423\\
-0.414141414141414	0.0254381095595057\\
-0.393939393939394	0.0263784693747925\\
-0.373737373737374	0.0273717960157517\\
-0.353535353535353	0.0284221308658255\\
-0.333333333333333	0.0295339069655064\\
-0.313131313131313	0.0307119952991931\\
-0.292929292929293	0.0319617575683321\\
-0.272727272727273	0.0332891065066886\\
-0.252525252525252	0.0347005749901266\\
-0.232323232323232	0.0362033954311366\\
-0.212121212121212	0.0378055912372285\\
-0.191919191919192	0.0395160824644686\\
-0.171717171717172	0.0413448082283596\\
-0.151515151515151	0.0433028689636662\\
-0.131313131313131	0.0454026922778416\\
-0.111111111111111	0.0476582269517321\\
-0.0909090909090909	0.0500851706479046\\
-0.0707070707070707	0.0527012381453001\\
-0.0505050505050505	0.0555264784993741\\
-0.0303030303030303	0.0585836515217421\\
-0.0101010101010101	0.0618986765041831\\
0.0101010101010102	0.065501169339842\\
0.0303030303030303	0.0694250883344979\\
0.0505050505050506	0.0737095143411065\\
0.0707070707070707	0.0783995977806985\\
0.0909090909090908	0.0835477141610383\\
0.111111111111111	0.0892148815947649\\
0.131313131313131	0.0954725095476125\\
0.151515151515152	0.102404568996794\\
0.171717171717172	0.11011030227958\\
0.191919191919192	0.118707628875893\\
0.212121212121212	0.128337455036708\\
0.232323232323232	0.139169166011836\\
0.252525252525253	0.151407677480579\\
0.272727272727273	0.165302558919479\\
0.292929292929293	0.181159932320268\\
0.313131313131313	0.199358118377362\\
0.333333333333333	0.220368382742624\\
0.353535353535354	0.244782675126507\\
0.373737373737374	0.273351020282777\\
0.393939393939394	0.307032299427944\\
0.414141414141414	0.347063654965773\\
0.434343434343434	0.395055741989025\\
0.454545454545455	0.453123485038102\\
0.474747474747475	0.524064369979394\\
0.494949494949495	0.611596783863426\\
0.515151515151515	0.720664170590745\\
0.535353535353535	0.857782566534872\\
0.555555555555556	1.03132403123548\\
0.575757575757576	1.25140601463591\\
0.595959595959596	1.52854247647591\\
0.616161616161616	1.86923618603195\\
0.636363636363636	2.26561742519051\\
0.656565656565657	2.67790145449557\\
0.676767676767677	3.02009215342433\\
0.696969696969697	3.18017858765273\\
0.717171717171717	3.0919278438923\\
0.737373737373737	2.79297689748194\\
0.757575757575758	2.39061700726999\\
0.777777777777778	1.98331544468362\\
0.797979797979798	1.62402664991532\\
0.818181818181818	1.32812055959444\\
0.838383838383838	1.09196891651639\\
0.858585858585859	0.905589316251765\\
0.878787878787879	0.758510866639273\\
0.898989898989899	0.641792880988959\\
0.919191919191919	0.54838375716072\\
0.939393939393939	0.472905137863776\\
0.95959595959596	0.411305892483498\\
0.97979797979798	0.360540297525405\\
1	0.318309886183791\\
};
\end{axis}
\end{tikzpicture}%
\end{center}
\vspace*{12pt}
\begin{center}
%
%
\definecolor{mycolor1}{rgb}{0.00000,0.44700,0.74100}%
\begin{tikzpicture}[rotate=90,oneDSample/.style={minimum size=0.2cm,draw=magenta,fill=magenta,circle,inner sep=0pt},
        yscale=1, xscale = 1.1]

\begin{axis}[
width=0.60\textwidth,
height=0.60\textwidth,
scale only axis,
line width = 2pt,
axis lines = left,
ymin=0,
axis background/.style={fill=white},
xtick=\empty,
ytick=\empty
]
\addplot [color=mycolor1,solid,forget plot,very thick]
  table[row sep=crcr]{%
-1	0.280861664279815\\
-0.97979797979798	0.298223442439697\\
-0.95959595959596	0.317011779925821\\
-0.939393939393939	0.337351066684761\\
-0.919191919191919	0.3593718786127\\
-0.898989898989899	0.383208954668965\\
-0.878787878787879	0.408998032786299\\
-0.858585858585859	0.436871158801217\\
-0.838383838383838	0.466949997678138\\
-0.818181818181818	0.499336597600329\\
-0.797979797979798	0.534101006851524\\
-0.777777777777778	0.571265157626744\\
-0.757575757575758	0.610782566758161\\
-0.737373737373737	0.652513740358771\\
-0.717171717171717	0.696197805872175\\
-0.696969696969697	0.741421929390021\\
-0.676767676767677	0.787591563312015\\
-0.656565656565657	0.833906443988631\\
-0.636363636363636	0.879349228955221\\
-0.616161616161616	0.922695110462172\\
-0.595959595959596	0.962550711012814\\
-0.575757575757576	0.997427966777714\\
-0.555555555555556	1.02585281621566\\
-0.535353535353535	1.04649971861507\\
-0.515151515151515	1.05833339931045\\
-0.494949494949495	1.0607323235331\\
-0.474747474747475	1.05356795635252\\
-0.454545454545455	1.03722162194539\\
-0.434343434343434	1.01253495288119\\
-0.414141414141414	0.980705574895949\\
-0.393939393939394	0.943151095739565\\
-0.373737373737374	0.901368105193096\\
-0.353535353535353	0.856808830900818\\
-0.333333333333333	0.810789332732297\\
-0.313131313131313	0.764433538352255\\
-0.292929292929293	0.718649936534392\\
-0.272727272727273	0.674133539584107\\
-0.252525252525252	0.63138453954302\\
-0.232323232323232	0.590735926850423\\
-0.212121212121212	0.552384148604294\\
-0.191919191919192	0.516418860890451\\
-0.171717171717172	0.482849530188821\\
-0.151515151515151	0.451627898098864\\
-0.131313131313131	0.42266614808304\\
-0.111111111111111	0.39585108670758\\
-0.0909090909090909	0.371054876957983\\
-0.0707070707070707	0.348142927734661\\
-0.0505050505050505	0.326979519674881\\
-0.0303030303030303	0.307431679229718\\
-0.0101010101010101	0.289371729115121\\
0.0101010101010102	0.27267886000402\\
0.0303030303030303	0.257239993608679\\
0.0505050505050506	0.242950144158148\\
0.0707070707070707	0.229712434000648\\
0.0909090909090908	0.217437878593745\\
0.111111111111111	0.206045024887217\\
0.131313131313131	0.195459503365718\\
0.151515151515152	0.185613536244323\\
0.171717171717172	0.176445431165084\\
0.191919191919192	0.167899080132572\\
0.212121212121212	0.159923476480553\\
0.232323232323232	0.152472257702604\\
0.252525252525253	0.145503278487487\\
0.272727272727273	0.138978215882507\\
0.292929292929293	0.132862206870509\\
0.313131313131313	0.127123517569984\\
0.333333333333333	0.121733242591535\\
0.353535353535354	0.11666503268949\\
0.373737373737374	0.111894848648692\\
0.393939393939394	0.107400739281024\\
0.414141414141414	0.103162641429466\\
0.434343434343434	0.0991621999574289\\
0.454545454545455	0.0953826058153508\\
0.474747474747475	0.0918084504093644\\
0.494949494949495	0.0884255946375683\\
0.515151515151515	0.0852210511007854\\
0.535353535353535	0.0821828781320368\\
0.555555555555556	0.0793000844193778\\
0.575757575757576	0.0765625431185814\\
0.595959595959596	0.0739609144646184\\
0.616161616161616	0.0714865759937643\\
0.636363636363636	0.0691315595816178\\
0.656565656565657	0.0668884945867667\\
0.676767676767677	0.0647505564658227\\
0.696969696969697	0.0627114202937081\\
0.717171717171717	0.0607652186840571\\
0.737373737373737	0.0589065036590481\\
0.757575757575758	0.0571302120665424\\
0.777777777777778	0.0554316341856537\\
0.797979797979798	0.0538063852003698\\
0.818181818181818	0.0522503792550945\\
0.838383838383838	0.0507598058364364\\
0.858585858585859	0.0493311082526562\\
0.878787878787879	0.0479609640062743\\
0.898989898989899	0.0466462668767674\\
0.919191919191919	0.045384110549346\\
0.939393939393939	0.0441717736427771\\
0.95959595959596	0.0430067060043267\\
0.97979797979798	0.0418865161533614\\
1	0.0408089597671527\\
};
\end{axis}
\end{tikzpicture}%
\end{center}
\end{minipage}
\begin{minipage}[t]{0.45\textwidth}
\begin{center}
%
%
\definecolor{mycolor1}{rgb}{0.00000,0.44700,0.74100}%
\begin{tikzpicture}[rotate=90,oneDSample/.style={minimum size=0.2cm,draw=magenta,fill=magenta,circle,inner sep=0pt},
        yscale=1, xscale = 1.1]

\begin{axis}[
width=0.64\textwidth,
height=0.64\textwidth,
scale only axis,
line width = 2pt,
axis lines = left,
ymin=0,
axis background/.style={fill=white},
xtick=\empty,
ytick=\empty
]
\addplot [color=mycolor1,solid,forget plot,very thick]
  table[row sep=crcr]{%
-1	0.0109762029718549\\
-0.97979797979798	0.0112408622795783\\
-0.95959595959596	0.0115151986508714\\
-0.939393939393939	0.0117996890783316\\
-0.919191919191919	0.0120948402715634\\
-0.898989898989899	0.0124011909038369\\
-0.878787878787879	0.0127193140591549\\
-0.858585858585859	0.0130498199003925\\
-0.838383838383838	0.0133933585815979\\
-0.818181818181818	0.0137506234302887\\
-0.797979797979798	0.0141223544286893\\
-0.777777777777778	0.0145093420263855\\
-0.757575757575758	0.0149124313208926\\
-0.737373737373737	0.0153325266472079\\
-0.717171717171717	0.0157705966226403\\
-0.696969696969697	0.0162276796991783\\
-0.676767676767677	0.0167048902824919\\
-0.656565656565657	0.0172034254845038\\
-0.636363636363636	0.0177245725854757\\
-0.616161616161616	0.0182697172919304\\
-0.595959595959596	0.0188403528887023\\
-0.575757575757576	0.0194380903972494\\
-0.555555555555556	0.0200646698683946\\
-0.535353535353535	0.0207219729562834\\
-0.515151515151515	0.0214120369420068\\
-0.494949494949495	0.0221370704006083\\
-0.474747474747475	0.0228994707347294\\
-0.454545454545455	0.0237018438327623\\
-0.434343434343434	0.0245470261500423\\
-0.414141414141414	0.0254381095595057\\
-0.393939393939394	0.0263784693747925\\
-0.373737373737374	0.0273717960157517\\
-0.353535353535353	0.0284221308658255\\
-0.333333333333333	0.0295339069655064\\
-0.313131313131313	0.0307119952991931\\
-0.292929292929293	0.0319617575683321\\
-0.272727272727273	0.0332891065066886\\
-0.252525252525252	0.0347005749901266\\
-0.232323232323232	0.0362033954311366\\
-0.212121212121212	0.0378055912372285\\
-0.191919191919192	0.0395160824644686\\
-0.171717171717172	0.0413448082283596\\
-0.151515151515151	0.0433028689636662\\
-0.131313131313131	0.0454026922778416\\
-0.111111111111111	0.0476582269517321\\
-0.0909090909090909	0.0500851706479046\\
-0.0707070707070707	0.0527012381453001\\
-0.0505050505050505	0.0555264784993741\\
-0.0303030303030303	0.0585836515217421\\
-0.0101010101010101	0.0618986765041831\\
0.0101010101010102	0.065501169339842\\
0.0303030303030303	0.0694250883344979\\
0.0505050505050506	0.0737095143411065\\
0.0707070707070707	0.0783995977806985\\
0.0909090909090908	0.0835477141610383\\
0.111111111111111	0.0892148815947649\\
0.131313131313131	0.0954725095476125\\
0.151515151515152	0.102404568996794\\
0.171717171717172	0.11011030227958\\
0.191919191919192	0.118707628875893\\
0.212121212121212	0.128337455036708\\
0.232323232323232	0.139169166011836\\
0.252525252525253	0.151407677480579\\
0.272727272727273	0.165302558919479\\
0.292929292929293	0.181159932320268\\
0.313131313131313	0.199358118377362\\
0.333333333333333	0.220368382742624\\
0.353535353535354	0.244782675126507\\
0.373737373737374	0.273351020282777\\
0.393939393939394	0.307032299427944\\
0.414141414141414	0.347063654965773\\
0.434343434343434	0.395055741989025\\
0.454545454545455	0.453123485038102\\
0.474747474747475	0.524064369979394\\
0.494949494949495	0.611596783863426\\
0.515151515151515	0.720664170590745\\
0.535353535353535	0.857782566534872\\
0.555555555555556	1.03132403123548\\
0.575757575757576	1.25140601463591\\
0.595959595959596	1.52854247647591\\
0.616161616161616	1.86923618603195\\
0.636363636363636	2.26561742519051\\
0.656565656565657	2.67790145449557\\
0.676767676767677	3.02009215342433\\
0.696969696969697	3.18017858765273\\
0.717171717171717	3.0919278438923\\
0.737373737373737	2.79297689748194\\
0.757575757575758	2.39061700726999\\
0.777777777777778	1.98331544468362\\
0.797979797979798	1.62402664991532\\
0.818181818181818	1.32812055959444\\
0.838383838383838	1.09196891651639\\
0.858585858585859	0.905589316251765\\
0.878787878787879	0.758510866639273\\
0.898989898989899	0.641792880988959\\
0.919191919191919	0.54838375716072\\
0.939393939393939	0.472905137863776\\
0.95959595959596	0.411305892483498\\
0.97979797979798	0.360540297525405\\
1	0.318309886183791\\
};
\end{axis}
\end{tikzpicture}%
\end{center}
\vspace*{12pt}
\begin{center}
\input{multivariateGaussianShifted2.tex}
\end{center}
\vspace{10pt}
\begin{center}
%
%
\definecolor{mycolor1}{rgb}{0.00000,0.44700,0.74100}%
\begin{tikzpicture}[baseline,oneDSample/.style={minimum size=0.2cm,draw=blue,fill=blue,circle,inner sep=0pt},
        yscale=1, xscale = 1.1]

\begin{axis}[%
width=0.6\textwidth,
height=0.6\textwidth,
scale only axis,
axis lines= left,
ymin=0,
line width =2pt,
axis background/.style={fill=white},
xtick=\empty,
ytick=\empty
]
\addplot [color=mycolor1,solid,forget plot,very thick]
  table[row sep=crcr]{%
-1	0.331590462642496\\
-0.97979797979798	0.370134615241341\\
-0.95959595959596	0.411289844212404\\
-0.939393939393939	0.454953363958235\\
-0.919191919191919	0.500975383990722\\
-0.898989898989899	0.549156975593069\\
-0.878787878787879	0.599248883749375\\
-0.858585858585859	0.65095141349983\\
-0.838383838383838	0.703915497032365\\
-0.818181818181818	0.757745018771203\\
-0.797979797979798	0.812000441121054\\
-0.777777777777778	0.866203734345221\\
-0.757575757575758	0.919844571567051\\
-0.737373737373737	0.972387705615475\\
-0.717171717171717	1.02328140011459\\
-0.696969696969697	1.07196674469803\\
-0.676767676767677	1.11788764540414\\
-0.656565656565657	1.16050124801635\\
-0.636363636363636	1.19928852604208\\
-0.616161616161616	1.23376474761423\\
-0.595959595959596	1.26348952796446\\
-0.575757575757576	1.28807617696523\\
-0.555555555555556	1.30720006482618\\
-0.535353535353535	1.32060575313111\\
-0.515151515151515	1.32811267230449\\
-0.494949494949495	1.32961916912272\\
-0.474747474747475	1.32510479744693\\
-0.454545454545455	1.3146307800164\\
-0.434343434343434	1.29833862672258\\
-0.414141414141414	1.27644695295563\\
-0.393939393939394	1.2492465980213\\
-0.373737373737374	1.21709419599786\\
-0.353535353535353	1.18040439766851\\
-0.333333333333333	1.13964098055376\\
-0.313131313131313	1.09530711318874\\
-0.292929292929293	1.0479350586939\\
-0.272727272727273	0.998075610903666\\
-0.252525252525252	0.946287553862521\\
-0.232323232323232	0.893127422860809\\
-0.212121212121212	0.839139823280798\\
-0.191919191919192	0.784848533644616\\
-0.171717171717172	0.730748582974799\\
-0.151515151515151	0.677299451670746\\
-0.131313131313131	0.624919501448109\\
-0.111111111111111	0.573981695367632\\
-0.0909090909090909	0.524810625396146\\
-0.0707070707070707	0.477680823931664\\
-0.0505050505050505	0.432816298692498\\
-0.0303030303030303	0.390391198442482\\
-0.0101010101010101	0.35053149101441\\
0.0101010101010102	0.313317515491416\\
0.0303030303030303	0.278787257381724\\
0.0505050505050506	0.24694018904899\\
0.0707070707070707	0.21774151714747\\
0.0909090909090908	0.191126683749377\\
0.111111111111111	0.167005977463881\\
0.131313131313131	0.145269124243218\\
0.151515151515152	0.125789743802133\\
0.171717171717172	0.108429575685181\\
0.191919191919192	0.0930423980824404\\
0.212121212121212	0.0794775816734762\\
0.232323232323232	0.0675832393322401\\
0.252525252525253	0.0572089498374323\\
0.272727272727273	0.0482080493254473\\
0.292929292929293	0.0404394977576075\\
0.313131313131313	0.0337693389460401\\
0.333333333333333	0.0280717816137279\\
0.353535353535354	0.023229935583696\\
0.373737373737374	0.0191362416206187\\
0.393939393939394	0.0156926358770778\\
0.414141414141414	0.0128104905688122\\
0.434343434343434	0.01041037169251\\
0.454545454545455	0.00842165259367977\\
0.474747474747475	0.006782019276936\\
0.494949494949495	0.00543689979848966\\
0.515151515151515	0.00433884613882971\\
0.535353535353535	0.00344689284005235\\
0.555555555555556	0.00272591259016744\\
0.575757575757576	0.00214598499308955\\
0.595959595959596	0.00168179108954534\\
0.616161616161616	0.00131204286907698\\
0.636363636363636	0.00101895408492685\\
0.656565656565657	0.000787756174117749\\
0.676767676767677	0.000606260995268282\\
0.696969696969697	0.000464470410449896\\
0.717171717171717	0.000354231425998118\\
0.737373737373737	0.00026893463341847\\
0.757575757575758	0.000203253013176632\\
0.777777777777778	0.000152917737024689\\
0.797979797979798	0.000114527384761479\\
0.818181818181818	8.53869373171276e-05\\
0.838383838383838	6.33729816244044e-05\\
0.858585858585859	4.68217300087542e-05\\
0.878787878787879	3.44366886044056e-05\\
0.898989898989899	2.52130811785188e-05\\
0.919191919191919	1.83764269372351e-05\\
0.939393939393939	1.33329679080554e-05\\
0.95959595959596	9.62993169790184e-06\\
0.97979797979798	6.92389050567992e-06\\
1	4.95573171578099e-06\\
};
\end{axis}
\end{tikzpicture}%
\end{center}
\end{minipage}

\begin{tikzpicture}[overlay]
\draw[<->,line width=0.5mm,dashed,draw=black!50](5.6,7.0)to[out=180,in=0]node[left]{}(2.5,7.0);
\draw[<->,line width=0.5mm,dashed,draw=black!50](4.1,3.0)to[out=270,in=90]node[left]{}(4.1,0.1);  
\draw[<->,line width=0.5mm,dashed,draw=black!50](4.1,3.0)to[out=180,in=0]node[left]{}(2.4,3.0);      
\end{tikzpicture}

\begin{tikzpicture}[overlay]
\node[anchor=north,text width=4.0cm, text centered] at (1.5,8.6) {\bf \scriptsize $p(k_{1,1} \vert M_{1},\mathcal{D})$};
\node[anchor=north,text width=5.0cm, text centered] at (4.5,8.6) {\bf \scriptsize $q(u_{1,1}\vert k_{2,1},k_{2,2})$};
\node[anchor=north,text width=4.0cm, text centered] at (5.2,2.5) {\bf \scriptsize $q(u_{2,2} \vert k_{1,1})$};
\node[anchor=north,text width=4.0cm, text centered] at (4.6,5.6) {\bf \scriptsize $p(k_{2,1},k_{2,2} \vert M_{2},\mathcal{D})$};
\node[anchor=north,text width=4.0cm, text centered] at (1.2,4.5) {\bf \scriptsize $q(u_{2,1} \vert k_{1,1})$};
    
\draw[-,line width=0.1mm,dashed,draw=black!50](0.0,8.5)to[out=270,in=90]node[left]{}(0.0,0.5);  

\end{tikzpicture}
\label{DenA}
\end{subfigure}
\caption{Sensitivity-based move determination (right) leads to the posterior densities of Model 1\ ($p(k_{1,1} \vert M_{1},\mathcal{D})$) and Model 2 ($p(k_{2,1},k_{2,2} \vert M_{2},\mathcal{D})$) being aligned with proposal densities $q(u_{1,1}\vert k_{2,1},k_{2,2})$ and $q(u_{2,1},u_{2,2} \vert k_{1,1})$ respectively; produces higher MCMC acceptance rates. Without sensitivity-based proposals (left), densities are unaligned.}
\label{aligningDensitiesBySensitivityAnalysis}
\end{figure}

For instance, consider the example on the left of Figure \ref{aligningDensitiesBySensitivityAnalysis}, where the observable is highly sensitive to the value of the rate constant of reaction 1 ($k_{1,1}$ in Model 1 and $k_{2,1}$ in Model 2). Moves that keep the value of the rate constant fixed ($k_{1,1}=k_{2,1}$) when transitioning between Model 1 and Model 2 would have very low acceptance probabilities. Here we propose to further improve sampling efficiency by identifying critical reactions common to the current and the proposed networks, and updating these reactions' rate constants in moves between the two networks. As a result, the reverse move from a high-dimensional effective network to a low-dimensional effective network will no longer be deterministic.
Figure \ref{aligningDensitiesBySensitivityAnalysis} (right) shows pictorially how the aligning of densities is improved by including the high-sensitivity rate constant in proposal construction. When a move is proposed between Model 1 and Model 2, the value of the common rate constant is updated. In particular, the proposal distribution $q(u_{2,1},u_{2,2}|k_{1,1})$ for the forward move (from Model 1 to Model 2) is designed to approximate the conditional posterior $p(k_{2,1},k_{2,2}|k_{2,\sim\{1,2\}},M_{2},\mathcal{D})$, and the proposal $q(u_{1,1}|k_{2,1},k_{2,2})$ for the reverse move (from Model 2 to Model 1) is designed to approximate the conditional posterior $p(k_{1,1}|k_{1,\sim1},M_{1},\mathcal{D})$. 

The next question is how to identify the critical reactions whose inclusion in the proposal would improve MCMC mixing, and how to do so with limited computational overhead. We cannot simply include \textit{all} the common reactions, as the tuning of MCMC proposals becomes increasingly demanding as dimension increases; indeed, there is a tension between including too many reactions and too few. 
Also, we want a relatively cheap criterion for identifying reactions; a more expensive criterion might not sufficiently improve mixing to overcome its cost.
Given a set of observables and the current and proposed networks, our strategy is to identify the reactions to which the posterior density is most sensitive. Suppose that we have network $M$ with reactions $R_{1}, R_{2},\ldots,R_{M}$. We can rank the reactions according to their expected local sensitivity index, 
$ \mathbb{E} \left[\left\vert \partial_{k_{M,i}} \log p(k_{M,i} \vert \mathcal{D},k^{*}_{M,\sim\{i\}}, M) \right\vert\right] $
 for reaction $i$, with $k^{*}_{M,\sim\{i\}}$ set to a nominal value and the expectation taken with respect to the prior distribution $p(k_{M,i}|M)$. In practice, since the expectation is usually not analytically tractable, we settle for a noisy estimate by evaluating the local sensitivity at a few realizations from the prior distribution and taking their average. Note the similarity with the Morris method for global sensitivity analysis \cite{Morris1991}, except that we condition on nominal values of $k^{*}_{M,\sim\{i\}}$. 

Having determined the sensitivities of the log-posterior of the current and the proposed reaction network, we select a random number of highest-sensitivity reactions common to the two networks and include proposals for their rate constants in the forward and the reverse moves. The number of reactions to be included in the proposals is a draw from a Poisson distribution whose mean is kept at a small value. Choosing to include only a few common rate constants in the proposals is again based on the understanding that constructing effective proposals in high dimensions is generally hard. 

To illustrate this idea, consider a move from a lower-dimensional model $M_m$ to a higher-dimensional model $M_n$. The mapping $f$ from $(k_{m},u_{m})$, the rate constants and proposal parameters of network $M_m$, to $(k_{n},u_{n})$, the rate constants and proposal parameters of network $M_n$, is given by
\begin{align}
k_{n,1:i}=k_{m,{1:i}},\ k_{n,i+1:i+c}=u_{m,1:c},\  k_{n,{i+c+1:i+a}}=u_{m,c+1:a},\  u_{n,1:c}=k_{m,{i+1:i+c}}.
\end{align}
Here, ${\{1:i\}}$ are indices of reactions that are common to the two networks and whose rate constants are kept fixed during moves between networks $M_m$ and $M_n$, ${\{i+c+1:i+a\}}$ are indices of reactions that are in network $M_n$ but not in $M_m$, and ${\{i+1:i+c\}}$ are the highest-sensitivity reactions that are present in both networks, and whose rate constants are updated according a proposal distribution rather than kept fixed during the between-model move. The parameter proposal densities $q(u_{m,1:a} \vert k_{m,1:i})$ and $q(u_{n,1:c} \vert k_{n,1:i})$ are constructed to approximate the conditional posteriors  $p(k_{n,i+1:i+a} \vert k_{m,1:i},M_n,\mathcal{D})$ and $p(k_{m,i+1:i+c} \vert k_{n,1:i},M_m,\mathcal{D})$, respectively. The reverse move in this case is non-deterministic and involves generating a proposal $u_{n,1:c} \sim q(u_{n,1:c} \vert k_{n,1:i})$ for the rate constants of the common critical reactions ${\{i+1:i+c\}}$. 

For simplicity, the discussion in this subsection has not explicitly considered knowledge of the effective networks underlying a given move. But these sensitivity-based parameter updates can be combined with the ideas of Section \ref{sec:networkaware} on network-aware proposals. In other words, when changing ENs, we construct a proposal that approximates the conditional posterior of the rate constants associated with the difference between the current and proposed ENs. But we simultaneously update the parameters of the highest-sensitivity reactions \textit{common} to the two ENs. The resulting updates are called \emph{sensitivity-based network-aware proposals}. Details of our sensitivity-based NA sampler, along with pseudocode for the steps involved, are given in the supplementary material. 

\subsection{Derandomization of conditional expectations}
The identification of \emph{clusters} of models sharing an identical effective network can be used for additional variance reduction, by post-processing the MCMC output.  All models in a cluster have identical model evidence; we can use this knowledge to compute certain expectations analytically and thereby obtain posterior estimates with lower variances. Details are given in the supplementary material. 

\section{Results}
\label{sec:results}
We present two example problems to evaluate the efficiency of our network-aware (NA) sampling approaches, compared to a network-unaware (NuA) method. The examples we consider are a subset of reactions proposed for a protein-signalling network of the activation of extracellular signal-regulated kinase (\texttt{ERK}) by epidermal growth factor (\texttt{EGF}) \cite{Xu2010}. The network diagram is shown in Figure \ref{fig:sub1}. The observable in our examples is the time-dependent concentration of \texttt{BRaf}, whose dynamics are modeled using the law of mass action/Michaelis-Menten functionals. The corresponding ODE model governing the evolution of all species concentrations, along with the numerical simulation methodology, are detailed in the supplementary material. 

\subsection{Likelihood function}
The Bayesian approach requires specifying a  likelihood function {$p(\mathcal{D}|k_M,M)$}, where {$\mathcal{D}$} are the data and {$k_M$} are the reaction parameters. We employ an i.i.d.\ additive Gaussian model for the difference between model predictions and observations; thus the data are represented as
\begin{equation}
\mathcal{D}=G(M, k_M)+\epsilon_{d},
\end{equation}
where $\epsilon_{d} \sim \mathcal{N}_{d}(0,\sigma^{2}I_{d})$, $d$ is the number of observations, $I_{d}$ is an {$d$}-by-{$d$} identity matrix, and $G(M,k_M)$ represents the predicted concentrations of \texttt{BRaf} given a reaction network $M$ with rate constant values $k_M$. The deterministic predictions $G(M, k_M)$ are obtained with the ODE integrator.

\subsection{Prior specification}
\label{sec:priorspecification}
Since reaction rate constants must be positive, while their uncertainties may span multiple orders of magnitude, we endow each rate constant with an independent log-normal prior distribution. That is,
\begin{equation}
{p(k_{M,i}):\log_{10} k_{M,i} \sim \mathcal{N}(\mu_{p,i},\sigma_{p,i}^{2})}.
\end{equation}
%
The prior distribution over models, $p(M)$, is taken to be uniform in the following examples.

\begin{figure}[h]
\makeatletter
\def\@captype{figure}
\makeatother
{\begin{minipage}[h]{0.49\textwidth}
\begin{subfigure}[b]{1.0\textwidth}
  \includegraphics[width=1.0\linewidth]{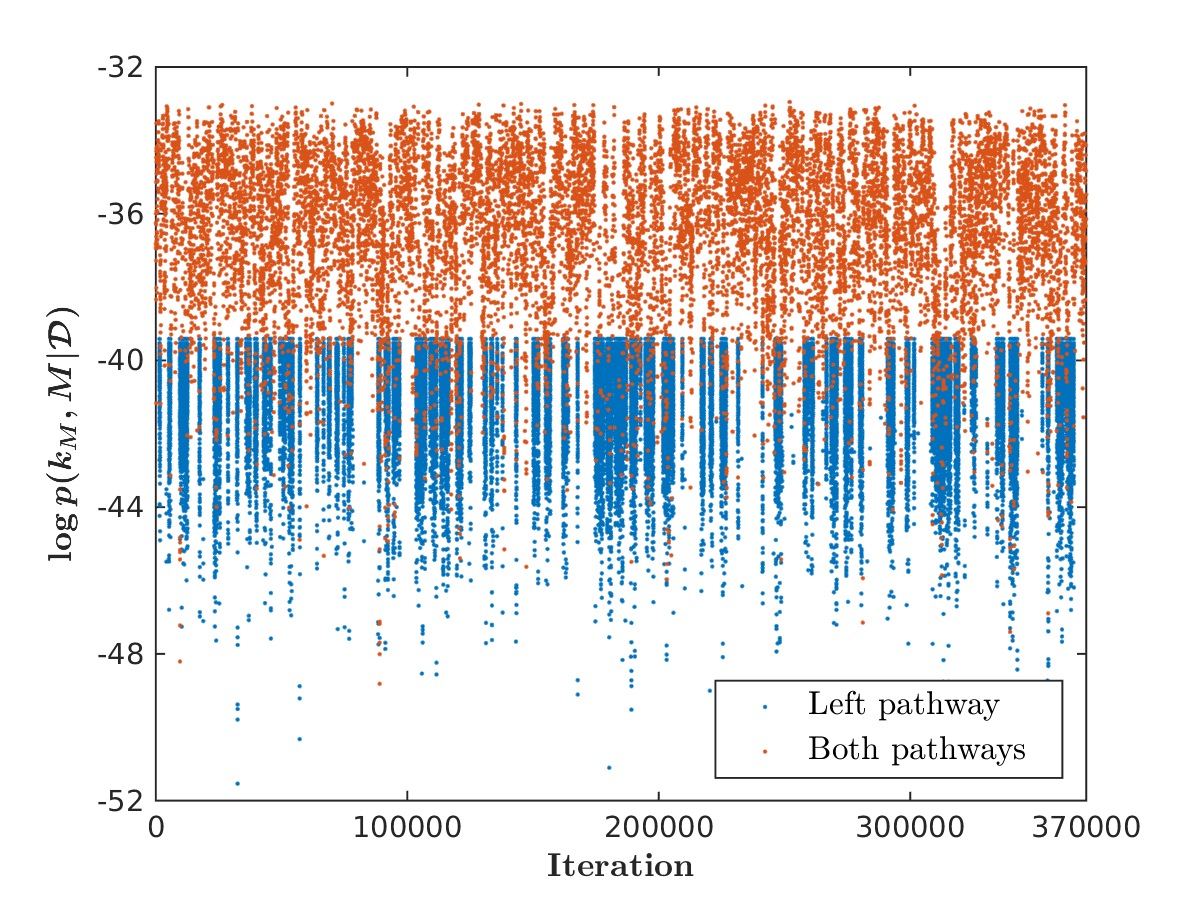}
  \caption{Standard network-unaware}
  \label{fig:Example1NetworkUnaware}
\end{subfigure}%
\end{minipage}
\begin{minipage}[h]{0.49\textwidth}
\begin{subfigure}[b]{1.0\textwidth}
  \includegraphics[width=1.0\linewidth]{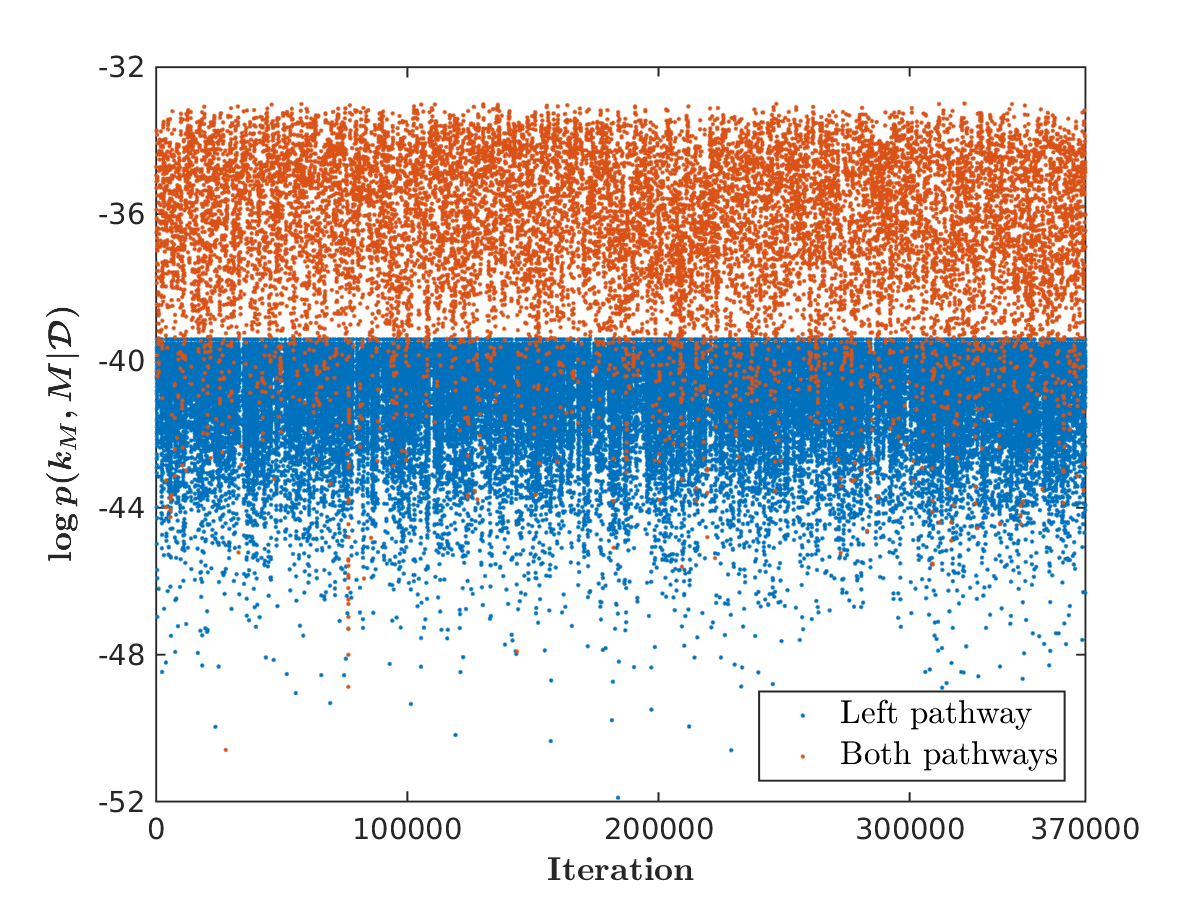}
  \caption{Network-aware}
  \label{fig:Example1NetworkAware}
\end{subfigure}
\end{minipage}
}
\caption{MCMC trace plots for Example 1}
\label{fig:traceplotsExample1}
\end{figure}
\subsection{Example 1: five-dimensional nonlinear network inference}
In our first example, we fix reactions 1, 2, 8, 9, 10, 11, and 12, i.e., we include them in all candidate models. The rate constants of all fixed reactions and Michaelis constants of all reactions are set to their base values (see supplementary material). The presence of Reactions 3, 4, 5, 6, and 7, along with their rate constants, are taken to be uncertain. With these uncertain reactions, the number of potential models is 32. And with only \texttt{BRaf} as the observable, the number of effective networks is five (see network diagrams in supplementary material). 
We generate 20 data points by simulating \texttt{BRaf} concentrations at different times with rate constants and Michaelis constants set to their base values  and adding i.i.d noise with mean $\mu=0$ and variance $\sigma^{2} =4$.
 The simulated data are available in the supplementary material. The noise variance in the likelihood function matches that of the data generating process. The prior uncertainties of the rate constants are listed in the supplementary material. We perform five replicate MCMC runs, each with 400\,000 steps, using both NuA and NA (Section \ref{sec:networkaware}) approaches. 
 30000 samples each were discarded as burn-in. MCMC trace plots of the log-posterior density value using the two approaches are shown in Figure \ref{fig:traceplotsExample1}. 

Each dot in the MCMC traceplots is color-coded according to whether the associated model operates with only the left pathway (subsets of the reaction network in Figure \ref{fig:sub1} that contain reactions 1, 2, 8, 10, and 12, and exclude at least one of reactions 3, 5, and 6; colored blue) 
or both pathways (subsets of the reaction network in Figure \ref{fig:sub1} that contain reactions 1, 2, 3, 5, 6, 8, 10, and 12; colored orange). The frequency of moves between these two model types is higher with our NA approach, indicating superior mixing. 
 \begin{figure}
 \makeatletter
 \def\@captype{figure}
 \makeatother
{\centering  
\includegraphics[width=0.5\linewidth]{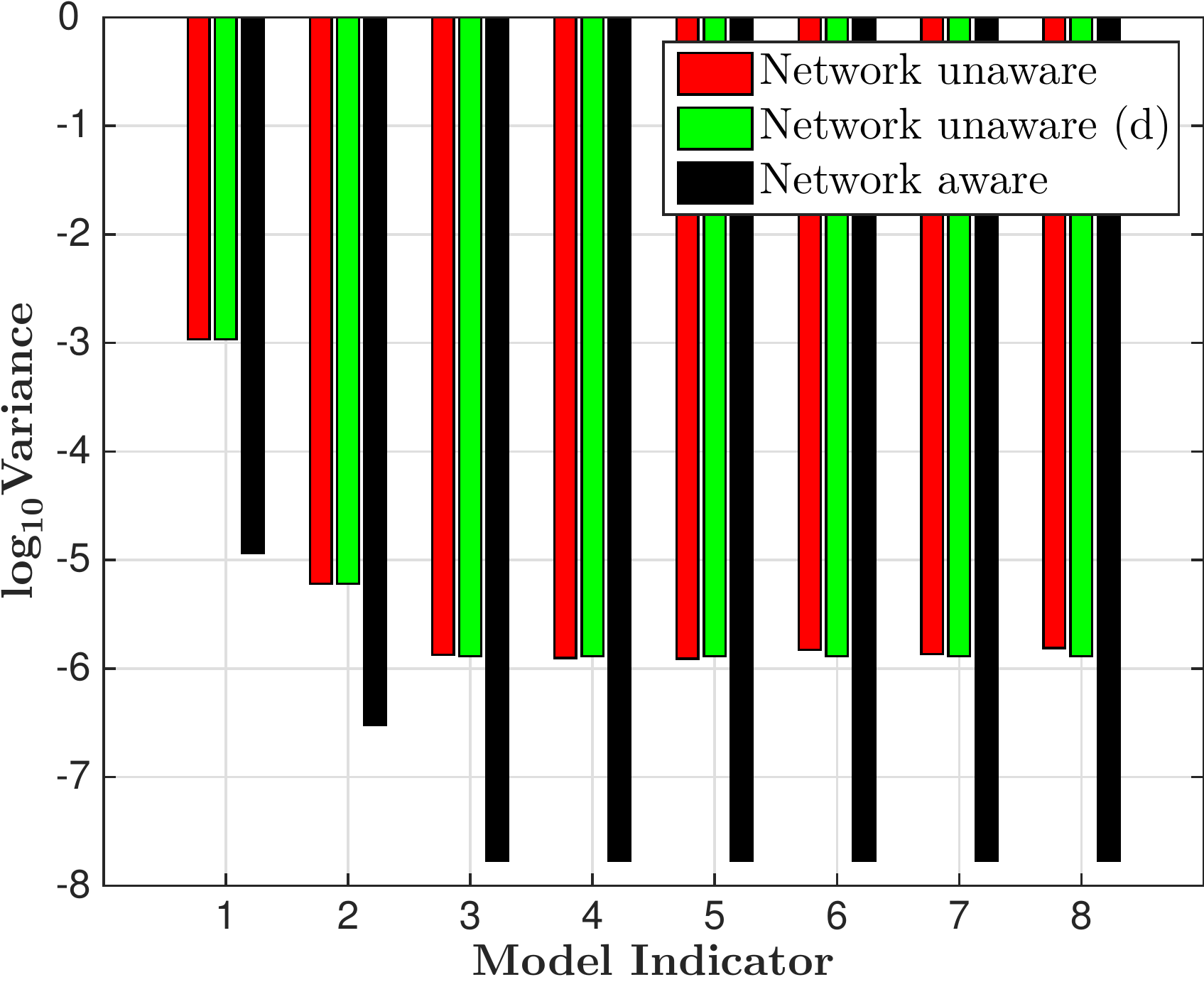}
\caption{Variance comparison for the eight highest-probability models in Example 1: network-unaware without derandomization, network-unaware (d) with derandomization, and network-aware with derandomization.}
\label{fig:variancesExample1}
}
\end{figure}
We can also evaluate the performance of the two approaches by computing the variance of each posterior model probability estimate. In Figure \ref{fig:variancesExample1}, we show variance estimates with the NuA approach, NuA approach with derandomization applied as a post-processing step, and our NA approach with derandomization. Our NA approach produces model probability estimates whose variances are two orders of magnitude lower.
Table \ref{Example2Table} shows the acceptance rates of between-model and between-cluster moves for the two approaches. Higher model-switching and cluster-switching acceptance rates are 
another indication of superior mixing of the NA approach. 

Another useful MCMC diagnostic is the effective sample size (ESS) for a statistic that retains interpretation throughout the simulation. (The ESS of a set of correlated samples is the number of independent samples that would yield expectation estimates with the same variance \cite{Gilks1996}.) To this end, we choose our statistic to be the number of reactions in the model and report its ESS in Table \ref{Example2Table}.  The NA scheme has an  ESS ten times higher than that of the NuA approach. A more complete comparison of the two schemes should also incorporate relative differences in computational time; hence the ESS per minute diagnostic in the last column of Table \ref{Example2Table}, which again favors the NA approach. Note that the absolute value of ESS per minute depends quite strongly on the relative and absolute tolerance settings of the ODE solver. Here we chose very tight tolerances, but higher ESS per minute can be obtained with looser tolerances. 
%
\begin{table}[h]
\begin{center}
\begin{tabular}{|ccc|ccc|ccc|ccc|ccc|ccc|}
\hline
\multicolumn{3}{|c|}{Method$^{\dagger}$} & \multicolumn{3}{c|}{$p(M)^{a}$} & \multicolumn{3}{c|}{$\bar{\alpha}_{M}^{b}$} & \multicolumn{3}{c|}{$\bar{\alpha}_{C}^{c}$}  & \multicolumn{3}{c|}{ESS$^{d}$} & \multicolumn{3}{c|}{ESS/min$^{e}$}\\
\hline
\multicolumn{3}{|c|}{Network-unaware} & \multicolumn{3}{c|}{0.7545} & \multicolumn{3}{c|}{0.19} & \multicolumn{3}{c|}{0.015} & \multicolumn{3}{c|}{10} & \multicolumn{3}{c|}{0.175}\\
\multicolumn{3}{|c|}{Network-aware} & \multicolumn{3}{c|}{0.7544} & \multicolumn{3}{c|}{0.22} & \multicolumn{3}{c|}{0.034} & \multicolumn{3}{c|}{110} & \multicolumn{3}{c|}{0.301}\\
\hline 
\multicolumn{1}{l}{$\dagger$: Performance is averaged over 5 simulation runs}\\
\multicolumn{1}{l}{$a$: Posterior probability of the data-generating model}\\
\multicolumn{1}{l}{$b$: Between-model move acceptance rate}\\
\multicolumn{1}{l}{$c$: Between-cluster move acceptance rate}\\
\multicolumn{1}{l}{$d$: Effective sample size for 10000 samples}\\
\multicolumn{1}{l}{$e$: Timings depend on the tolerances chosen for the ODE solver}
\end{tabular}
\end{center}
{
\caption{Summary statistics of MCMC simulation performance (Example 1).} 
\label{Example2Table}
}
\end{table} 
\subsection{Example 2: ten-dimensional nonlinear network inference}
\noindent
In our second example, we keep only reactions 1 and 2 fixed. The rate constants of the fixed reactions and Michaelis constants of all reactions are set to their base values (see supplementary material). Reactions 3--12 are uncertain, producing a total of 1024 potential models. With \texttt{BRaf} as the observable, the number of ENs is 24 (see network diagrams in supplementary material). We set all rate constants and Michaelis constants to their base values and generate 30 data points perturbed with i.i.d.\ zero-mean Gaussian noise of variance $\sigma^{2}=0.04$. A smaller noise variance in this example makes the likelihood more sensitive to the rate constants common to the current and proposed networks of between-model moves. As a result, we will see that even the NA approach (Section \ref{sec:networkaware}), in which the rate constants of the common reactions are kept fixed, has difficulty exploring the model space. 
In contrast, sensitivity-based samplers (Section \ref{sec:sensitivitynetworkaware}) are able to identify reactions to which the observable is most sensitive, and then build proposals that approximate the posterior distributions of associated rate parameters. The simulated data are available in the supplementary material. The noise variance in the likelihood function matches that of the data generating process. The prior uncertainties of the rate constants are listed in the supplementary material. 

\begin{figure}[h]
\begin{minipage}[h]{0.49\textwidth}
\begin{subfigure}[b]{1.0\textwidth}
  \includegraphics[width=1.0\linewidth]{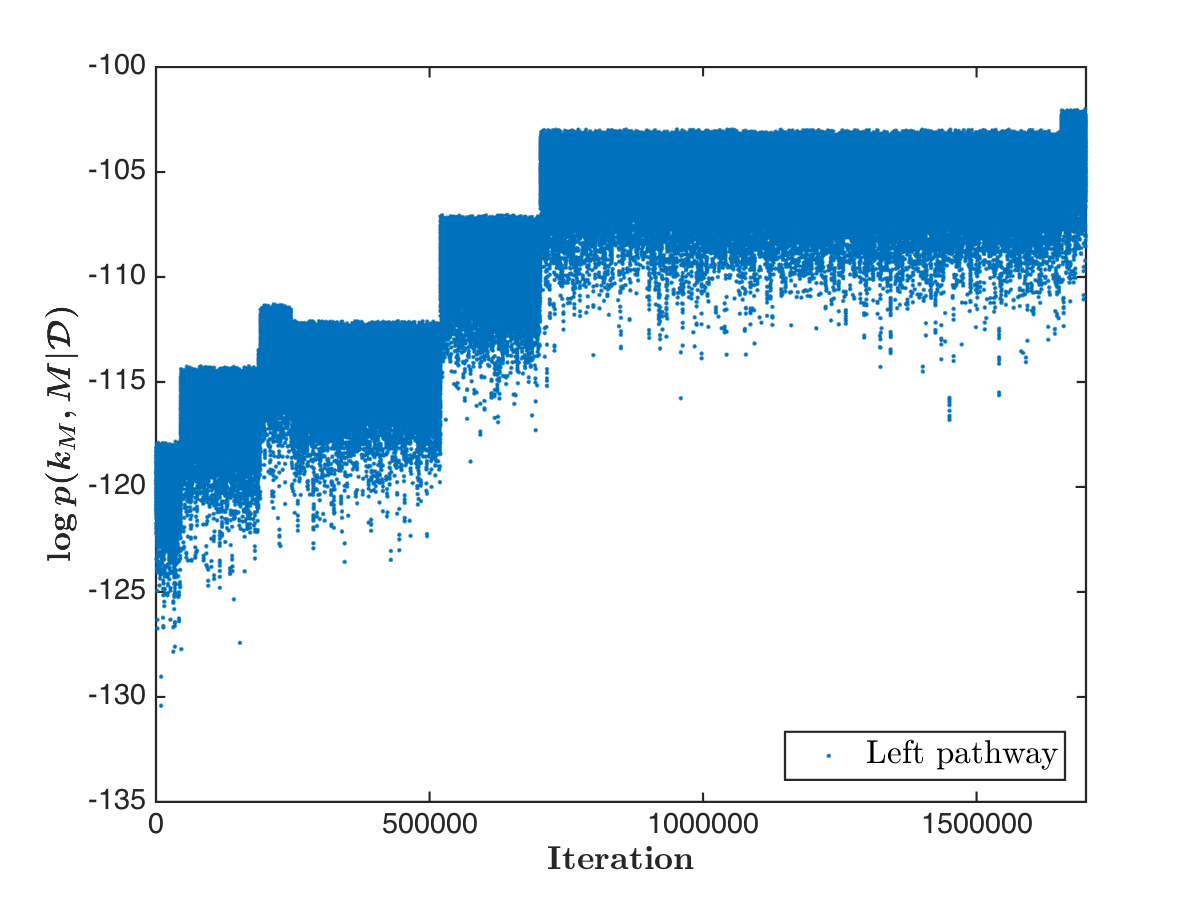}
  \caption{Standard network-unaware}
  \label{fig:Example2NetworkUnaware}
\end{subfigure}%
\end{minipage}
\begin{minipage}[h]{0.49\textwidth}
\begin{subfigure}[b]{1.0\textwidth}
  \includegraphics[width=1.0\linewidth]{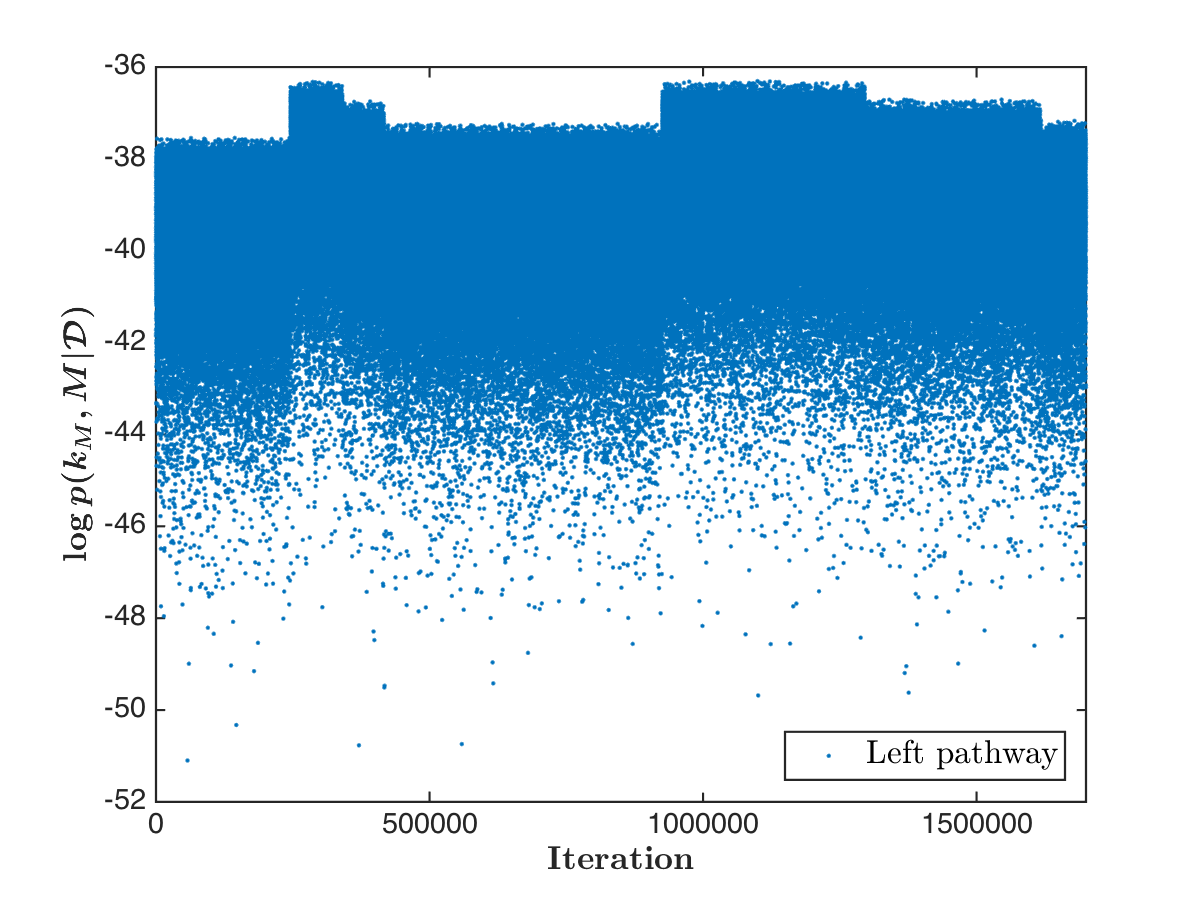}
  \caption{Network-aware}
  \label{fig:Example2NetworkAware}
\end{subfigure}%
\end{minipage}
\begin{minipage}[h]{0.49\textwidth}
\begin{subfigure}[b]{1.0\textwidth}
  \includegraphics[width=1.0\linewidth]{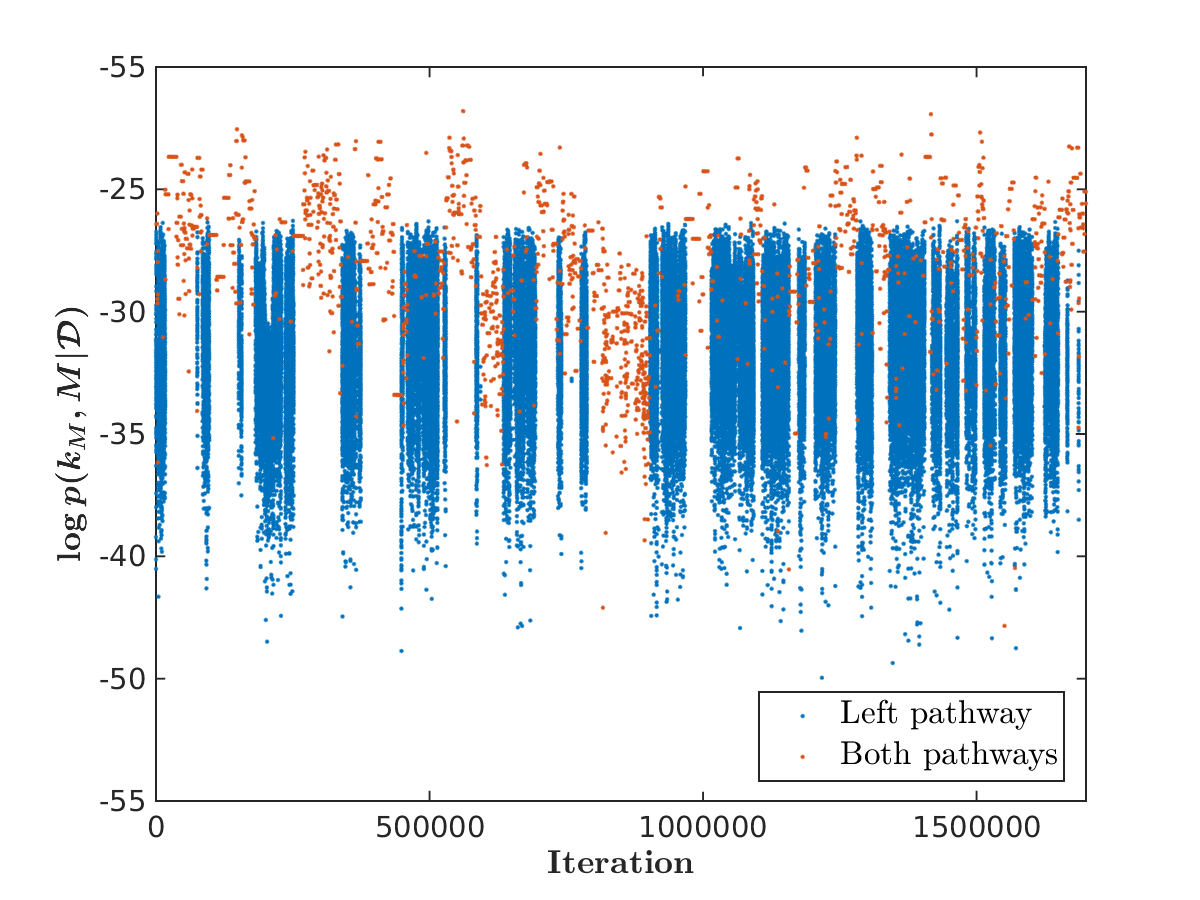}
  \caption{Sensitivity-based network-unaware}
  \label{fig:Example2SensBasedNetworkUnaware}
\end{subfigure}%
\end{minipage}
\begin{minipage}[h]{0.49\textwidth}
\begin{subfigure}[b]{1.0\textwidth}
  \includegraphics[width=1.0\linewidth]{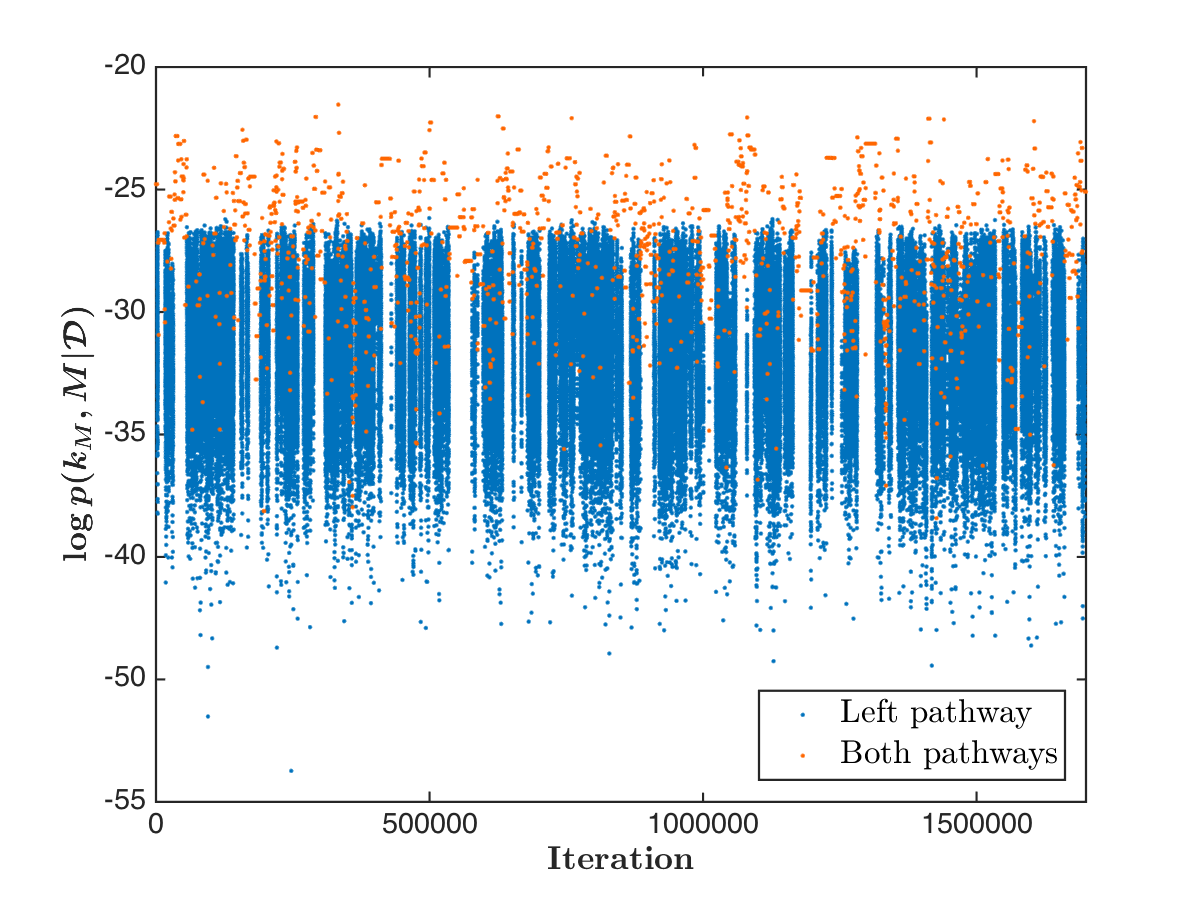}
  \caption{Sensitivity-based network-aware}
  \label{fig:Example2SensBasedNetworkAware}
\end{subfigure}
\end{minipage}
\caption{MCMC trace plots for Example 2}
\label{fig:traceplotsExample2}
\end{figure}

We simulate four MCMC chains, each 2 million steps long, using the standard NuA approach, the NA approach (Section \ref{sec:networkaware}), and the sensitivity-based NuA and NA approaches (Section \ref{sec:sensitivitynetworkaware}). 300\,000 samples were discarded as burn-in from each chain. MCMC trace plots of the four approaches are shown in Figure \ref{fig:traceplotsExample2}. We see that MCMC sampling of this 10-dimensional problem is in fact intractable using the standard approach, even our NA approach (Section \ref{sec:networkaware}): this is demonstrated by the inability of the samplers to switch between models belonging to left pathway  (subsets of the reaction network in Figure \ref{fig:sub1} that contain reactions 1, 2, 8, 10, and 12, and exclude at least one of reactions 3, 5, and 6) and models that incorporate both pathways (subsets of the reaction network in Figure \ref{fig:sub1} that contain reactions 1, 2, 3, 5, 6, 8, 10, and 12). Indeed, both chains at the top of Figure~\ref{fig:traceplotsExample2} are, by visual inspection, very far from stationarity.

In contrast, our sensitivity-based samplers are able to move between the two pathways and explore the posterior distribution over all models and parameters. We can then compare the sampling efficiency of the sensitivity-based NA algorithm to that of the sensitivity-based NuA approach. We observe that the frequency of moves between the left-pathway models and both-pathways models is higher with the NA approach, indicating faster posterior exploration. Table \ref{Example4Table} shows the acceptance rates of between-model moves and between-pathway moves for the two approaches. High model-switching and pathway-switching acceptance rates with the same posterior inference is an indication of the superior mixing of the network-aware approach. In Table \ref{Example4Table}, we also present the ESS for the number-of-reactions-in-model statistic. The network-aware approach has an ESS that is roughly three times the ESS obtained using the network-unaware approach. The ESS per minute diagnostic in the last column of Table \ref{Example4Table} also supports the use of the network-aware approach. Again, we chose very tight tolerances for the ODE solver, and higher ESS per minute can be obtained with looser tolerances.  

\begin{figure}[h]
{\centering
  \includegraphics[width=1.0\linewidth]{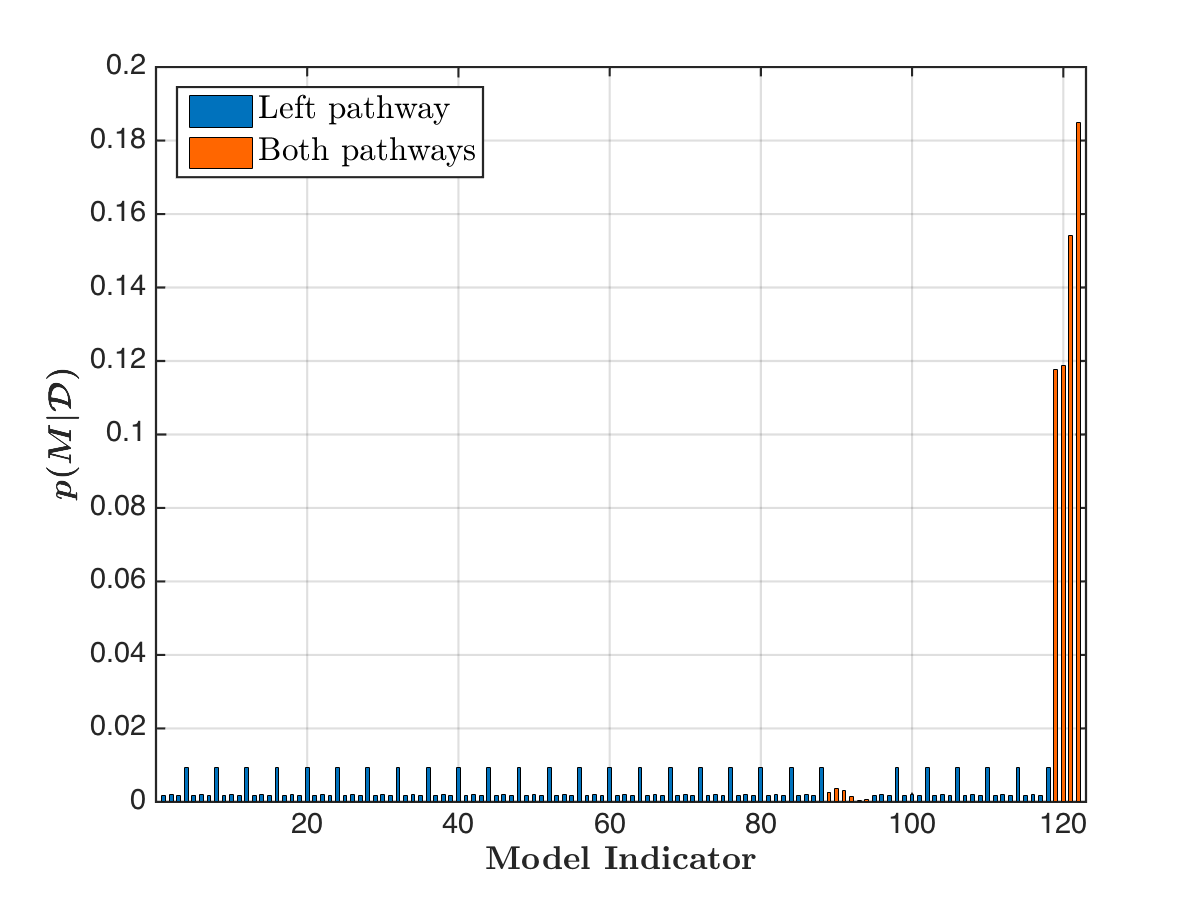}
  \caption{Example 2: Posterior probabilities of all models with non-zero probability, the network diagram of the model with the highest posterior probability, and the one- and two-dimensional marginal densities of the parameters of the highest-probability model.}
\label{fig:Example2}} 
\begin{tikzpicture}[overlay]
\node[anchor=north,text width=4.0cm, text centered] at (-2.5,9.8) {\includegraphics[width=1.4\linewidth]{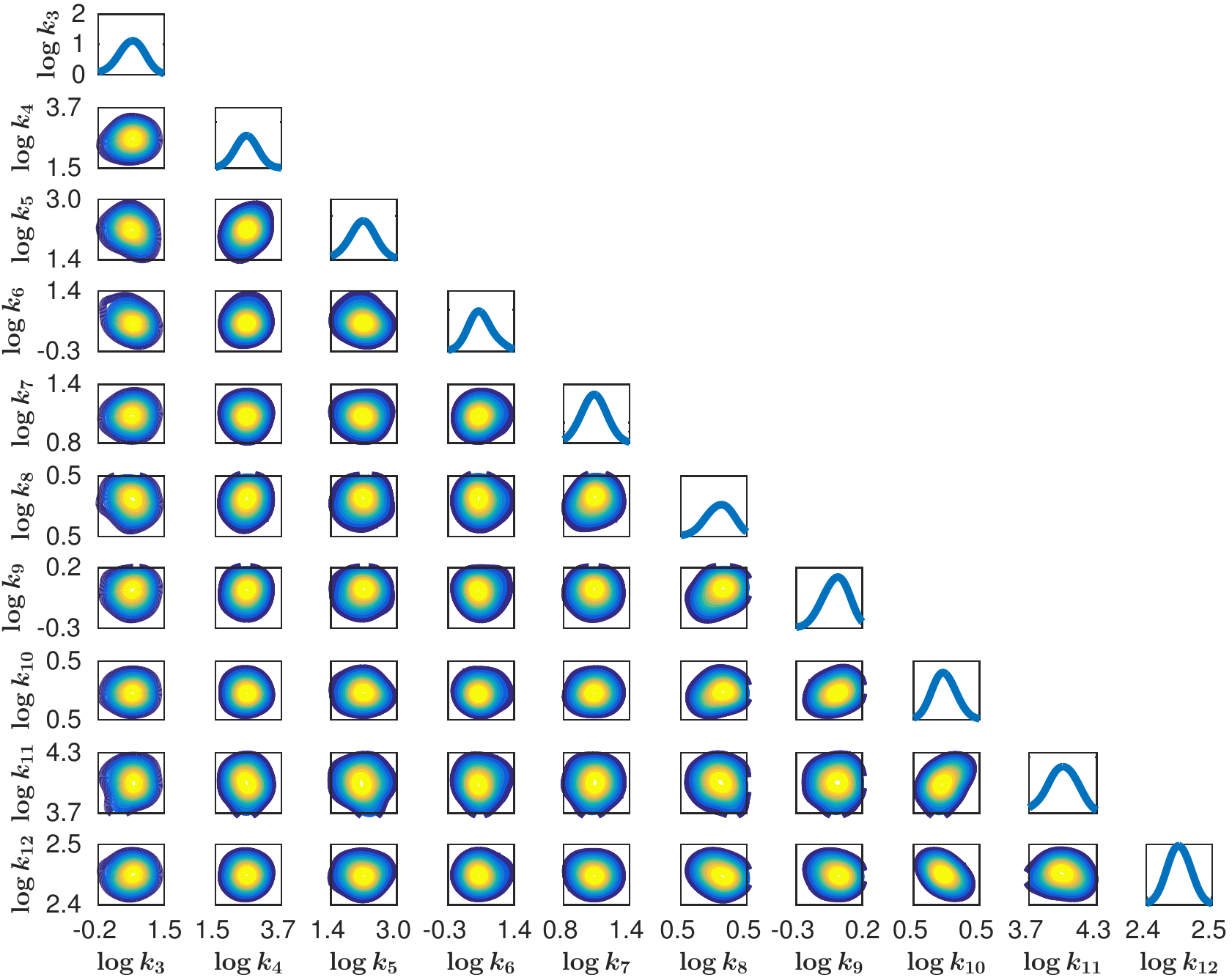}};   
\end{tikzpicture}
\begin{tikzpicture}[overlay]
\node[anchor=north,text width=4.0cm, text centered] at (2.0,13.5) {\includegraphics[width=1.4\linewidth]{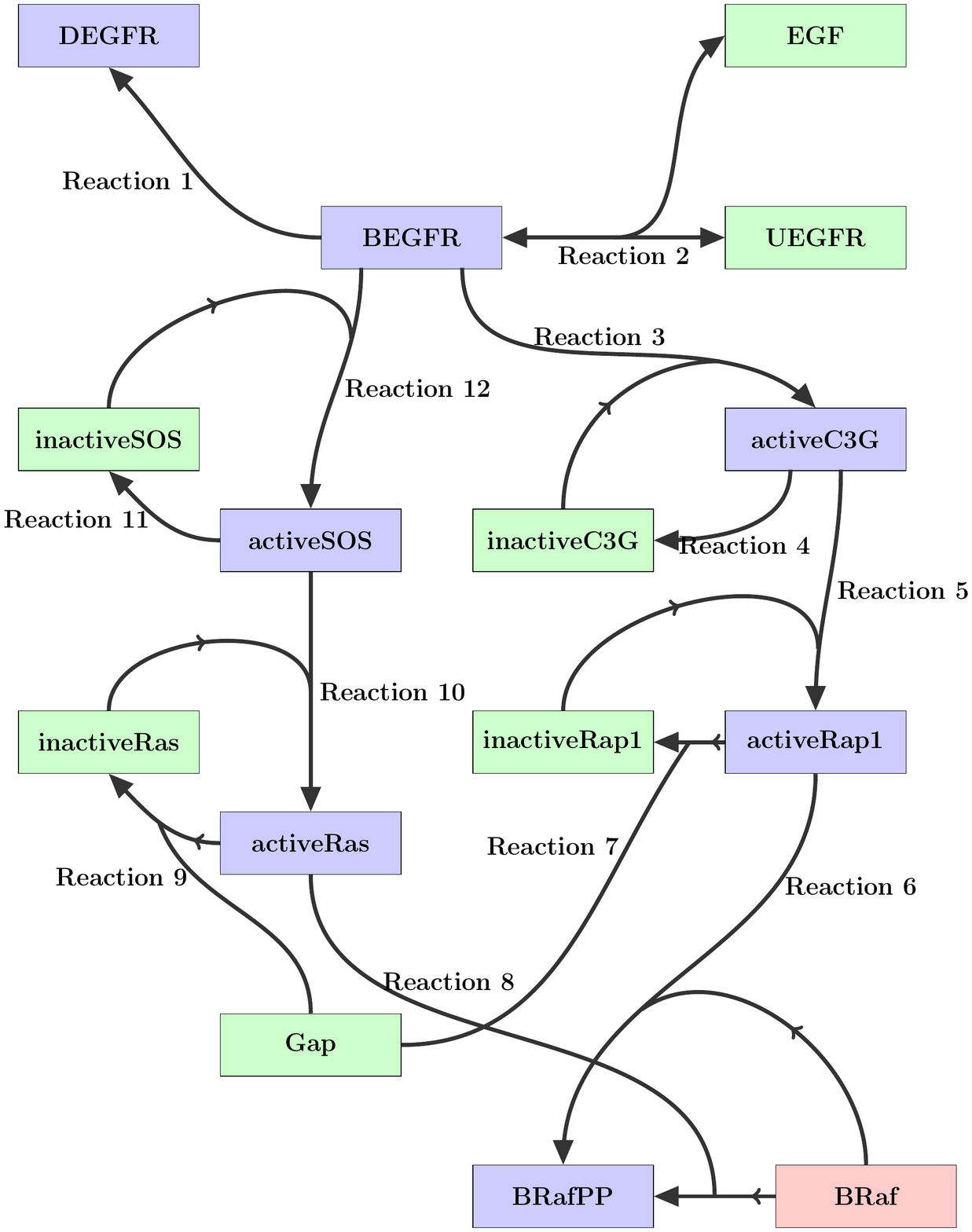}};   
\end{tikzpicture}
\begin{tikzpicture}[overlay]
\draw[->,line width=0.2mm,draw=black](5.8,12.5)to[out=180,in=0]node[left]{}(5.0,12.5);
\draw[->,line width=0.2mm,draw=black](-0.0,8.4)to[out=180,in=0]node[left]{}(-0.8,7.6);  
\end{tikzpicture}
\end{figure}

Overall, the sensitivity-based NA approach is a sampler that allows efficient large-scale inference of nonlinear reaction networks and provides myriad different types of information: posterior probabilities of all models, posterior probabilities of particular pathways, parameter posterior densities for any EN, etc. Several of these posterior characterizations are given in Figure \ref{fig:Example2}.

\begin{table}[h]
\begin{center}
\begin{tabular}{|ccc|ccc|ccc|ccc|ccc|ccc|}
\hline
\multicolumn{3}{|c|}{Method$^{\dagger}$} & \multicolumn{3}{c|}{$p(M)^{a}$} & \multicolumn{3}{c|}{$\bar{\alpha}_{M}^{b}$} & \multicolumn{3}{c|}{$\bar{\alpha}_{P}^{c}$}  & \multicolumn{3}{c|}{ESS$^{d}$} & \multicolumn{3}{c|}{ESS/min$^{e}$}\\
\hline
\multicolumn{3}{|c|}{Sensitivity-based network unaware} & \multicolumn{3}{c|}{0.145} & \multicolumn{3}{c|}{0.095} & \multicolumn{3}{c|}{0.0013} & \multicolumn{3}{c|}{86} & \multicolumn{3}{c|}{3.37$\times 10^{-3}$}\\
\multicolumn{3}{|c|}{Sensitivity-based network aware} & \multicolumn{3}{c|}{0.157} & \multicolumn{3}{c|}{0.145} & \multicolumn{3}{c|}{0.0027} & \multicolumn{3}{c|}{275} & \multicolumn{3}{c|}{8.66$\times 10^{-3}$}\\
\hline 
\multicolumn{1}{l}{$\dagger$: Performance is averaged over 5 simulation runs}\\
\multicolumn{1}{l}{$a$: Posterior probability of the data-generating model}\\
\multicolumn{1}{l}{$b$: Between-model move acceptance rate}\\
\multicolumn{1}{l}{$c$: Between-pathway move acceptance rate}\\
\multicolumn{1}{l}{$d$: Effective sample size for 1000000 samples}\\
\multicolumn{1}{l}{$e$: Timings depend on the tolerances chosen for the ODE solver}
\end{tabular}

\end{center}
{
\caption{Summary statistics of MCMC simulations (Example 2).} 
\label{Example4Table}
}
\end{table}

\section{Conclusions}
\label{sec:concs}

The inference of chemical reaction networks---particularly networks encoding the structure of differential equation models, as associated with mass action kinetics---is important to mechanistic insight and quantitative performance predictions in many applications. The essential challenge of inference is to efficiently explore the combinatorially large model space of possible network structures and, simultaneously, the associated model parameters. Systematic Bayesian inference of reaction networks has typically been infeasible due to the difficulties of such model-space sampling.
%
%
%
The algorithms developed in this paper exploit \emph{structural properties} of reaction networks to achieve more efficient sampling and inference, in nonlinear settings where likelihoods and posteriors are analytically intractable. 

Future work might incorporate the NA approaches of this paper into across-model sampling methods that do not require mapping between parameter spaces. The analysis proposed here could also accelerate the goal-oriented simulation of large networks by identifying the effective network of a reaction model \textit{a priori} and only solving the ODE system corresponding to the effective network of the given model and observables. 

%
%
\section{Data accessibility}
The datasets supporting this article and the code implementing the developed methodology can be found at \url{https://github.com/nikhilgalagali/networkinference}.

\section{Funding}
We acknowledge support from BP through the BP-MIT Energy Conversion Program.

\newpage
\section{Appendix}

\begin{algorithm}[h]
\begin{algorithmic}[1]
\State {\bf Definitions}: $R_{e}$: reactions in the effective network; $S_{e}$: species in the effective network; $r_{i}$: reactants of reaction $i$; $p_{i}$: products of reaction $i$; $a_{i}$: enzymes of reaction $i$
\State {\bf Given}: $R_{prop}$: proposed reactions; $S_{in}$: species initially present \\ 
$R_{e}=\varnothing$, $S_{e}=S_{in}$, $n_{e}=0$, $t_{e}=0$
\While{$n_{e} \neq |R_{e}|$ or $t_{e} \neq |S_{e}|$}
\State $n_{e}=|R_{e}|$ and $t_{e}=|S_{e}|$
\For {$i=1$ to $|R_{prop}|$}

\If {Reaction $R_{i}$ irreversible}
\If {($r_{i} \cup a_{i}) \in S_{e}$}
\State  $R_{e}=R_{e} \cup R_{i}$ and $S_{e}=S_{e} \cup p_{i}$ 
\EndIf
\ElsIf {Reaction $R_{i}$ is reversible}
\If {($r_{i} \cup a_{i}) \in S_{e}$ or ($p_{i} \cup a_{i}) \in S_{e}$}
\State $R_{e}=R_{e} \cup R_{i}$ and $S_{e}=S_{e} \cup p_{i} \cup r_{i}$ 
\EndIf
\EndIf
\EndFor
\EndWhile
\State $R_{active}=R_{e}$
\For {$i=1$ to $|R_{active}|$}
\State Inf $\leftarrow$ {\bf Algorithm \ref{alg:advancedreactionmanager} ($R_{active}$, $R_{i}$)}
\If {Inf==0}
\State $R_{e}=R_{e}\setminus\{R_{i}\}$
\EndIf  
\EndFor
\end{algorithmic}
\caption{Effective reaction network from a set of reactions}
\label{alg:effectivereactions}
\end{algorithm}

\begin{algorithm}[h]
\begin{algorithmic}[1]
\State {\bf Given}: $R_{active}$: active reactions; $O$: observables 
\State $S_{inc}$: collection of species, $S_{inc}=r_{i} \cup p_{i}$; $R_{inc}$: collection of reactions, $R_{inc}=R_{i}$\\
$t_{inc}=0$
\While{$t_{inc} \neq |S_{inc}|$}
\State $t_{inc}=|S_{inc}|$
\For {$j=1$ to $|R_{active}|$}
\If {$R_{j} \not \in R_{inc}$}
\If {$r_{j} \cup a_{j} \in S_{inc}$ or $p_{j} \in S_{inc}$}
\State $R_{inc}=R_{inc} \cup R_{j}$   
\EndIf
\EndIf
\EndFor
\For {$j=1$ to $|R_{inc}|$}
\State $S_{inc}=S_{inc} \cup p_{j} \cup r_{j}$
\EndFor
\EndWhile
\If {$O \in S_{inc}$}
\State Inf=1
\EndIf
\end{algorithmic}
\caption{Algorithm to check if reaction $R_{i}$ influences the observables}
\label{alg:advancedreactionmanager}
\end{algorithm}

\newpage

\begin{center}
{\bf Supplementary Information}
\end{center}

\section{Network analysis for improved sampling efficiency}
\label{sec:supp_computation} 
\subsection{Centered Gaussian parameter proposals}
\noindent
In the main paper, we describe the network-unaware approach that does not recognize the effective networks during the between-model moves. We provide further details here. For nested models, as is the case in the  reaction network inference problem, a natural choice of the jump function $f$ is to choose the identity function. Thus, when proposing a move from a lower-dimensional model $M_m$ to a higher-dimensional model $M_n$, the rate constants of the newly added reactions are proposed according to $q(u \vert k_{m})$ and the values of the rate constants of reactions common to the two models are kept fixed. Suppose that model $M_m$ has $i$ reactions and that model $M_n$ has $a>i$ reactions, with the first $i$ reactions common; then an identity $f$ mapping $( {k}_{m,{1:i}}, {u} )$ to $( {k}_{n,{1:i}}, {k}_{n,{i+1:a}})$ is simply
\begin{equation}
{k}_{n,{1:i}} =  {k}_{m,{1:i}}, \  {k}_{n,{i+1:a}} = u
\end{equation}

\noindent
and the acceptance probability is given by

\begin{equation}
\alpha(k_{m},k_{n})=\mbox{min}\left \{1, \frac{p(M_n, k_{n} \vert \mathcal{D}) q(M_m \vert M_n)}{p(M_m, k_{m} \vert \mathcal{D}) q(M_n \vert M_m)q(u|k_{m})}\right \}.
\label{identityAcceptanceRatio}
\end{equation}

\noindent
The reverse move in this case is deterministic. Let the proposal $q(u \vert k_{m})$ be given by

\begin{equation}
q(u \vert k_{m})=\mathcal{N}(u; \mu,\Sigma).
\end{equation}

\noindent
To improve the chance of proposal acceptance we center the proposal distribution at the conditional mode of the posterior distribution. Next, we construct an approximation to the posterior distribution by setting the covariance of the Gaussian to be the negative inverse-Hessian of the conditional posterior density at $\mu$. In other words, we construct a Gaussian approximation to the conditional posterior distribution. In the framework of Brooks et al., the above construction is equivalent to the centered second-order conditions \cite{Brooks2003}. In the scheme described above, the mean vector $\mu$ is set to the conditional maximum:
\begin{equation}
\mu=\underset{k_{n,{i+1:a}}}{\arg\max}\mbox{  }p(k_{n,{i+1:a}} \vert k_{m,{1:i}}, M_n,\mathcal{D}). 
\label{supp_firstordercondition}
\end{equation}

\noindent
A proposal centered at the posterior conditional maximum satisfies the first order condition:
\begin{align}
\nabla_{u}\log A(M_m,k_{m} \rightarrow M_n,k_{n})\big|_{\mu}&=\nabla[ \log \mathcal{L}(\mathcal{D}; k_{m,{1:i}},u) + \log p(k_{m,{1:i}},u) \nonumber \\ &\qquad{        } -\log \mathcal{N}(u; \mu,\Sigma) ]\big|_{u=\mu} \nonumber \\ &=0.                                                
\end{align}
\noindent
Here, $\mathcal{L}(\mathcal{D}; k_{m,{1:i}},u)$ refers to the likelihood function of the models and $p(k_{m,{1:i}},u)$ to the prior probability distribution of the rate constants. Further, setting the second-derivative of the acceptance ratio at the conditional maximum to be $0$, we obtain the second order condition as:
\begin{align}
\nabla_{u}^2\log A((M_m,k_{m}) \rightarrow (M_n,k_{n}))\big|_{\mu}&=\nabla^2[\log \mathcal{L}(\mathcal{D}|k_{m,{1:i}},u) + \log p(k_{m,{1:i}},u) \nonumber \\ &\qquad{} - \log \mathcal{N}(u; \mu,\Sigma) ]\big|_{u=\mu} \nonumber \\ &=0.  
\label{supp_secondordercondition}
\end{align}
\noindent
Taking $\mathcal{H}$ to be the Hessian of the conditional posterior density $p(k_{n,{i+1:a}} \vert k_{m,{1:i}}, M_n,\mathcal{D})$ at $\mu$, (\ref{supp_secondordercondition}) yields

\begin{equation}
\mathcal{H}+\Sigma^{-1}=0 \implies \Sigma=-\mathcal{H}^{-1}.
\end{equation}

\subsection{Network-aware parameter proposals}
\label{sec:supp_networkaware}
\noindent
As discussed in the paper, many reaction networks can have the same effective network. In such a case, if the proposed move is between two networks with the same effective network (i.e., the two networks belong to the same cluster), the parameter proposal adapts to the prior distribution of the newly added reaction. We propose a network-aware approach in which, because we have determined the effective networks, we design parameter proposals that adapt to the difference between the effective networks of the two networks. When the proposed move is between two networks belonging to different clusters, we construct a proposal that approximates the conditional posterior distribution of the rate constants of all reactions \emph{not included} in the two effective networks. Formally, suppose that the sampler proposes a move from a lower-dimensional model $M_m$ to a higher-dimensional model $M_n$. Let the effective networks of the two models $M_m$ and $M_n$ be $M_{m_e}$ and $M_{n_e}$, respectively. Suppose the proposed move is such that $M_{n_e} \neq M_{m_e}$, i.e., the effective networks of the current and the proposed networks are different. Further, suppose that model $M_{m_e}$ has $i$ reactions and that model $M_{n_e}$ has $a>i$ reactions, with the first $i$ reactions common; then an identity $f$ mapping $( {k}_{{m_e},{1:i}}, {u} )$ to $( {k}_{n_{e},{1:i}}, {k}_{n_{e},{i+1:a}})$ is simply

\begin{equation}
(k_{n_{e},1:i},k_{n_{e},i+1:a})=(k_{m_{e},1:i},u),
\end{equation}

\noindent
where $u \sim \mathcal{N}(\mu_{m_e},\Sigma_{{m_e}})$. The proposal mean

\begin{equation}
\mu_{{m_e}} =\underset{k_{{n_e},i+1:a}}{\arg\max}\mbox{  }p(k_{{n_e},i+1:a} \vert k_{{m_e},1:i},M_{n_e},\mathcal{D})
\end{equation}
\noindent
is obtained by solving an $(a-i)$-dimensional optimization problem. The proposal covariance

\begin{equation}
\Sigma_{{m_e}} =-[\nabla^2 \log p(k_{{n_e},i+1:a} \vert k_{m_{e},1:i},M_{n_e},\mathcal{D})]^{-1}\big|_{k_{{n_e},i+1:a}=\mu_{{m_e}}}
\end{equation}
\noindent
is determined numerically using a finite-difference approximation at the proposal mean. The acceptance probability of the proposed move is given by

\begin{equation}
\alpha((M_{m_e},k_{m_e}),(M_{n_e},k_{n_e}))=\mbox{min}\left\{1,A\right\},
\end{equation}

\noindent
where

\begin{equation}
A=\frac{p(M_{n_e}, k_{n_{e}} \vert \mathcal{D}) q(M_m \vert M_n)}{p(M_{m_e}, k_{m_{e}} \vert \mathcal{D}) q(M_n \vert M_m)\mathcal{N}(u;\mu_{m_e},\Sigma_{m_e})}.
\end{equation}

\noindent
The reverse move is deterministic and has an acceptance probability $\mbox{min}\{1,A^{-1}\}$. The idea behind the construction of our network aware proposals is that by solving for the maximum of the joint conditional posterior density of reactions that are in $M_{n_e}$ but not in $M_{m_e}$, and determining the Hessian approximation at that point, we are building a Gaussian approximation of the conditional probability density $p(k_{n_e,{i+1:a}} \vert k_{m_e,{1:i}},M_{n_e},\mathcal{D})$. In contrast, the standard network-unaware approach would not; in particular, it produces a proposal that is the product of prior densities for $dim(M_{n_e})-dim(M_{m_e})-1$ rate constants of the reactions that do not change the effective network and the conditional posterior density of the rate constant of the final reaction that ultimately leads to the change in effective network from $M_{m_e}$ to $M_{n_e}$.

\begin{table}[H]
\label{proposalTable}
\begin{center}
\begin{tabular}{|ccc|ccc|}
\hline
\multicolumn{3}{|c|}{Method} & \multicolumn{3}{c|}{$Proposal$} \\
\hline
\multicolumn{3}{|c|}{Network unaware}  & \multicolumn{3}{c|}{$q_{nu}(u \vert k_{m_e}) \approx \prod^{a-1}_{j=i+1} p(k_{n_e,j})p(k_{{n_e},a} \vert k_{n_e,1:a-1},M_{n_e},\mathcal{D}) $}\\
\multicolumn{3}{|c|}{Network aware}    & \multicolumn{3}{c|}{$q_{na}(u \vert k_{m_e}) \approx p(k_{n_e} \vert k_{m_e},M_{n_e},\mathcal{D})$}\\
\hline 
\end{tabular}
\end{center}
{
\normalsize
\caption{Cluster switching parameter proposals} 
\label{supp_Proposals}
}
\end{table}
\noindent
Mathematically, the two proposals are shown in Table \ref{supp_Proposals}. In moves where the effective networks of the current and the proposed network are the same, both the network-unaware and our network-aware approaches use the prior distribution of the newly added reaction as the proposal and have acceptance probabilities of 1. The steps of our network-aware reversible jump MCMC algorithm are given in Algorithm \ref{alg:networkawareRJMCMC}. 
\begin{algorithm}
\begin{algorithmic}[1]
\State \textbf{Given}: A set of models ${M_m \in \mathcal{M}}$ with corresponding parameter vectors $k_{m}$, posterior densities $p(M_m,k_{m} \vert \mathcal{D})$.
\State $\beta \in (0,1)$: probability of within-model move
\State Initialize chain: $\{(M^{0}, k_{M^{0}})$, $M^0_{e} \leftarrow$ effective network of $M^0$\}
\For {$s\mbox{ = }0 \mbox{ to }N_{iter}$}
\State Sample $b\sim\mathcal{U}_{[0,1]}$
\If {$b \leq \beta$}
\State Metropolis-Hastings within-model move
\Else
\State Sample $M'\sim q(M' \vert M^s)$;  $M'_{e} \leftarrow$ effective network of $M'$
\vskip 5pt
\If {$|M_{e}'|>|M^{s}_{e}|$}
\begin{align}
\mbox{    }\mu_{M_e} ={\arg\max}\mbox{  }p(k_{M'_e,i+1:a} \vert k_{M^{s}_{e},1:i},M'_{e},\mathcal{D}),\mbox{  } \Sigma_{M_e}& =-[\nabla^{2} \log p(k_{M'_e,i+1:a} \vert k_{M^{s}_{e},1:i},M'_{e},\mathcal{D})]^{-1}\big|_{\mu_{M_e}} \nonumber
\end{align}
\State Sample $u_{M_e}\sim\mathcal{N}(\mu_{M_e},\Sigma_{M_e})$ 
\vskip 5pt
\State Set $(k_{M'_{e},{1:i}},k_{M'_{e},{i+1:a}})=(k_{M_{e},{1:i}},u_{M_e})$
\State {$\alpha((M^s,k_{M^s}),(M',k_{M'}))=\mbox{min}\left\{1, \frac{p(M'_{e}, k_{M'_{e}} \vert \mathcal{D}) q(M^s \vert M')}{p(M^{s}_{e}, k_{M^{s}_{e}} \vert \mathcal{D}) q(M' \vert M^{s})\mathcal{N}(u_{M_e};\mu_{M_e},\Sigma_{M_e})}\right\}$}
\ElsIf{$|M'_{e}|<|M^s_{e}|$}
\begin{align}
\mbox{    }{\mu}_{M'_e} ={\arg\max}\mbox{  }p(k_{M^{s}_e,i+1:a} \vert k_{M'_{e},{1:i}},M^{s}_{e},\mathcal{D}),\mbox{  } \Sigma_{M'_e}& =-[\nabla^{2} \log p(k_{M^{s}_e,i+1:a} \vert k_{M'_{e},{1:i}},M^{s}_{e},\mathcal{D})]^{-1}\big|_{\mu_{M'_e}} \nonumber
\end{align}
\State Set $(k_{M'_{e},{1:i}},u_{M'_e})=(k_{M^{s}_{e},{1:i}},k_{M^{s}_{e},{i+1:a}})$
\State {$\alpha((M^s,k_{M^s}),(M',k_{M'}))=\mbox{min}\left\{1, \frac{p(M'_{e}, k_{M'_{e}} \vert \mathcal{D}) q(M^s \vert M')\mathcal{N}(u_{M'_e};\mu_{M'_e},\Sigma_{M'_e})}{p(M^s_{e}, k_{M^s_{e}} \vert \mathcal{D}) q(M' \vert M^s)}\right \}$}
\Else
\State Sample $k_{M'} \setminus k_{M^s} \sim p(k_{M'} \setminus k_{M^s})$ 
\State {$\alpha((M^{s},k_{M^s}),(M',k_{M'}))=1$}
\EndIf
\State Sample $p \sim \mathcal{U}_{[0,1]}$
\If {$p< \alpha((M^s,k_{M^s}),(M',k_{M'}))$}
\State $\{(M^{s+1},k^{s+1}_{M^{s+1}}),M^{s+1}_{e}\}=\{(M',k_{M'}),M'_e\}$
\Else
\State $\{(M^{s+1},k^{s+1}_{M^{s+1}}),M^{s+1}_e\}=\{(M^{s},k^{s}_{M^{s}}),M^{s}_e\}$
\EndIf
\EndIf
\EndFor
\end{algorithmic}
\caption{Network-aware reversible jump MCMC}
\label{alg:networkawareRJMCMC}
\end{algorithm}

\subsection{Sensitivity-based network-aware proposals} 
\label{sec:supp_sensitivitynetworkaware}
\noindent
In the paper, we presented the sensitivity-based network aware algorithm and explained its construction. Here we provide technical details and a pseudocode for the application of the sampler. Suppose, the sampler move is proposed between a lower-dimensional model $M_m$ to a higher-dimensional model $M_n$. Let the effective networks of the two models $M_m$ and $M_n$ be $M_{m_e}$ and $M_{n_e}$, respectively. Suppose the proposed move is such that $M_{n_e} \neq M_{m_e}$, i.e., the effective networks of the current and the proposed networks are different. The identity mapping $f$ between $(k_{m_e},u_{m_e})$, the rate constants and proposal parameters of network $M_{m_e}$ to $(k_{n_e},u_{n_e})$, the rate constants and proposal parameters of network $M_{n_e}$ for sensitivity-based move proposals is given by
\begin{align}
k_{n_e,1:i}=k_{m_e,{1:i}},\ &k_{n_e,i+1:i+c}=u_{m_e,1:c},\ u_{n_e,1:c}=k_{m_e,{i+1:i+c}},\mbox{ and } \nonumber \\ &k_{n_e,{i+c+1:i+a}}=u_{m_e,c+1:a}.
\end{align}
\noindent
Here, ${\{1:i\}}$ are indices of reactions that are common to the two networks and whose rate constant values are kept fixed during moves between networks $M_{m_e}$ and $M_{n_e}$, ${\{i+c+1:i+a\}}$ are indices of reactions that are in network $M_{n_e}$ but not in $M_{m_e}$, and ${\{i+1:i+c\}}$ are the critical reactions that are present in both networks, but whose values are proposed according to a proposal distribution rather than keeping them fixed during the between-model move.  The proposal distributions are again taken to be Gaussian ($u_{m_e} \sim \mathcal{N}(\mu_{m_e},\Sigma_{{m_e}})$ and $u_{n_e} \sim \mathcal{N}(\mu_{n_e},\Sigma_{{n_e}})$). The proposal means 

\begin{equation}
\mu_{{m_e}} =\underset{k_{n_e,i+1:i+a}}{\arg\max}\mbox{  }p(k_{n_e,i+1:i+a} \vert k_{{m_e},1:i},M_{n_e},\mathcal{D})
\end{equation}
\noindent
and 
\begin{equation}
\mu_{{n_e}} =\underset{k_{m_e,i+1:i+c}}{\arg\max}\mbox{  }p(k_{m_e,i+1:i+c} \vert k_{{n_e},1:i},M_{m_e},\mathcal{D})
\end{equation}
\noindent
are obtained by solving $a$-dimensional and $c$-dimensional optimization problems. The proposal covariances

\begin{equation}
\Sigma_{{m_e}} =-[\nabla^2 \log p(k_{n_e,i+1:i+a} \vert k_{m_{e},1:i},M_{n_e},\mathcal{D})]^{-1}\big|_{\mu_{{m_e}}}
\end{equation}
\noindent
and 

\begin{equation}
\Sigma_{{n_e}} =-[\nabla^2 \log p(k_{m_e,i+1:i+c} \vert k_{n_{e},1:i},M_{m_e},\mathcal{D})]^{-1}\big|_{\mu_{{n_e}}}
\end{equation}
\noindent
are determined numerically using a finite-difference approximation at the proposal means. The acceptance probability of the proposed move is given by

\begin{equation}
\alpha((M_{m_e},k_{m_e}),(M_{n_e},k_{n_e}))=\mbox{min}\left\{1,A\right\},
\end{equation}

\noindent
where

\begin{equation}
A=\frac{p(M_{n_e}, k_{n_{e}} \vert \mathcal{D}) q(M_m \vert M_n)\mathcal{N}(u_{n_e};\mu_{n_e},\Sigma_{n_e})}{p(M_{m_e}, k_{m_{e}} \vert \mathcal{D}) q(M_n \vert M_m)\mathcal{N}(u_{m_e};\mu_{m_e},\Sigma_{m_e})}.
\end{equation}

\noindent
The reverse move is nondeterministic and has an acceptance probability $\mbox{min}\{1,A^{-1}\}$. The pseudocode for sensitivity-based network aware algorithm is given in Algorithm \ref{alg:sensitivityRJMCMC}.
 
\begin{algorithm}
\begin{algorithmic}[1]
{
\State \textbf{Given}: A set of models ${M \in \mathcal{M}}$ with corresponding parameter vectors $k_{M}$, posterior densities $p(M,k_{M} \vert \mathcal{D})$.
\State $\beta \in (0,1)$: probability of within-model move
\State Initlialize chain: $\{(M^{0}, k_{M^{0}}),M^{0}_e \leftarrow$ effective network of $M^0 \}$
\For {$s\mbox{ = }0 \mbox{ to }N_{iter}$}
\State Sample $b\sim\mathcal{U}_{[0,1]}$
\If {$b \leq \beta$}
\State Metropolis-Hastings within-model move
\Else
\State Sample $M'\sim q(M' \vert M^s)$; $M'_{e} \leftarrow$ effective network of $M' $
\State $r_{1} \sim Poisson(1.5)$ and $r_{2} \sim Poisson(1.5)$
\State ${\{1:i\}}$=reactions that are common to $M^{s}_{e}$ and $M_{e}'$  and whose rate constant values are kept fixed during moves
\State $\{i+1:i+c\}$=reactions common to $M^{s}_{e}$ and $M_{e}'$ and with top $r_{1}$ and $r_{2}$ sensitivities of $M^{s}_{e}$ and $M_{e}'$, respectively
\State ${\{i+c+1:i+a\}}$=reactions only present in $M^{s}_{e}$ or $M_{e}'$
\If {$|M_{e}'|>|M^{s}_{e}|$}
\begin{align}
\mbox{    }\mu_{M_{e}} ={\arg\max}\mbox{  }p(k^{i+1:i+a}_{M'_e} \vert k^{1:i}_{M^s_{e}},M'_{e},\mathcal{D}),\mbox{  } \Sigma_{M_{e}}& =-[\nabla^{2} \log p(k^{i+1:i+a}_{M'_e} \vert k^{1:i}_{M^s_{e}},M'_{e},\mathcal{D})]^{-1}\big|_{\mu_{M_e}} \nonumber \\ 
\mbox{    }\mu_{M_{e}'}  ={\arg\max}\mbox{  }p(k_{M^{s}_e}^{i+1:i+c} \vert k^{1:i}_{M'_{e}}, M^s_{e},\mathcal{D}),\mbox{  } \Sigma_{M'_{e}}& =-[\nabla^{2} \log p(k_{M^s_e}^{i+1:i+c} \vert k^{1:i}_{M'_{e}},M^s_{e},\mathcal{D})]^{-1}\big|_{\mu_{M'_e}} \nonumber
\end{align}
\State Sample $u \sim \mathcal{N}(\mu_{M_{e}},\Sigma_{M_{e}})$
\vskip 3pt
\State Set $(k_{M_{e}'}^{1:i},k_{M'_{e}}^{i+1:i+c},u'^{1:c},k_{M'_{e}}^{i+c+1:i+a})=(k_{M^s_{e}}^{1:i},u^{1:c},k_{M^s_{e}}^{i+1:i+c},u^{c+1:a})$
\vskip 3pt
\ElsIf {$|M_{e}'|<|M^{s}_{e}|$}
\begin{align}
\mbox{    }\mu_{M_{e}} ={\arg\max}\mbox{  }p(k^{i+1:i+c}_{M'_e} \vert k^{1:i}_{M^s_{e}},M'_{e},\mathcal{D}),\mbox{  } \Sigma_{M_{e}}& =-[\nabla^{2} \log p(k^{i+1:i+c}_{M'_{e}} \vert k^{1:i}_{M^s_{e}},M_{e}',\mathcal{D})]^{-1}\big|_{\mu_{M_e}} \nonumber \\ 
\mbox{    }\mu_{M'_{e}}  ={\arg\max}\mbox{  }p(k^{i+1:i+a}_{M^s_e} \vert k^{1:i}_{M'_{e}}, M^s_{e},\mathcal{D}),\mbox{  } \Sigma_{M_{e}'}& =-[\nabla^{2} \log p(k^{i+1:i+a}_{M^s_{e}} \vert k^{1:i}_{M'_{e}},M^s_{e},\mathcal{D})]^{-1}\big|_{\mu_{M'_e}} \nonumber
\end{align}
\State Sample $u \sim \mathcal{N}(\mu_{M_{e}},\Sigma_{M_{e}})$ 
\vskip 3pt
\State Set $(k_{M'_{e}}^{1:i},k_{M'_e}^{i+1:i+c},u'^{1:c},u'^{c+1:a})=(k_{M^s_{e}}^{1:i},u^{1:c},k_{M^s_{e}}^{i+1:i+c},k_{M^s_{e}}^{i+c+1:i+a})$
\vskip 3pt
\Else 
\EndIf
\vskip 3pt
\State Sample $k_{M'} \setminus k_{M^s} \sim p(k_{M'} \setminus k_{M^s})$
\vskip 3pt
\State Sample $p \sim \mathcal{U}_{[0,1]}$
\vskip 3pt
\If {$p<\mbox{min}\left\{1, \frac{p(M'_{e}, k_{M'_{e}} \vert \mathcal{D}) q(M^{s} \vert M')\mathcal{N}(u';\mu_{M_{e}'},\Sigma_{M_{e}'})}{p(M^s_{e}, k_{M^s_{e}} \vert \mathcal{D}) q(M' \vert M^{s})\mathcal{N}(u;\mu_{M_{e}},\Sigma_{M_{e}})}\right\}$}
\State $\{(M^{s+1},k^{s+1}_{M^{s+1}}),M^{s+1}_{e}\}=\{(M',k_{M'}),M'_e\}$
\Else
\State $\{(M^{s+1},k^{s+1}_{M^{s+1}}),M^{s+1}_e\}=\{(M^{s},k^{s}_{M^{s}}),M^{s}_e\}$
\EndIf
\EndIf
\EndFor
}
\end{algorithmic}
\caption{Sensitivity-based network-aware reversible jump MCMC}
\label{alg:sensitivityRJMCMC}
\end{algorithm}
\noindent
\subsection{Derandomization of conditional expectations}
\noindent
The above Algorithms \ref{alg:networkawareRJMCMC} and \ref{alg:sensitivityRJMCMC} lead to gains in sampling efficiency compared to a reversible jump MCMC algorithm that does not use information on network structure in designing between-model moves and parameter proposals. Identifying clusters of models can be further used for additional variance reduction. With the knowledge that all models belonging to the same cluster have identical model evidence, we can compute some expectations analytically and thereby obtain posterior averages of features with lower variances.
\subsubsection{General formulation}
\noindent
Let us assume we are performing model inference with $F$ as one the quantities of interest. Generally, we may be interested in quantities such as the posterior model probabilities, reaction inclusion probabilities of reactions, or pathway probabilities. The Monte Carlo estimate of $F$ from posterior samples can be written as:

\begin{align*}
\hat{F}&=p(F=1|\mathcal{D}) \nonumber \\
       &=\int p(F=1|C)p(C|\mathcal{D})dC \nonumber \\
       &=\int p(F=1|M)p(M|C)p(C|\mathcal{D})dMdC \nonumber \\
       &=\int \mathbb E_{p(M \vert C)}\left[p(F=1 \vert M)\right]p(C \vert \mathcal{D})dC \nonumber \\
       &=\frac{1}{N_s}\sum_{i=1}^{N_s}\mathbb E_{p(M \vert C^{i})}\left[p(F=1 \vert M)\right],
\end{align*}
\noindent
where $C$ refers to model clusters, $N_{s}$ is the number of posterior samples and $\mathcal{D}$ the available data. In the above equation, $\mathbb E_{p(M \vert C^{i})}[p(F=1 \vert M)]$ is the expected value of $p(F=1 \vert M)$ conditioned on the generated sample $C^{i}$. Knowing the cluster to which each sample belongs and the dependence of the feature on the models included in the cluster, the above expectation can be computed analytically and allows variance reduction. In contrast, in the network-unaware approach, the expectation is computed through Monte Carlo sampling.  

\subsubsection{Example: model probability estimates}
\noindent
Consider that the feature of interest is the probability of model $m$. Thus, applying the above formula to the estimation of model probability, we get

\begin{align}
\hat{M}_{m}&=p(M_{m}=1|\mathcal{D}) \nonumber \\
       &=\int p(M_{m}=1|C)p(C|\mathcal{D})dC \nonumber \\
       &=\int p(M_{m}=1|M)p(M|C)p(C|\mathcal{D})dMdC \nonumber \\
       &=\int \mathbb E_{p(M \vert C)}[p(M_{m}=1 \vert M)]p(C \vert \mathcal{D})dC \nonumber \\
       &=\frac{1}{N_s}\sum_{i=1}^{N_s}\mathbb E_{p(M \vert C^{i})}\left[p(M_{m}=1 \vert M)\right] \nonumber \\
       &=\frac{1}{N_s}\sum_{i=1}^{N_s}p(M_m|C_K)\mathbbm{1}_{C_K}(C^{i}),
\end{align}
\noindent
where $K: \mathbbm{1}_{M_m \in C_K}(M_m)=1$ and $\mathbbm{1}$ is the indicator function. In our network aware schemes, $p(M_{m} \vert C_{K})$ can be computed analytically. For example, for a cluster $C_{K}$ with $N_{K}$ models, taking the prior distribution over models to be uniform, the model probability estimate is

\begin{equation}
\hat{M}_{m}=\frac{1}{N_s}\sum_{i=1}^{N_s}\frac{1}{N_{K}}\mathbbm{1}_{C_K}(C^{i})
\end{equation}

\noindent
In contrast, with a standard reversible-jump algorithm, the model probability estimate is

\begin{equation}
\hat{M}_{m}=\frac{1}{N_s}\sum_{i=1}^{N_s}\mathbbm{1}_{M_{m}}(M^{i})\mathbbm{1}_{C_K}(C^{i})
\end{equation}
\begin{figure}[p]
\includegraphics[width=0.8\textwidth]{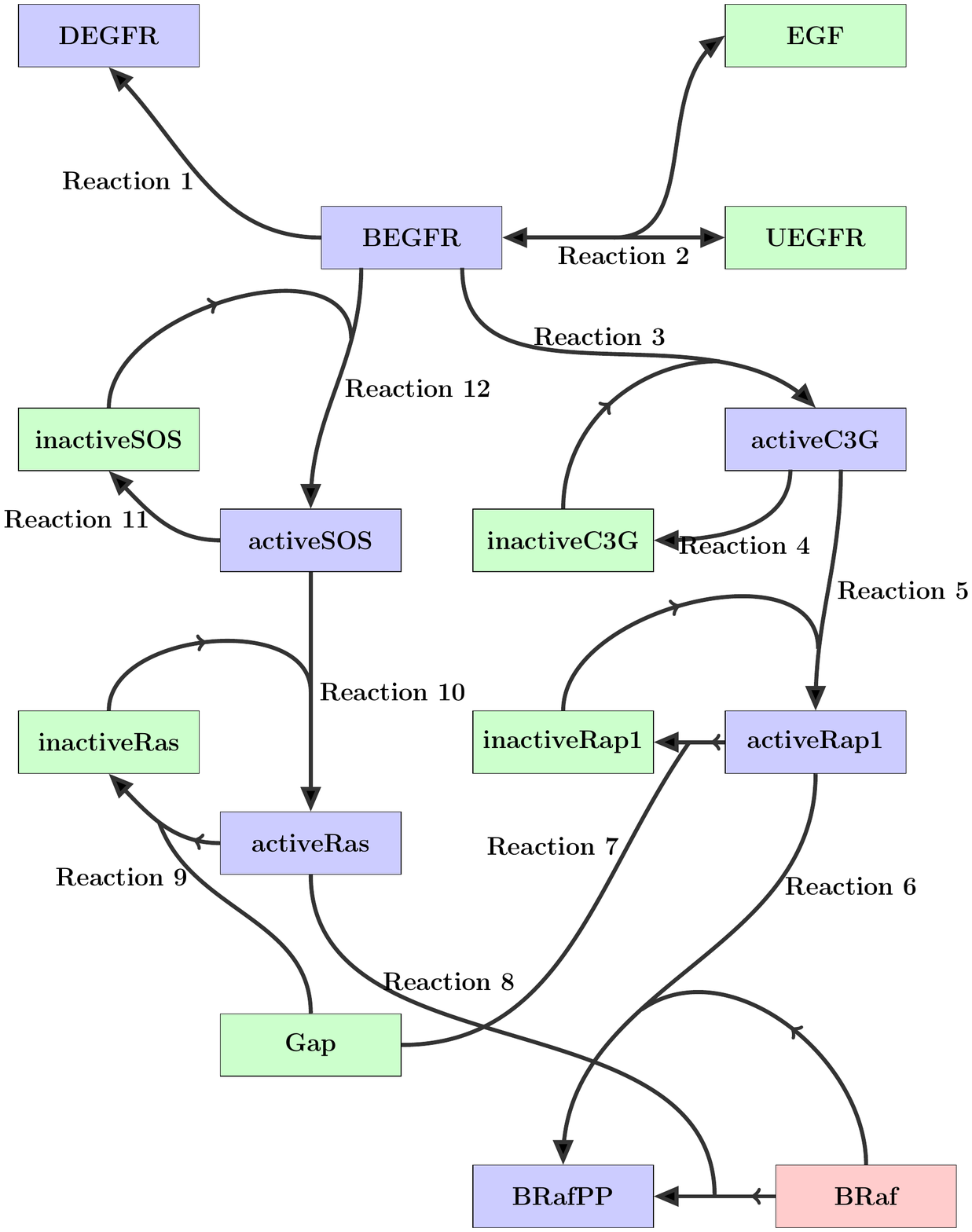}
\caption{Reaction network of Example 1 and Example 2}
\label{fig:supp_reactionNetwork1Allreactions}
\end{figure}
\section{Results}
\label{sec:supp_results}
\noindent
In our paper, we present two example problems and demonstrate the efficiency of our network-aware sampling approaches compared to the network-unaware approach. The observables in our examples are species concentrations and the concentration evolution is modeled using the law of mass action/Michaelis-Menten functionals. The law of mass action gives the rate of a chemical reaction (say $X+Y\rightarrow Z$) as the product of a reaction-specific rate constant $k$ with reactant concentrations $[X]$ and $[Y]$:

\begin{equation}
\text{Rate}=-k [X] [Y].
\label{supp_massactionkineticsrate}
\end{equation}

\noindent
Under some assumptions, the law of mass action produces Michaelis-Menten reaction rate expression

\begin{equation}
Rate=\frac{k[S]}{k_{M}+[S]},
\end{equation}
\noindent
or when enzyme concentration is taken into account \cite{Murray2002}:

\begin{equation}
Rate=k[E]_{0}\frac{[S]}{k_{M}+[S]},
\end{equation}

\noindent
where $k$ denotes the rate constant, $[E]_{0}$ is the enzyme concentration, $[S]$ the substrate concentration, and $k_{M}$ the Michaelis constant. 

\begin{figure}[h]
{ \makeatletter
 \def\@captype{figure}
 \makeatother
 \begin{minipage}[h]{0.45\textwidth}
 \begin{subfigure}[b]{1.0\textwidth}
 \begin{center}
 \includegraphics[width=0.7\linewidth]{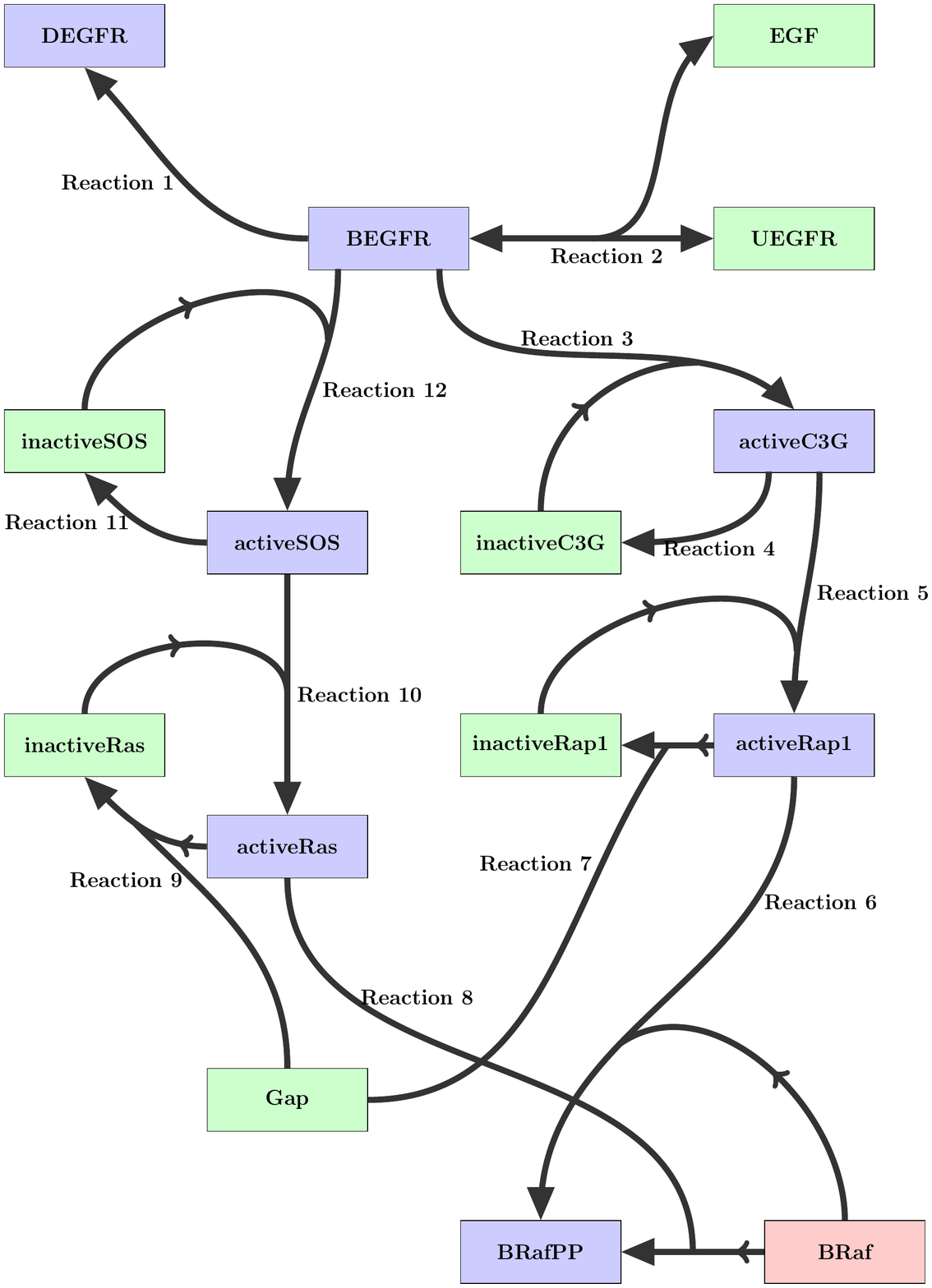}
 \includegraphics[width=0.7\linewidth]{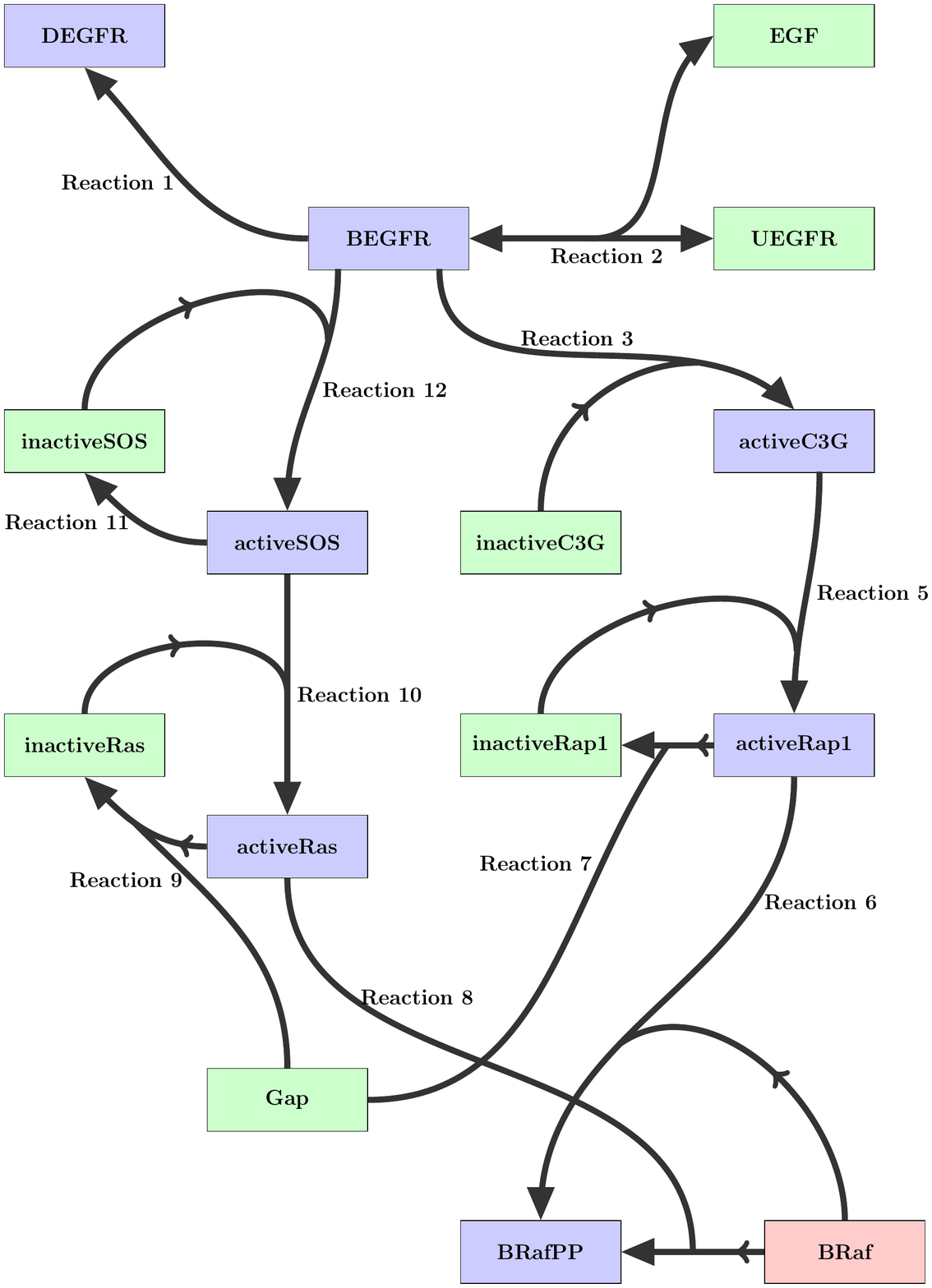}
 \includegraphics[width=0.7\linewidth]{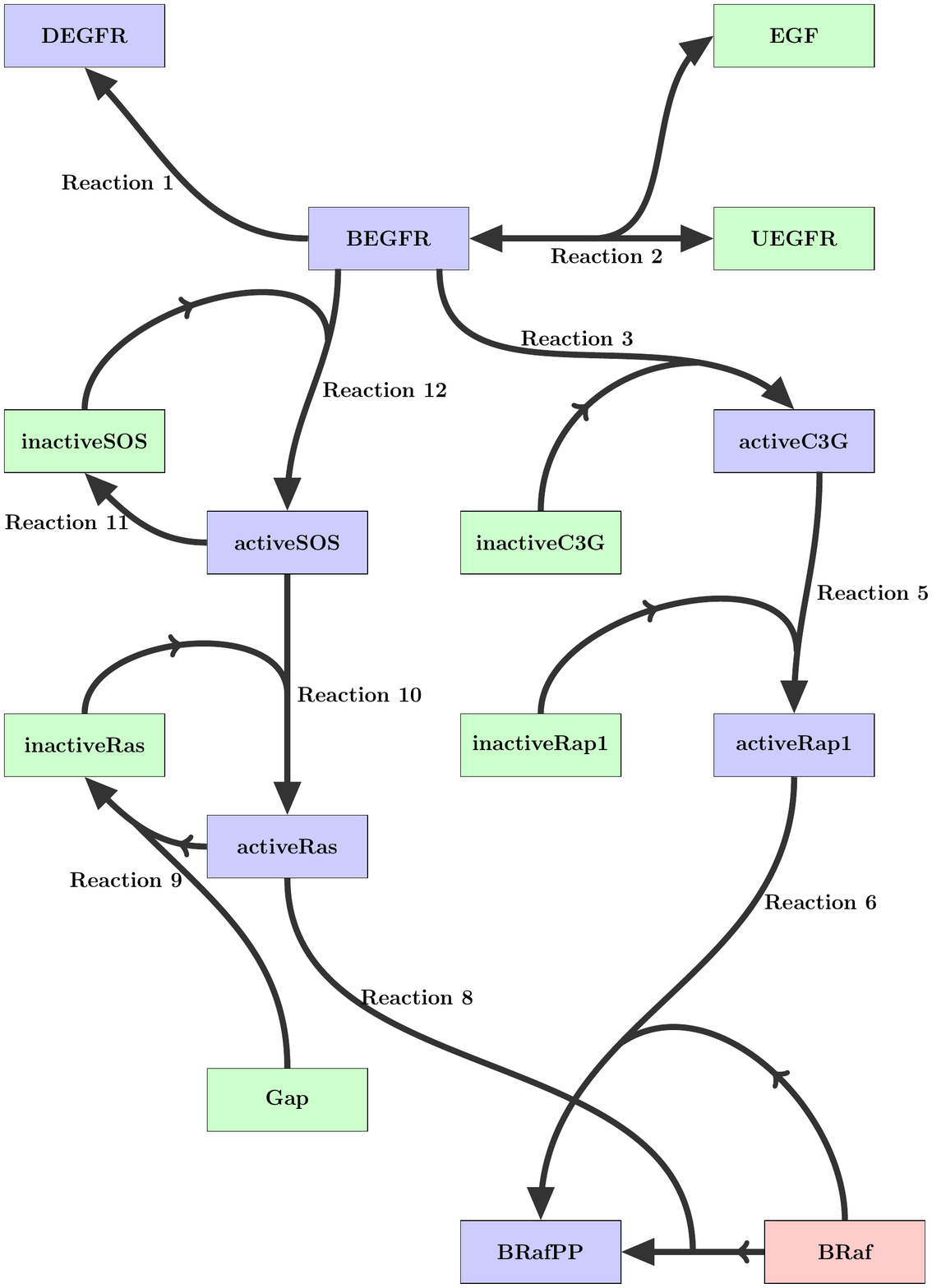}
 \end{center}  
  \label{fig:supp_sub1}
 \end{subfigure} 
 \end{minipage}
 \begin{minipage}[h]{0.45\textwidth} 
\begin{subfigure}[b]{1.0\textwidth}
\begin{center}
\includegraphics[width=0.7\linewidth]{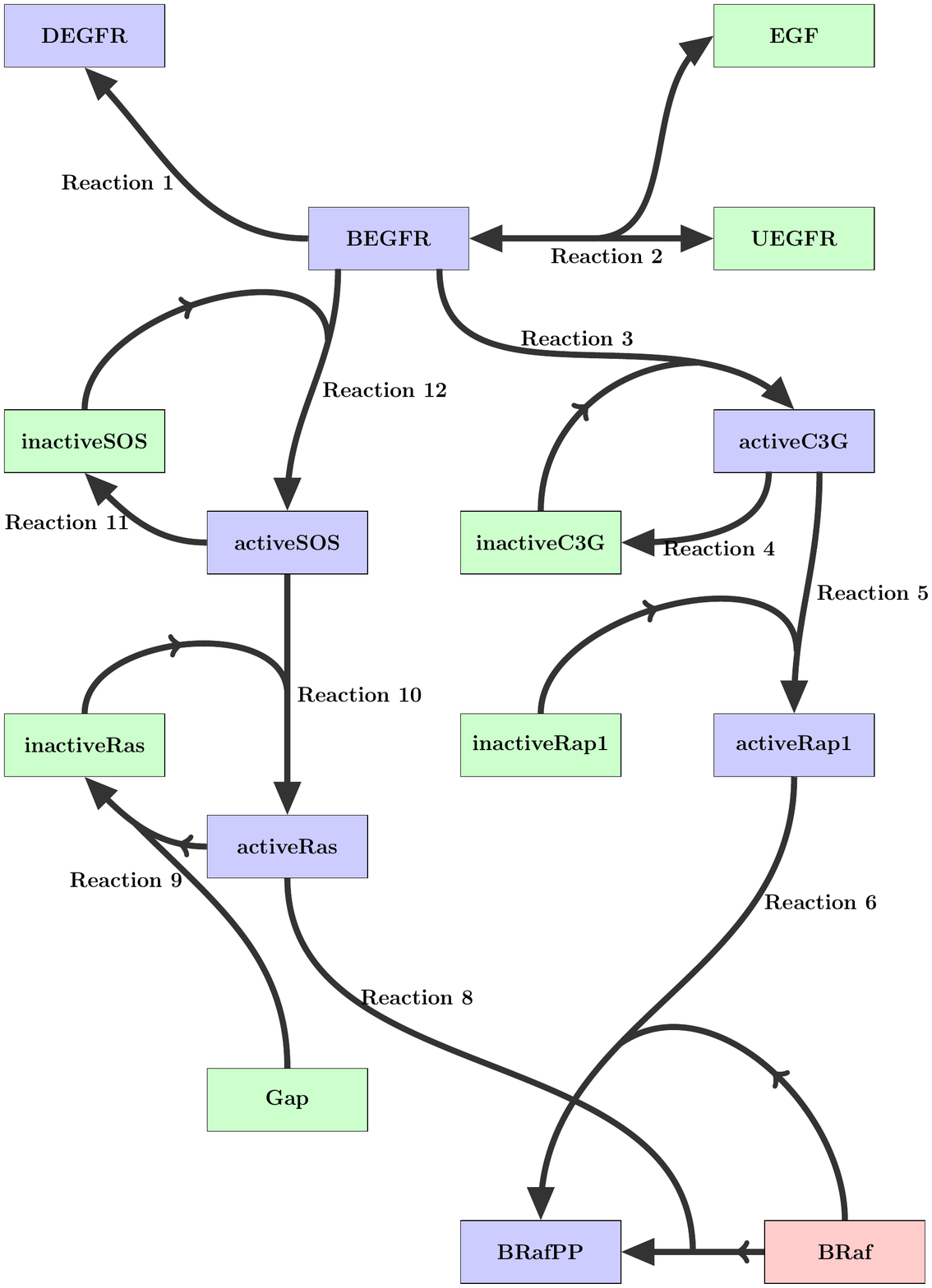}
\includegraphics[width=0.7\linewidth]{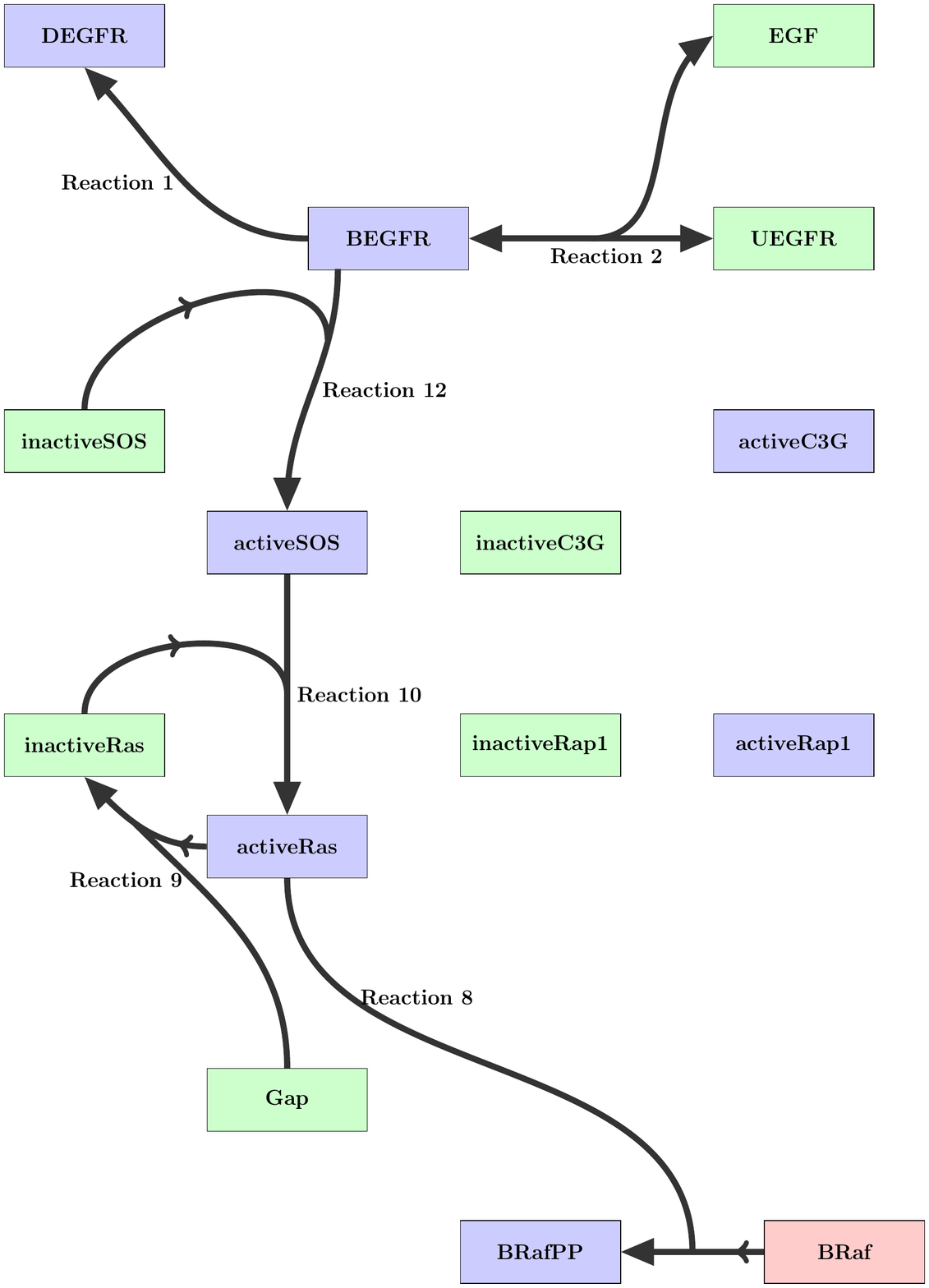}
\end{center}  
\label{fig:supp_sub2}
\end{subfigure}
\end{minipage}
\caption{Effective networks in Example 1}
\label{fig:effective_networks_example1}}
\end{figure}

\tikzstyle{SPECIES1}=[draw, fill=blue!20, text width=9em, text centered, minimum height=3.0em]
\tikzstyle{SPECIES2}=[draw, fill=blue!20, text width=9em, text centered, minimum height=3.0em]
\tikzstyle{SPECIES3}=[draw, fill=green!20, text width=9em, text centered, minimum height=3.0em]    
\tikzstyle{SPECIES4}=[draw, fill=green!20, text width=9em, text centered, minimum height=3.0em]    
\tikzstyle{SPECIES5}=[draw, fill=green!20, text width=9em, text centered, minimum height=3.0em]
\tikzstyle{SPECIES6}=[draw, fill=blue!20, text width=9em, text centered, minimum height=3.0em]
\tikzstyle{SPECIES7}=[draw, fill=blue!20, text width=9em, text centered, minimum height=3.0em]
\tikzstyle{SPECIES8}=[draw, fill=red!20, text width=9em, text centered, minimum height=3.0em]
\tikzstyle{SPECIES9}=[draw, fill=blue!20, text width=9em, text centered, minimum height=3.0em]    
\tikzstyle{SPECIES10}=[draw, fill=blue!20, text width=9em, text centered, minimum height=3.0em]    
\tikzstyle{SPECIES11}=[draw, fill=green!20, text width=9em, text centered, minimum height=3.0em]
\tikzstyle{SPECIES12}=[draw, fill=green!20, text width=9em, text centered, minimum height=3.0em]
\tikzstyle{SPECIES13}=[draw, fill=green!20, text width=9em, text centered, minimum height=3.0em]    
\tikzstyle{SPECIES14}=[draw, fill=blue!20, text width=9em, text centered, minimum height=3.0em]    
\tikzstyle{SPECIES15}=[draw, fill=green!20, text width=9em, text centered, minimum height=3.0em]
\begin{figure}[htp]
\begin{center}
\resizebox{1.0\textwidth}{!}{
\begin{tikzpicture}
\begin{scope}
\node[SPECIES1](ySPECIES1W) at (0,2) {\Large \bf BEGFR};
\node[SPECIES2](ySPECIES2W) at (-6,6) {\Large \bf DEGFR};
\node[SPECIES3](ySPECIES3W) at (8,6) {\Large \bf EGF};
\node[SPECIES4](ySPECIES4W) at (8,2) {\Large \bf UEGFR};
\node[SPECIES5](ySPECIES5W) at (3,-4) {\Large \bf inactiveC3G};
\node[SPECIES6](ySPECIES6W) at (8,-2) {\Large \bf activeC3G};
\node[SPECIES7](ySPECIES7W) at (8,-8) {\Large \bf activeRap1};
\node[SPECIES8](ySPECIES8W) at (9, -18) {\Large \bf BRaf};
\node[SPECIES9](ySPECIES9W) at (3, -18) {\Large \bf BRafPP};
\node[SPECIES10](ySPECIES10W) at (-2,-10) {\Large \bf activeRas};
\node[SPECIES11](ySPECIES11W) at (-2,-15) {\Large \bf Gap};
\node[SPECIES12](ySPECIES12W) at (3.0,-8) {\Large \bf inactiveRap1};
\node[SPECIES13](ySPECIES13W) at (-6, -8) {\Large \bf inactiveRas};
\node[SPECIES14](ySPECIES14W) at (-2,-4) {\Large \bf activeSOS};
\node[SPECIES15](ySPECIES15W) at (-6,-2) {\Large \bf inactiveSOS};

\draw [->,line width=1.2mm,draw=black!80,-triangle 45](ySPECIES1W.west)to[out=180,in=315]node[left]{\Large \bf Reaction 1}(ySPECIES2W.south) ;

\draw [->,line width=1.2mm,draw=black!80,-triangle 45](4,2)to[out=0,in=215] node[above]{}(ySPECIES3W.west);
\draw [->,line width=1.2mm,draw=black!80,-triangle 45](2.2,2)to[out=0,in=180]node[below]{\Large \bf Reaction 2}(ySPECIES4W.west);
\draw [->,line width=1.2mm,draw=black!80,-triangle 45](2.2,2)to[out=180,in=0](ySPECIES1W.east)node[below]{};

\draw  [->,line width=1.2mm,draw=black!80,-triangle 45](1,1.4)to[out=270,in=135] node[above]{\Large \bf Reaction 3}(ySPECIES6W.north);
\draw  [-,line width=1.2mm,draw=black!80,decoration={markings,mark=at position 0.5 with {\arrow{>}}},postaction={decorate}](ySPECIES5W.north)to[out=90,in=180](6.1,-0.45) node[above]{}(6,2);


\draw[->,line width=1.2mm,draw=black!80,-triangle 45](8.5,-2.6)to[out=270,in=90]node[right]{\Large \bf Reaction 5}(ySPECIES7W.north);
\draw[-,line width=1.2mm,draw=black!80,decoration={markings,mark=at position 0.5 with {\arrow{>}}},postaction={decorate}](ySPECIES12W.north)to[out=90,in=90]node[right]{}(8.05,-6.15);

\draw[->,line width=1.2mm,draw=black!80,-triangle 45](ySPECIES7W.south)to[out=270,in=90]node[near start,right]{\Large \bf Reaction 6}(ySPECIES9W.north);
\draw[-,line width=1.2mm,draw=black!80,decoration={markings,mark=at position 0.5 with {\arrow{>}}},postaction={decorate}](ySPECIES8W.north)to[out=90,in=35]node[right]{}(4.55,-13.9);


\draw[-,line width=1.2mm,draw=black!80](ySPECIES10W.south)to[out=270,in=90]node[near start,right]{\Large \bf Reaction 8}(6,-18);
\draw[->,line width=1.2mm,draw=black!80,-triangle 45,decoration={markings,mark=at position 0.2 with {\arrow{>}}},postaction={decorate}](ySPECIES8W.west)to[out=180,in=0]node[right]{}(ySPECIES9W.east);

\draw[->,line width=1.2mm,draw=black!80,-triangle 45,decoration={markings,mark=at position 0.2 with {\arrow{>}}},postaction={decorate}](ySPECIES10W.west)to[out=180,in=315]node[left]{}(ySPECIES13W.south);
\draw[-,line width=1.2mm,draw=black!80](ySPECIES11W.north)to[out=90,in=315]node[near end,left]{\Large \bf Reaction 9}(-5.0,-9.55);

\draw [->, line width=1.2mm,draw=black!80,-triangle 45](ySPECIES14W.south)to[out=270,in=90]node[right]{\Large \bf Reaction 10}(ySPECIES10W.north);
\draw [-, line width=1.2mm,draw=black!80,decoration={markings,mark=at position 0.5 with {\arrow{>}}},postaction={decorate}](ySPECIES13W.north)to[out=90,in=90]node[right]{}(-2,-7);

\draw [->, line width=1.2mm,draw=black!80,-triangle 45](ySPECIES14W.west)to[out=180,in=315]node[left]{\Large \bf Reaction 11}(ySPECIES15W.south);

\draw  [->,line width=1.2mm,draw=black!80,-triangle 45](-1,1.4)to[out=270,in=90] node[right]{\Large \bf Reaction 12}(ySPECIES14W.north);
\draw  [-,line width=1.2mm,draw=black!80,decoration={markings,mark=at position 0.5 with {\arrow{>}}},postaction={decorate}](ySPECIES15W.north)to[out=90,in=90] node[right]{}(-1.2,0);

\node[text width=8cm] at (3.0,-20) {\LARGE \bf Effective network 1};

\node[SPECIES1](ySPECIES1E) at (0+25,2) {\Large \bf BEGFR};
\node[SPECIES2](ySPECIES2E) at (-6+25,6) {\Large \bf DEGFR};
\node[SPECIES3](ySPECIES3E) at (8+25,6) {\Large \bf EGF};
\node[SPECIES4](ySPECIES4E) at (8+25,2) {\Large \bf UEGFR};
\node[SPECIES5](ySPECIES5E) at (3+25,-4) {\Large \bf inactiveC3G};
\node[SPECIES6](ySPECIES6E) at (8+25,-2) {\Large \bf activeC3G};
\node[SPECIES7](ySPECIES7E) at (8+25,-8) {\Large \bf activeRap1};
\node[SPECIES8](ySPECIES8E) at (9+25, -18) {\Large \bf BRaf};
\node[SPECIES9](ySPECIES9E) at (3+25, -18) {\Large \bf BRafPP};
\node[SPECIES10](ySPECIES10E) at (-2+25,-10) {\Large \bf activeRas};
\node[SPECIES11](ySPECIES11E) at (-2+25,-15) {\Large \bf Gap};
\node[SPECIES12](ySPECIES12E) at (3.0+25,-8) {\Large \bf inactiveRap1};
\node[SPECIES13](ySPECIES13E) at (-6+25, -8) {\Large \bf inactiveRas};
\node[SPECIES14](ySPECIES14E) at (-2+25,-4) {\Large \bf activeSOS};
\node[SPECIES15](ySPECIES15E) at (-6+25,-2) {\Large \bf inactiveSOS};

\draw [->,line width=1.2mm,draw=black!80,-triangle 45](ySPECIES1E.west)to[out=180,in=315]node[left]{\Large \bf Reaction 1}(ySPECIES2E.south) ;

\draw [->,line width=1.2mm,draw=black!80,-triangle 45](4+25,2)to[out=0,in=215] node[above]{}(ySPECIES3E.west);
\draw [->,line width=1.2mm,draw=black!80,-triangle 45](2.2+25,2)to[out=0,in=180]node[below]{\Large \bf Reaction 2}(ySPECIES4E.west);
\draw [->,line width=1.2mm,draw=black!80,-triangle 45](2.2+25,2)to[out=180,in=0](ySPECIES1E.east)node[below]{};

\draw  [->,line width=1.2mm,draw=black!80,-triangle 45](1+25,1.4)to[out=270,in=135] node[above]{\Large \bf Reaction 3}(ySPECIES6E.north);
\draw  [-,line width=1.2mm,draw=black!80,decoration={markings,mark=at position 0.5 with {\arrow{>}}},postaction={decorate}](ySPECIES5E.north)to[out=90,in=180](6.1+25,-0.45) node[above]{}(6+25,2);

\draw[->,line width=1.2mm,draw=black!80,-triangle 45](7.5+25,-2.6)to[out=270,in=0]node[below]{\Large \bf Reaction 4}(ySPECIES5E.east);

\draw[->,line width=1.2mm,draw=black!80,-triangle 45](8.5+25,-2.6)to[out=270,in=90]node[right]{\Large \bf Reaction 5}(ySPECIES7E.north);
\draw[-,line width=1.2mm,draw=black!80,decoration={markings,mark=at position 0.5 with {\arrow{>}}},postaction={decorate}](ySPECIES12E.north)to[out=90,in=90]node[right]{}(8.05+25,-6.15);

\draw[->,line width=1.2mm,draw=black!80,-triangle 45](ySPECIES7E.south)to[out=270,in=90]node[near start,right]{\Large \bf Reaction 6}(ySPECIES9E.north);
\draw[-,line width=1.2mm,draw=black!80,decoration={markings,mark=at position 0.5 with {\arrow{>}}},postaction={decorate}](ySPECIES8E.north)to[out=90,in=35]node[right]{}(4.55+25,-13.9);


\draw[-,line width=1.2mm,draw=black!80](ySPECIES10E.south)to[out=270,in=90]node[near start,right]{\Large \bf Reaction 8}(6+25,-18);
\draw[->,line width=1.2mm,draw=black!80,-triangle 45,decoration={markings,mark=at position 0.2 with {\arrow{>}}},postaction={decorate}](ySPECIES8E.west)to[out=180,in=0]node[right]{}(ySPECIES9E.east);

\draw[->,line width=1.2mm,draw=black!80,-triangle 45,decoration={markings,mark=at position 0.2 with {\arrow{>}}},postaction={decorate}](ySPECIES10E.west)to[out=180,in=315]node[left]{}(ySPECIES13E.south);
\draw[-,line width=1.2mm,draw=black!80](ySPECIES11E.north)to[out=90,in=315]node[near end,left]{\Large \bf Reaction 9}(-5.0+25,-9.55);

\draw [->, line width=1.2mm,draw=black!80,-triangle 45](ySPECIES14E.south)to[out=270,in=90]node[right]{\Large \bf Reaction 10}(ySPECIES10E.north);
\draw [-, line width=1.2mm,draw=black!80,decoration={markings,mark=at position 0.5 with {\arrow{>}}},postaction={decorate}](ySPECIES13E.north)to[out=90,in=90]node[right]{}(-2+25,-7);

\draw [->, line width=1.2mm,draw=black!80,-triangle 45](ySPECIES14E.west)to[out=180,in=315]node[left]{\Large \bf Reaction 11}(ySPECIES15E.south);

\draw  [->,line width=1.2mm,draw=black!80,-triangle 45](-1+25,1.4)to[out=270,in=90] node[right]{\Large \bf Reaction 12}(ySPECIES14E.north);
\draw  [-,line width=1.2mm,draw=black!80,decoration={markings,mark=at position 0.5 with {\arrow{>}}},postaction={decorate}](ySPECIES15E.north)to[out=90,in=90] node[right]{}(-1.2+25,0);

\node[text width=8cm] at (28,-20) {\LARGE \bf Effective network 2};
\draw [->,line width=1.2mm,draw=black!80,-triangle 45](13,-5/6-3)to[out=0,in=180]node[above]{}(15.0,-5/6-3);
\end{scope}
\end{tikzpicture}
}
\end{center}
\caption{A between-model move between two effective networks in Example 1 and Example 2}
\label{fig:supp_movesinmodelspace}
\end{figure}
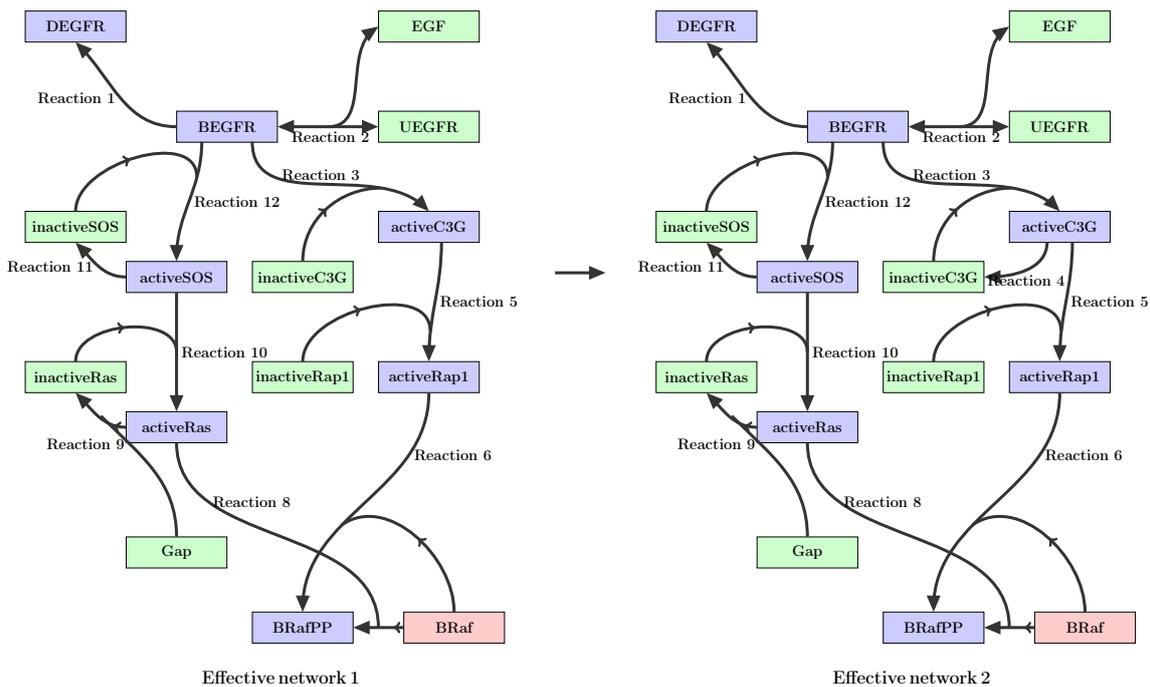

\subsection{Example 1: five-dimensional nonlinear network inference}
\noindent
In the first example, we consider a subset of reactions (15 species and 12 reactions) proposed for a protein-signalling network of the activation of extracellular signal-regulated kinase (\texttt{ERK}) by epidermal growth factor (\texttt{EGF}) \cite{Xu2010}. The reaction network used is shown in Figure \ref{fig:supp_reactionNetwork1Allreactions}. The ODE forward model governing the evolution of species concentrations is described in detail in Section \ref{sec:12Dnetwork}. We keep reactions 1, 2, 8, 9, 10, 11, and 12 fixed and thus they are included in all the inferred models. The rate constants of all fixed reactions (shaded pink in Table \ref{kineticModelExample2}) and Michaelis constants of all reactions are set to their base values (Table \ref{kineticModelExample2}).

\begin{table}[h]
\scriptsize
\begin{center}
\begin{tabular}{ c c c c c}
\hline
\mbox{} & Reaction & ${\log_{10}k^{*}}^a$ & $k_{M}^b$ & Prior uncertainty \\
\hline
\rowcolor{pink}
1 & BEGFR $\rightarrow$ DEGFR & 0.0  & - & $ - $\\
\rowcolor{pink}
2a & EGF + UEGFR $\rightarrow$ BEGFR & 1.5  & - & $ - $\\
\rowcolor{pink}
2b & BEGFR $\rightarrow$ EGF + UEGFR & 0.0  & - & $ - $\\
\rowcolor{LightCyan}
3 & inactiveC3G+BEGFR $\rightarrow$ activeC3G+BEGFR & 0.5  & 3386.3875 & $ \log_{10} k = \mathcal{N}(1.1,0.2) $\\
\rowcolor{LightCyan}
4 & activeC3G $\rightarrow$ inactiveC3G & 2.0  & - & $ \log_{10} k = \mathcal{N}(1.4,0.2) $\\
\rowcolor{LightCyan}
5 & inactiveRap1+activeC3G $\rightarrow$ activeRap1+activeC3G & 2.0  & 3566 & $ \log_{10} k = \mathcal{N}(2.6,0.2) $\\
\rowcolor{LightCyan}
6 & BRaf+activeRap1 $\rightarrow$ BRafPP+activeRap1 & 0.4  & 17991.179 & $ \log_{10} k = \mathcal{N}(1.0,0.2)$\\
\rowcolor{LightCyan}
7 & activeRap1+Gap $\rightarrow$ inactiveRap1+Gap & 1.0  & 6808.32 & $ \log_{10} k = \mathcal{N}(0.4,0.2) $\\
\rowcolor{pink}
8 & BRaf+activeRas $\rightarrow$ BRafPP+activeRas & 0.5  & 7631.63 & $ - $\\
\rowcolor{pink}
9 & activeRas+Gap $\rightarrow$ inactiveRas+Gap & 0.0  & 12457.816 & $ - $\\
\rowcolor{pink}
10 & inactiveRas+activeSOS $\rightarrow$ activeRas+activeSOS & 0.5  & 13.73 & $ - $\\
\rowcolor{pink}
11 & activeSOS $\rightarrow$ inactiveSOS & 4.0  & 9834.13 & $ - $\\
\rowcolor{pink}
12 & inactiveSOS+BEGFR $\rightarrow$ activeSOS+BEGFR & 2.5  & 8176.56 & $ - $\\
\hline
\multicolumn{5}{l}{$^{a}$ logarithm (base rate constant value)} \\ 
\multicolumn{5}{l}{$^{b}$ Base value of Michaelis constant (Obtained from Xu et al.\ \cite{Xu2010})}
\end{tabular}
\end{center}
{\normalsize
\caption{Proposed reactions for Example 1}
\label{kineticModelExample2}
}
\end{table}
\noindent
There are five effective networks in Example 1. Figure \ref{fig:effective_networks_example1} shows the network diagrams of all effective networks in Example 1. As an example of our network-aware proposal construction, we present the between-model proposal for the move between effective network 1 and effective network 2 in Figure \ref{fig:supp_movesinmodelspace}. The move from effective network 1 to effective network 2 involves the addition of reaction 4. The identify jump function $f$ for the between-model move between the two networks is:

\begin{equation}
k_{2,\{3,5,6\}}=k_{1,\{3,5,6\}}, \mbox{            } k_{2,4}=u,
\end{equation}

\noindent
where the proposal distribution $q(u \vert k_{1,\{3,5,6\}})$ is $\mathcal{N}(u;\mu,\Sigma)$, with

\begin{equation}
\mu =\underset{k_{2,4}}{\arg\max}\mbox{  }p(k_{2,4} \vert k_{1,\{3,5,6\}},M_{2},\mathcal{D})
\end{equation}

\noindent
and

\begin{equation}
\Sigma =-[\nabla^2 \log p(k_{{2},4} \vert k_{1,\{3,5,6\}},M_{2},\mathcal{D})]^{-1}\big|_{{k_{2,4}}=\mu}
\end{equation}

\subsection{Example 2: ten-dimensional nonlinear network inference}
\noindent
Our second example is a large scale nonlinear network inference problem with 10 uncertain reactions.  Once again, we consider a protein-signalling network consisting of 15 species and 12 potential species interactions (Figure \ref{fig:supp_reactionNetwork1Allreactions}). The ODE forward model governing the evolution of species concentrations is described in detail in Section \ref{sec:12Dnetwork}. The rate constants of all fixed reactions and Michaelis constants of all reactions are set to their base values (Table \ref{kineticModelExample4}). Reactions 3--12 are uncertain and the concentration of \texttt{BRaf} is again the observable.
\begin{table}[h]
\scriptsize
\begin{center}
\begin{tabular}{ c c c c c c }
\hline
\mbox{} & Reaction & ${\log_{10}k^{*}}^a$ & $k_{M}^b$ & Prior uncertainty \\
\hline
\rowcolor{pink}
1 & BEGFR $\rightarrow$ DEGFR & 0.0  & - & $ - $\\
\rowcolor{pink}
2a & EGF + UEGFR $\rightarrow$ BEGFR & 1.5  & - & $ - $\\
\rowcolor{pink}
2b & BEGFR $\rightarrow$ EGF + UEGFR & 0.0  & - & $ - $\\
\rowcolor{LightCyan}
3 & inactiveC3G+BEGFR $\rightarrow$ activeC3G+BEGFR & 0.5  & 3386.3875 & $ \log_{10} k = \mathcal{N}(1.2,0.1) $\\
\rowcolor{LightCyan}
4 & activeC3G $\rightarrow$ inactiveC3G & 2.0  & - & $ \log_{10} k = \mathcal{N}(2.0,0.1) $\\
\rowcolor{LightCyan}
5 & inactiveRap1+activeC3G $\rightarrow$ activeRap1+activeC3G & 2.0  & 3566 & $ \log_{10} k = \mathcal{N}(2.7,0.1) $\\
\rowcolor{LightCyan}
6 & BRaf+activeRap1 $\rightarrow$ BRafPP+activeRap1 & 0.4  & 17991.179 & $ \log_{10} k = \mathcal{N}(1.1,0.1)$\\
\rowcolor{LightCyan}
7 & activeRap1+Gap $\rightarrow$ inactiveRap1+Gap & 1.0  & 6808.32 & $ \log_{10} k = \mathcal{N}(1.0,0.01) $\\
\rowcolor{LightCyan}
8 & BRaf+activeRas $\rightarrow$ BRafPP+activeRas & 0.5  & 7631.63 & $ \log_{10} k = \mathcal{N}(0.5,0.1) $\\
\rowcolor{LightCyan}
9 & activeRas+Gap $\rightarrow$ inactiveRas+Gap & 0.0  & 12457.816 & $ \log_{10} k = \mathcal{N}(0.0,0.01) $\\
\rowcolor{LightCyan}
10 & inactiveRas+activeSOS $\rightarrow$ activeRas+activeSOS & 0.5  & 13.73 & $ \log_{10} k = \mathcal{N}(0.5,0.1) $\\
\rowcolor{LightCyan}
11 & activeSOS $\rightarrow$ inactiveSOS & 4.0  & 9834.13 & $ \log_{10} k = \mathcal{N}(4.0,0.01) $\\
\rowcolor{LightCyan}
12 & inactiveSOS+BEGFR $\rightarrow$ activeSOS+BEGFR & 2.5  & 8176.56 & $ \log_{10} k = \mathcal{N}(2.5,0.1) $\\
\hline
\multicolumn{5}{l}{$^{a}$ logarithm (base rate constant value)} \\ 
\multicolumn{5}{l}{$^{b}$ Base value of Michaelis constant (Obtained from Xu et al.\ \cite{Xu2010})} \\
\end{tabular}
\end{center}
{\normalsize
\caption{Proposed reactions for Example 2}
\label{kineticModelExample4}
}
\end{table}

\begin{figure}[h]
{ \makeatletter
 \def\@captype{figure}
 \makeatother
 \begin{minipage}[h]{0.3\textwidth}
 \begin{subfigure}[b]{1.0\textwidth}
 \begin{center}
 \includegraphics[width=0.8\linewidth]{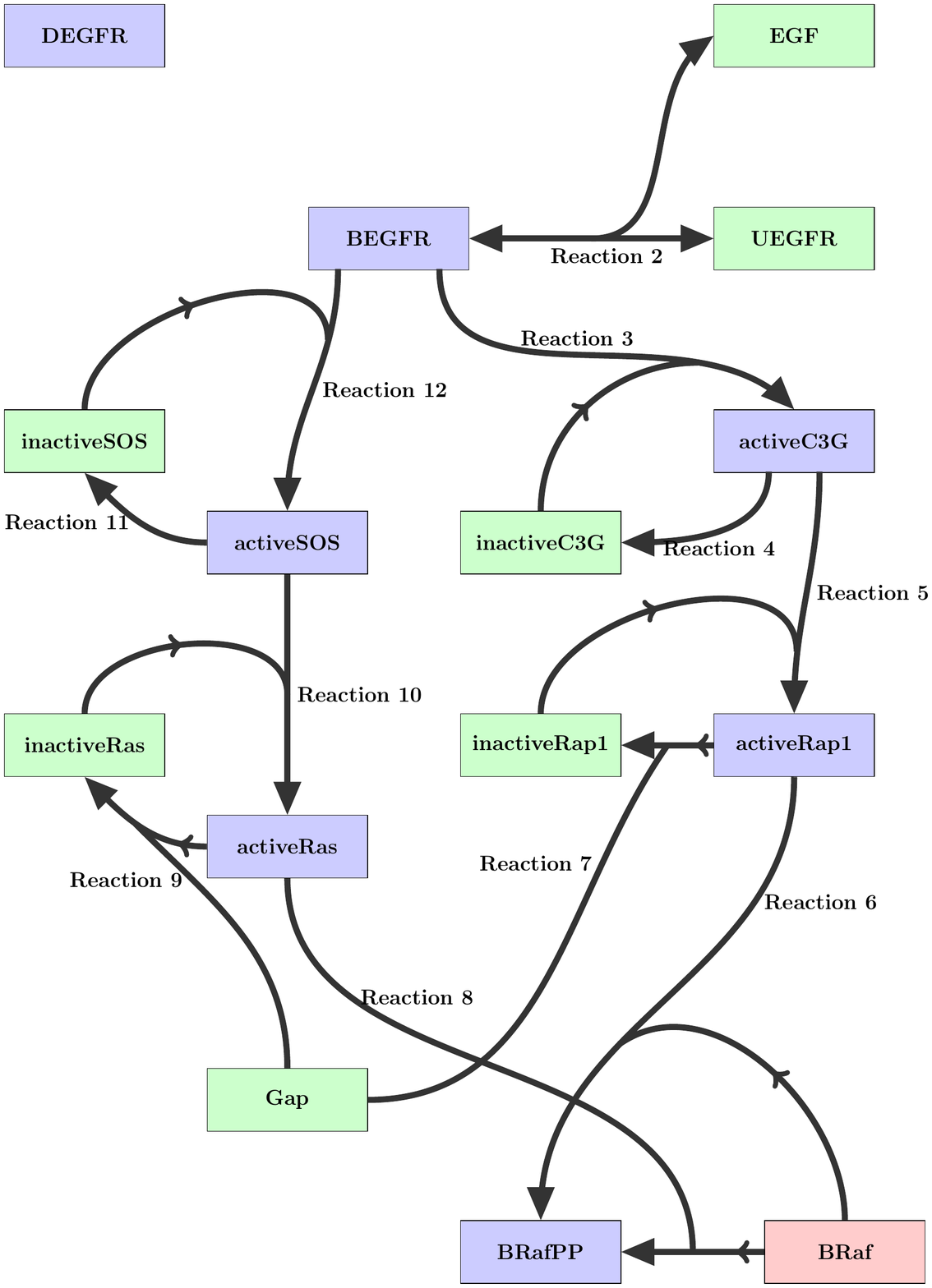}
 \includegraphics[width=0.8\linewidth]{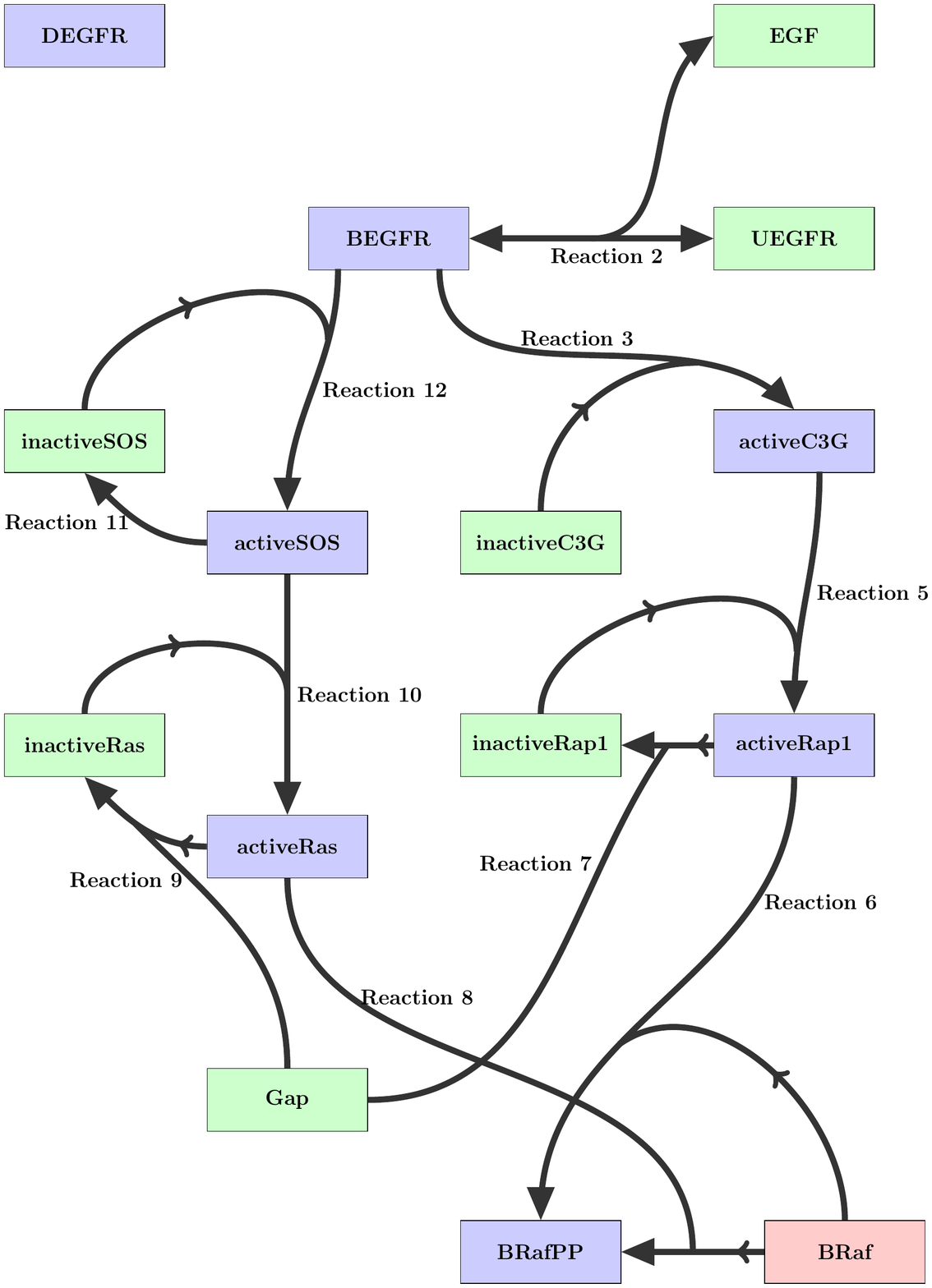}
 \includegraphics[width=0.8\linewidth]{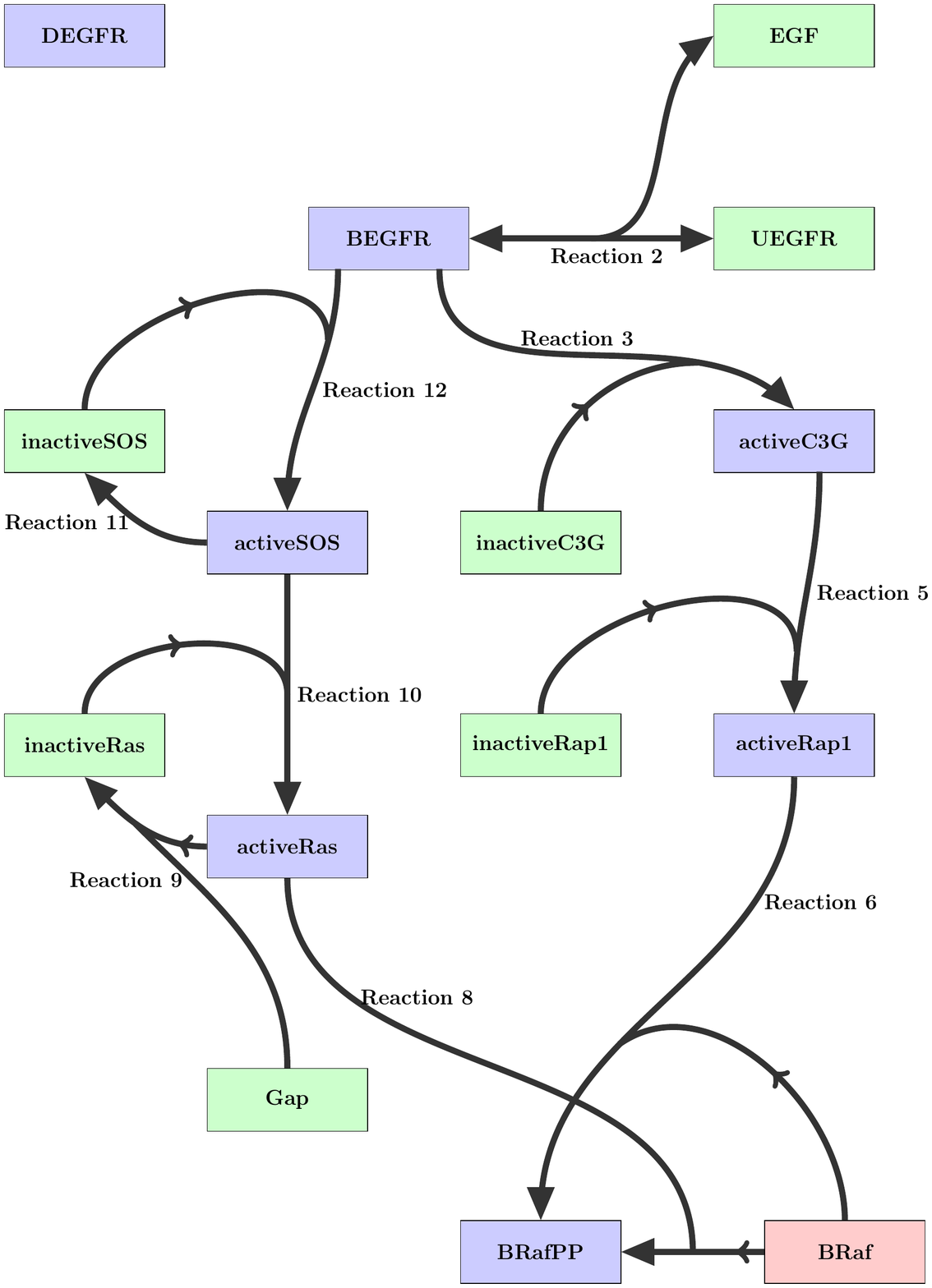}
 \includegraphics[width=0.8\linewidth]{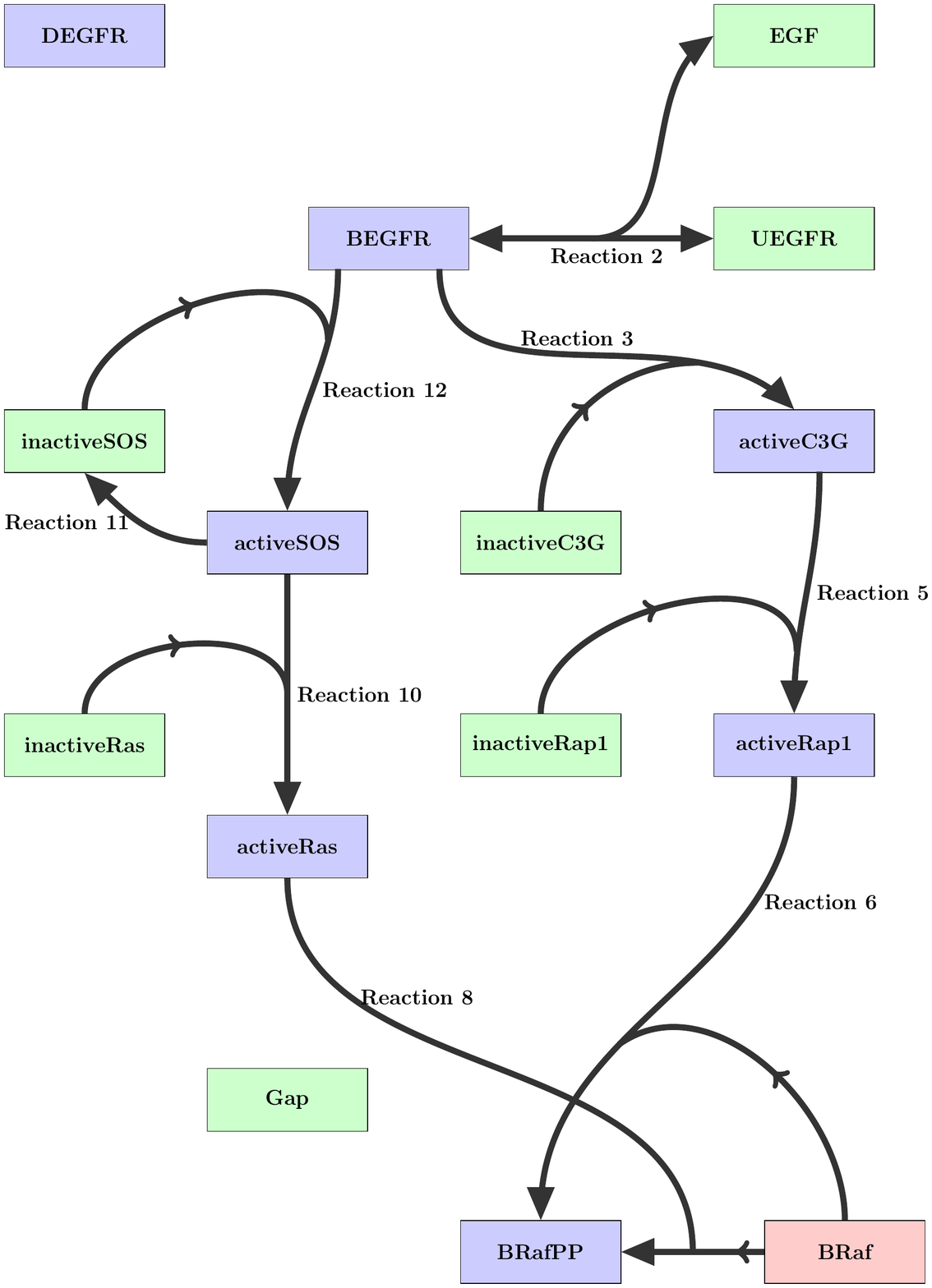}
 \end{center}  
  \label{fig:supp_sub1}
 \end{subfigure} 
 \end{minipage}
 \begin{minipage}[h]{0.3\textwidth} 
\begin{subfigure}[b]{1.0\textwidth}
\begin{center}
\includegraphics[width=0.8\linewidth]{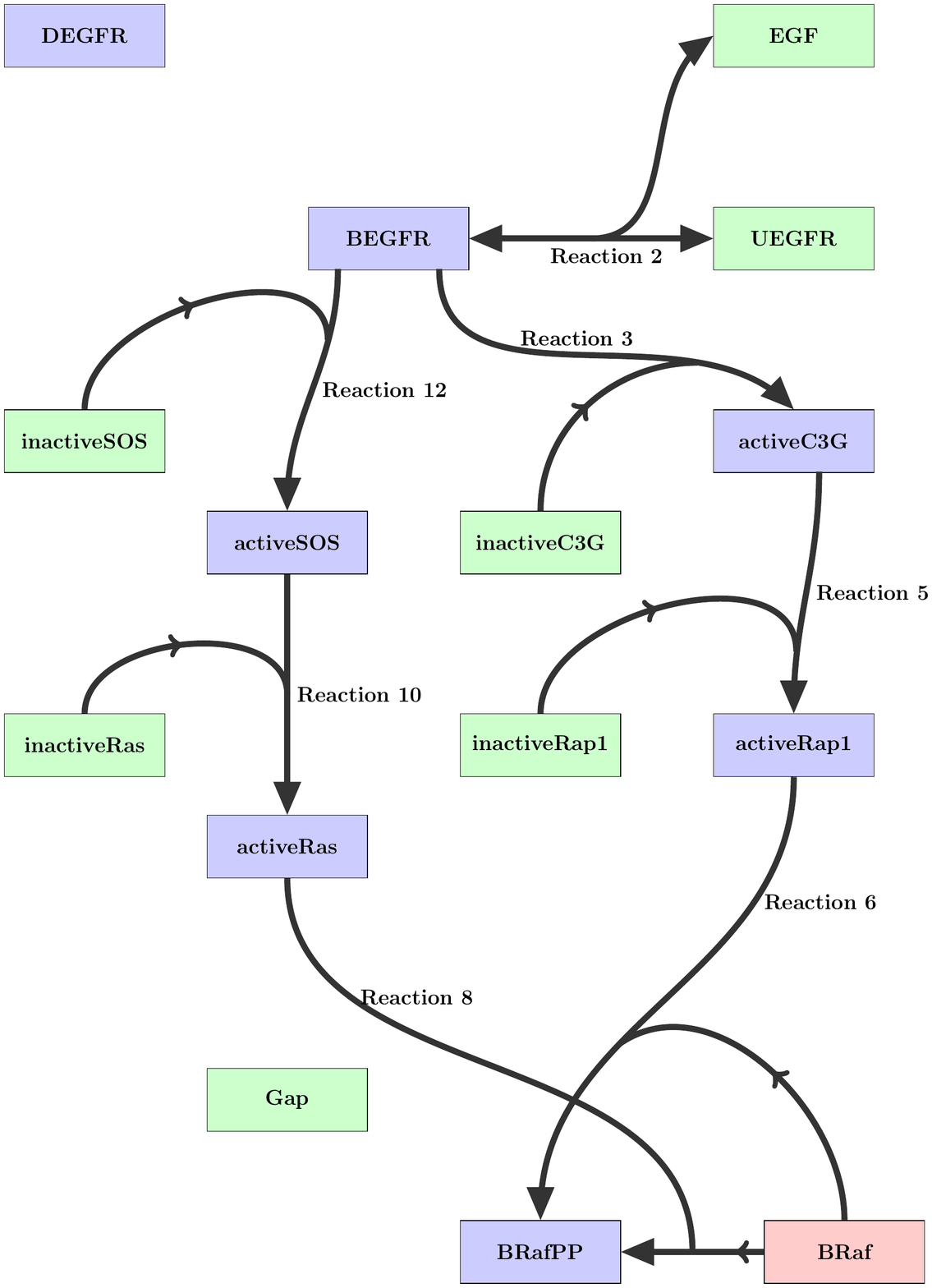}
\includegraphics[width=0.8\linewidth]{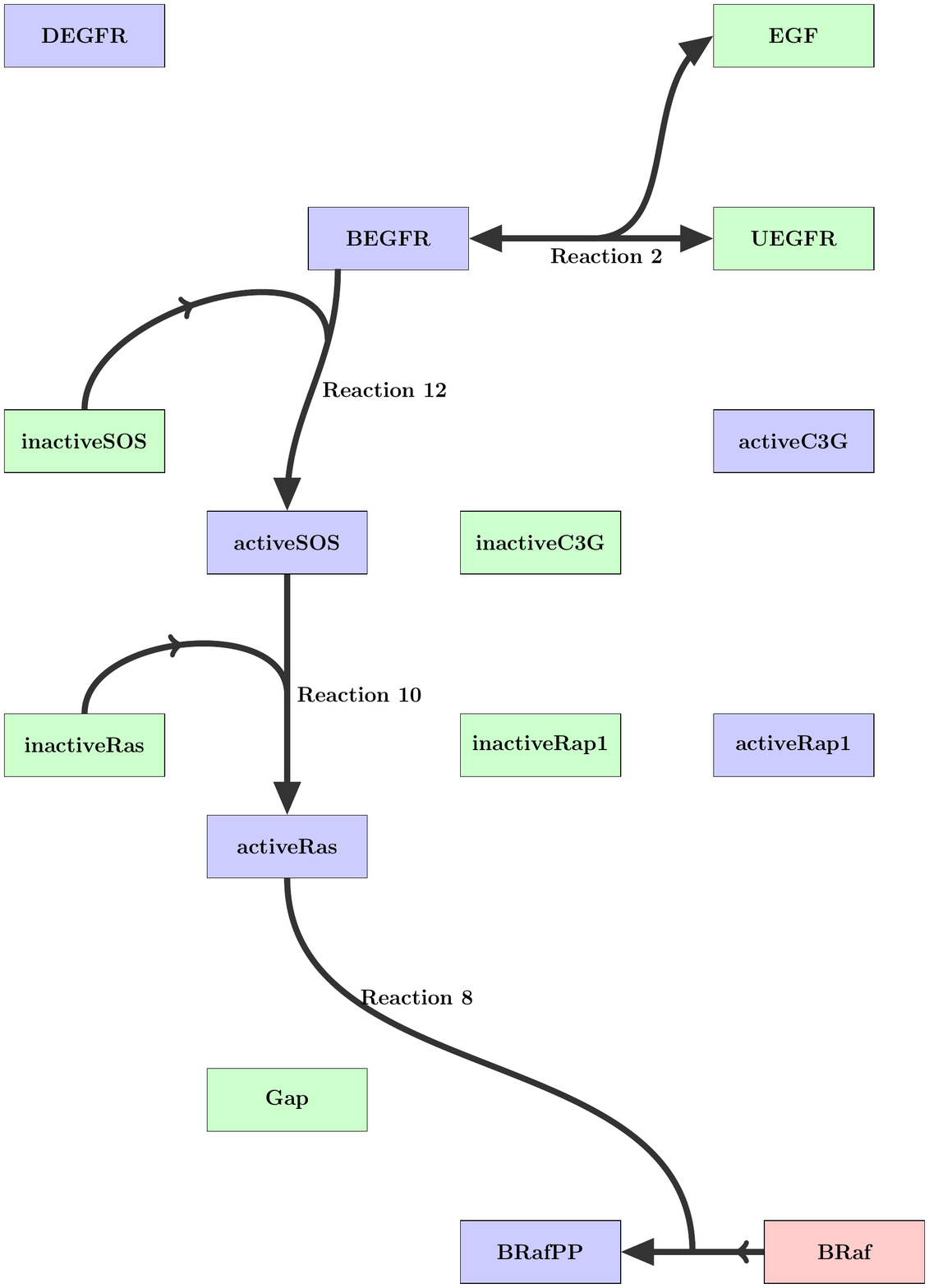}
\includegraphics[width=0.8\linewidth]{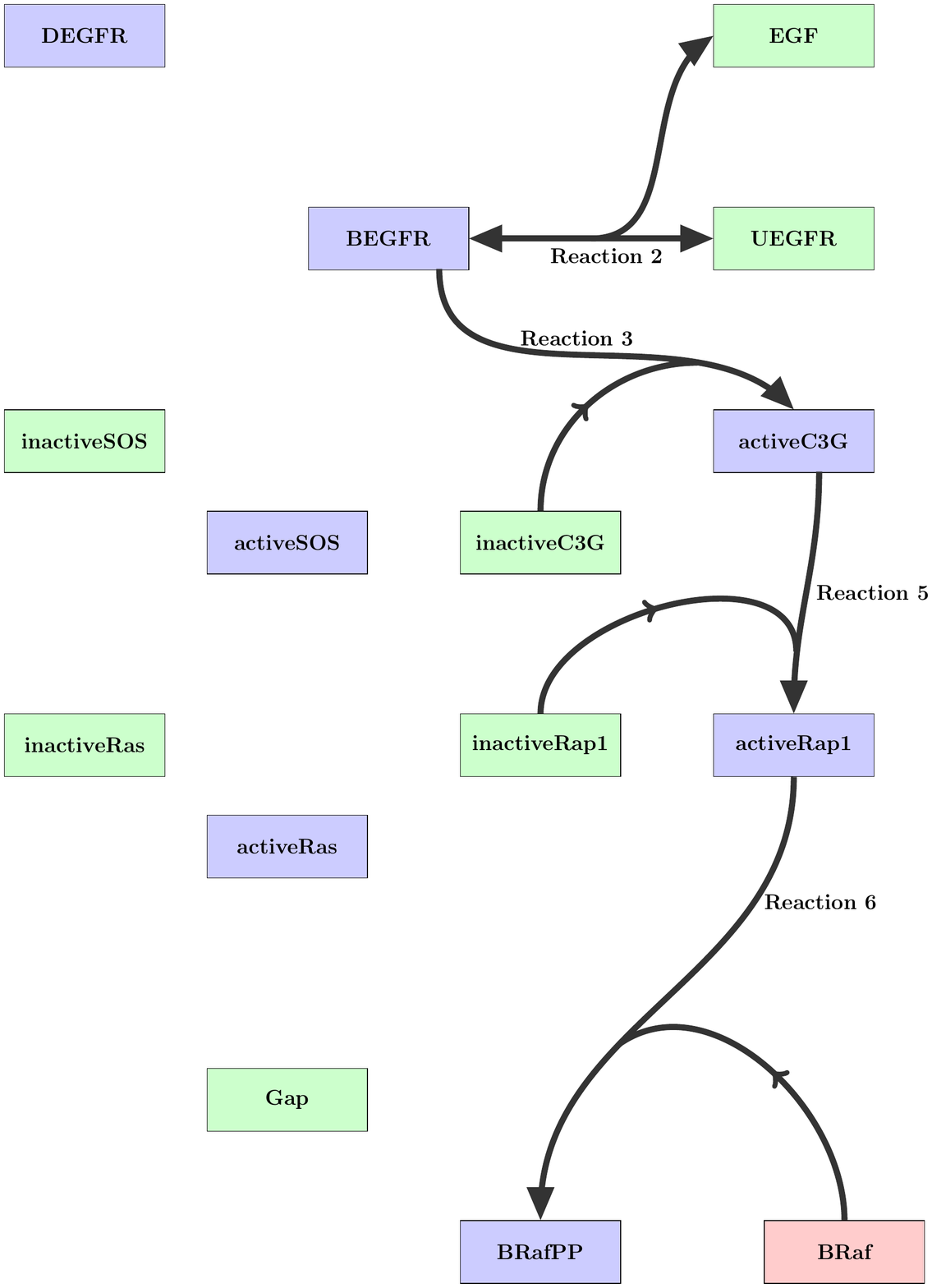}
\includegraphics[width=0.8\linewidth]{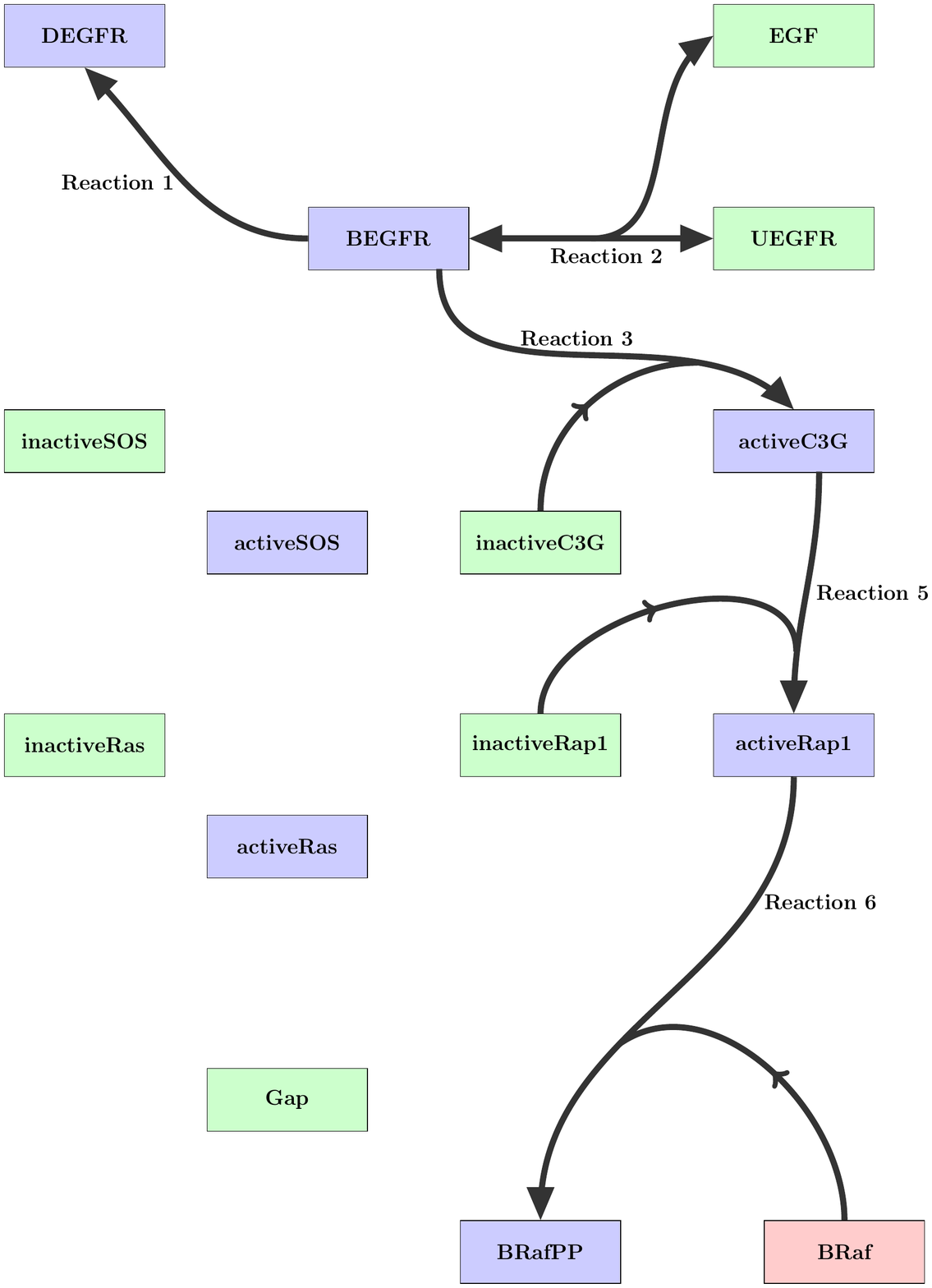}
\end{center}  
\label{fig:supp_sub2}
\end{subfigure}
\end{minipage}
 \begin{minipage}[h]{0.3\textwidth} 
\begin{subfigure}[b]{1.0\textwidth}
\begin{center}
\includegraphics[width=0.8\linewidth]{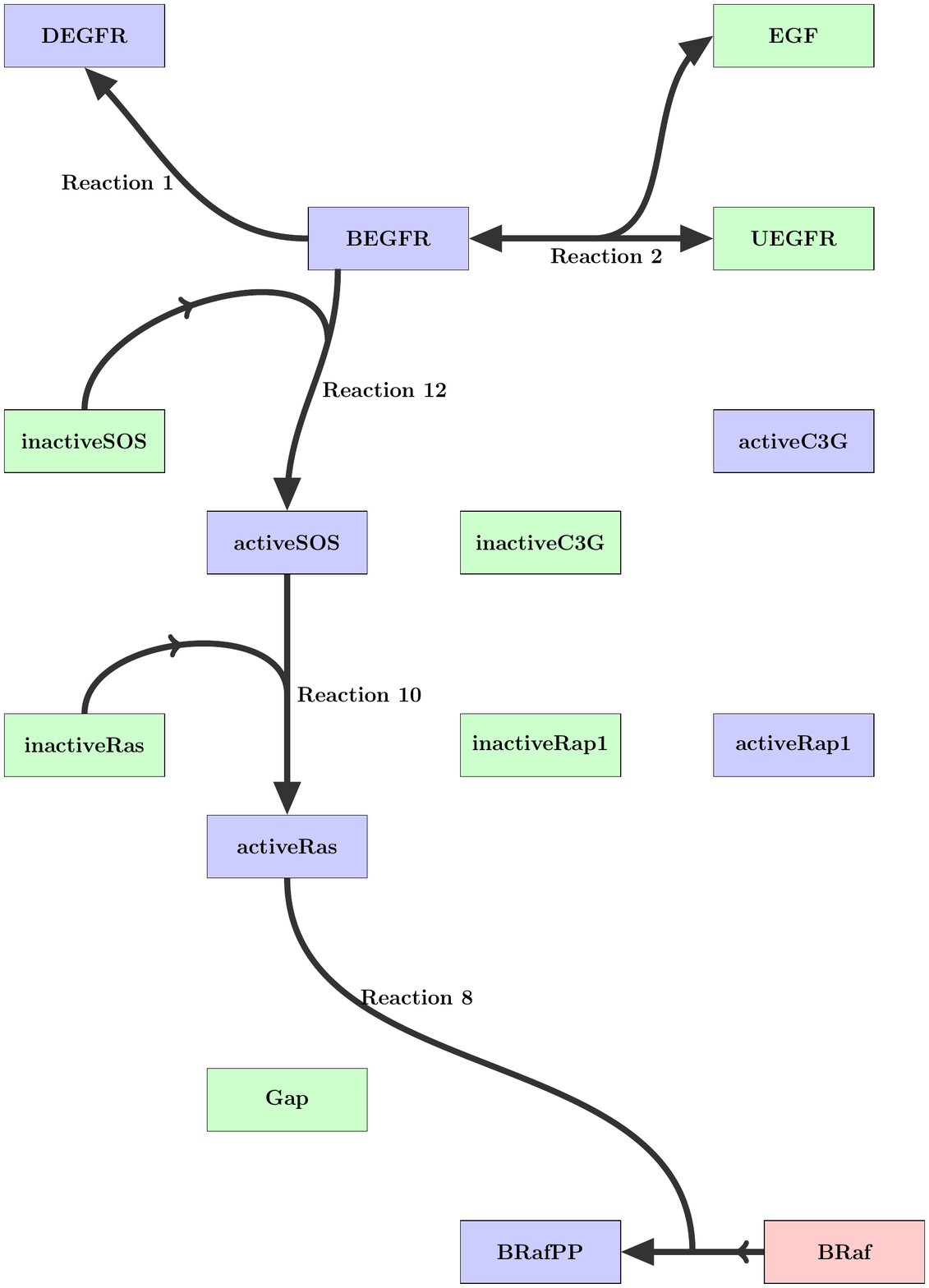}
\includegraphics[width=0.8\linewidth]{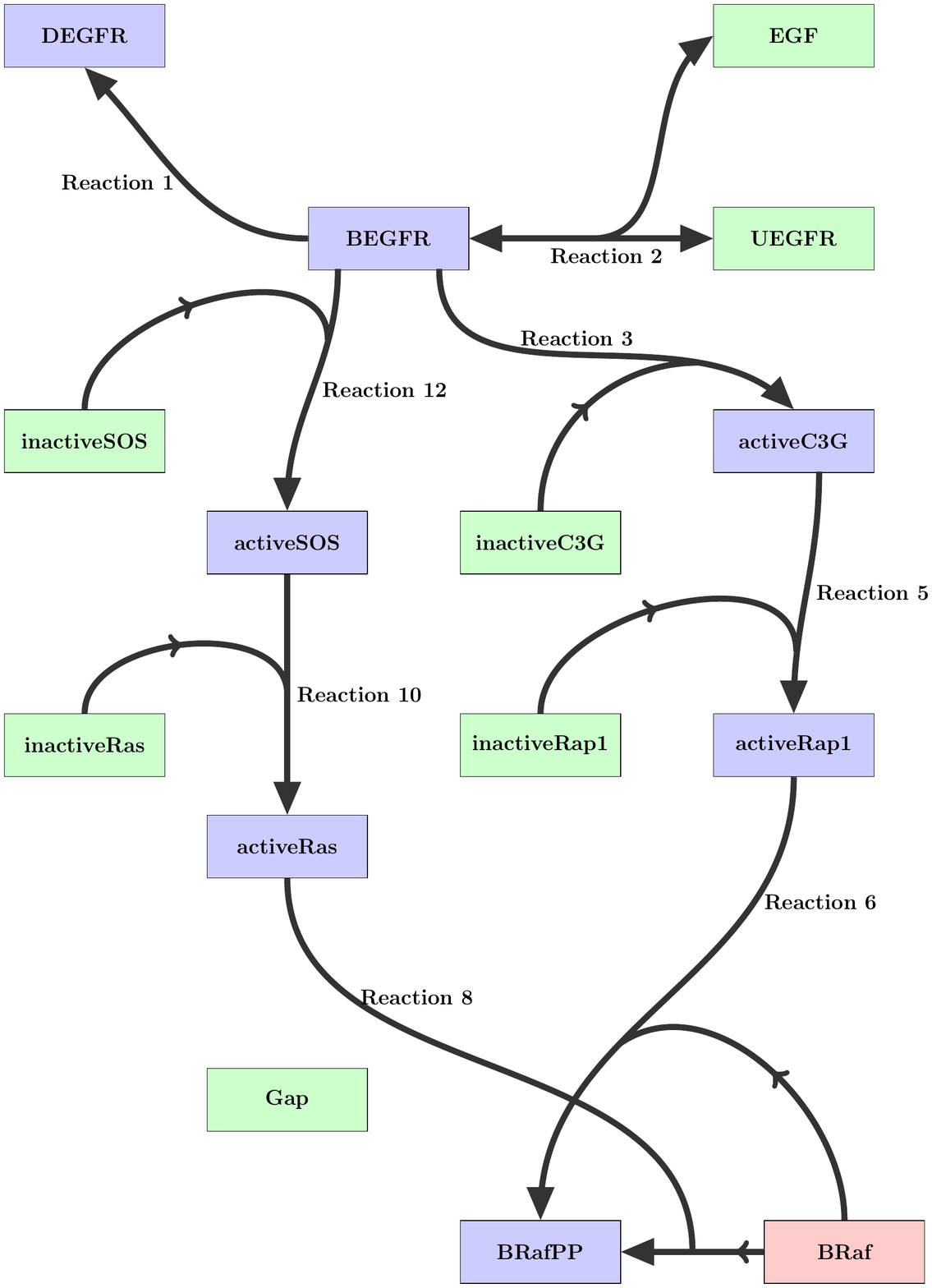}
\includegraphics[width=0.8\linewidth]{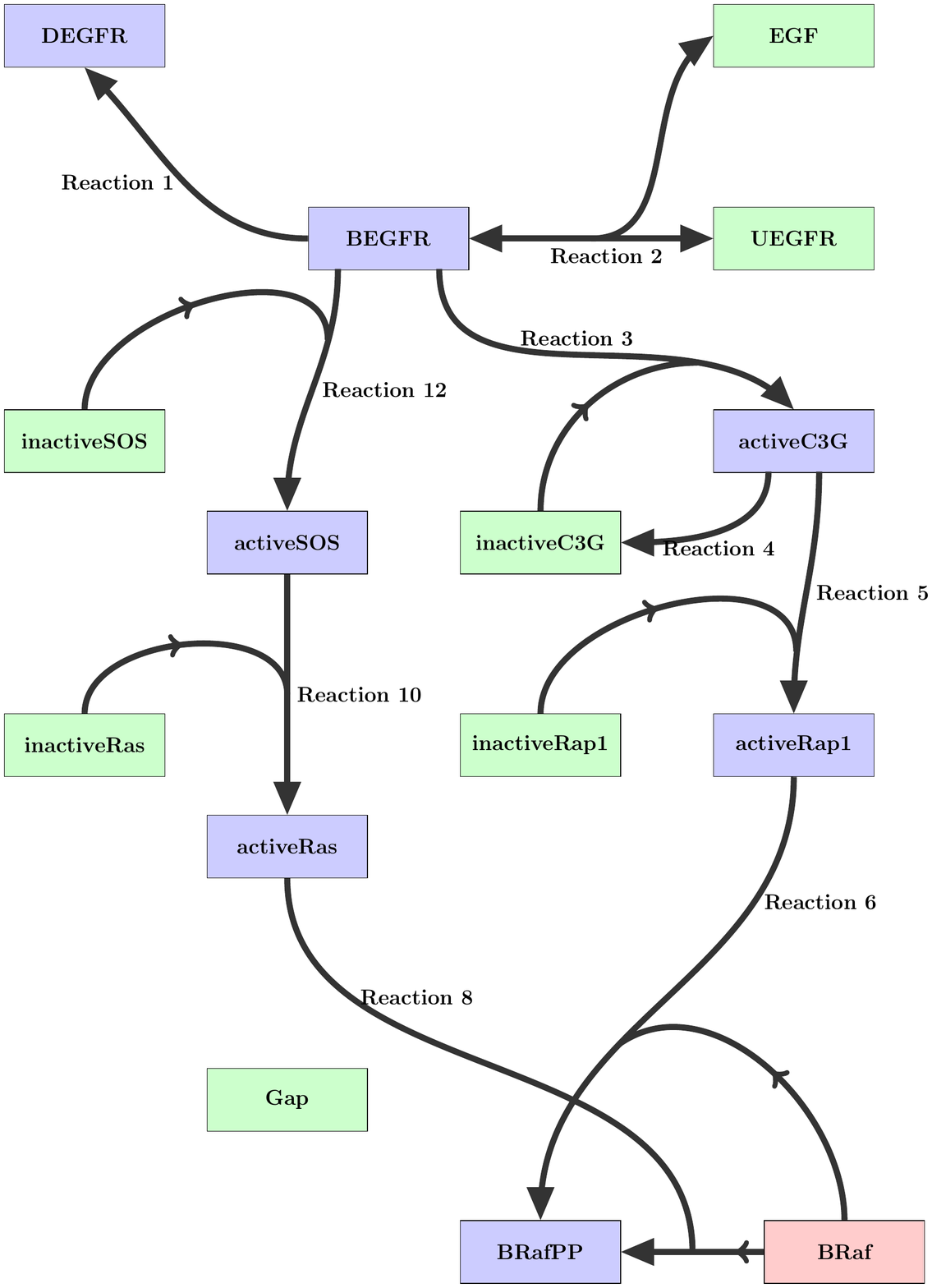}
\includegraphics[width=0.8\linewidth]{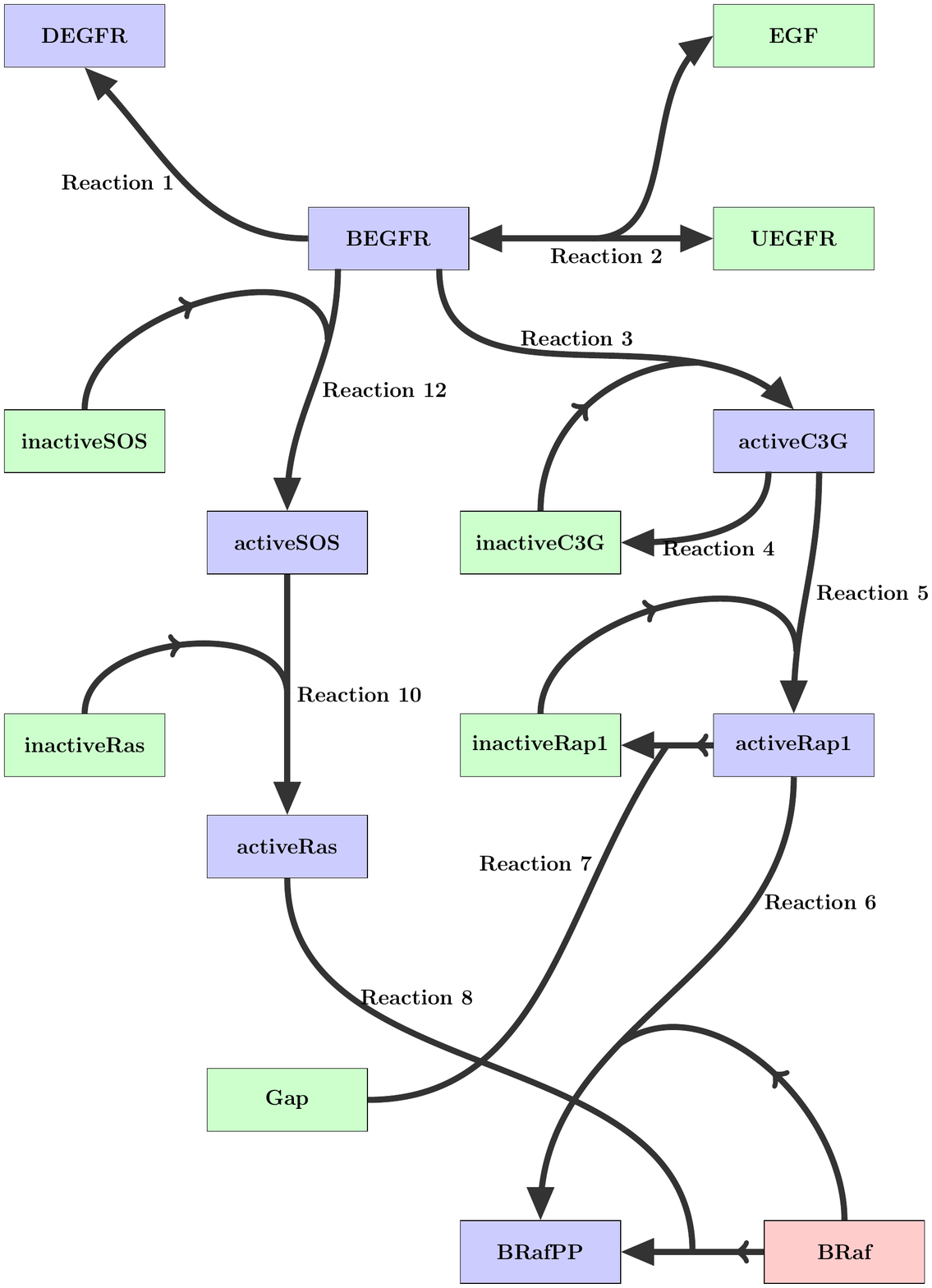}
\end{center}  
\label{fig:supp_sub3}
\end{subfigure}
\end{minipage}
\caption{Effective networks in Example 2}
\label{fig:effective_networks_example2}}
\end{figure}

\begin{figure}[h]
{ \makeatletter
 \def\@captype{figure}
 \makeatother
 \begin{minipage}[h]{0.3\textwidth}
 \begin{subfigure}[b]{1.0\textwidth}
 \begin{center}
 \includegraphics[width=0.8\linewidth]{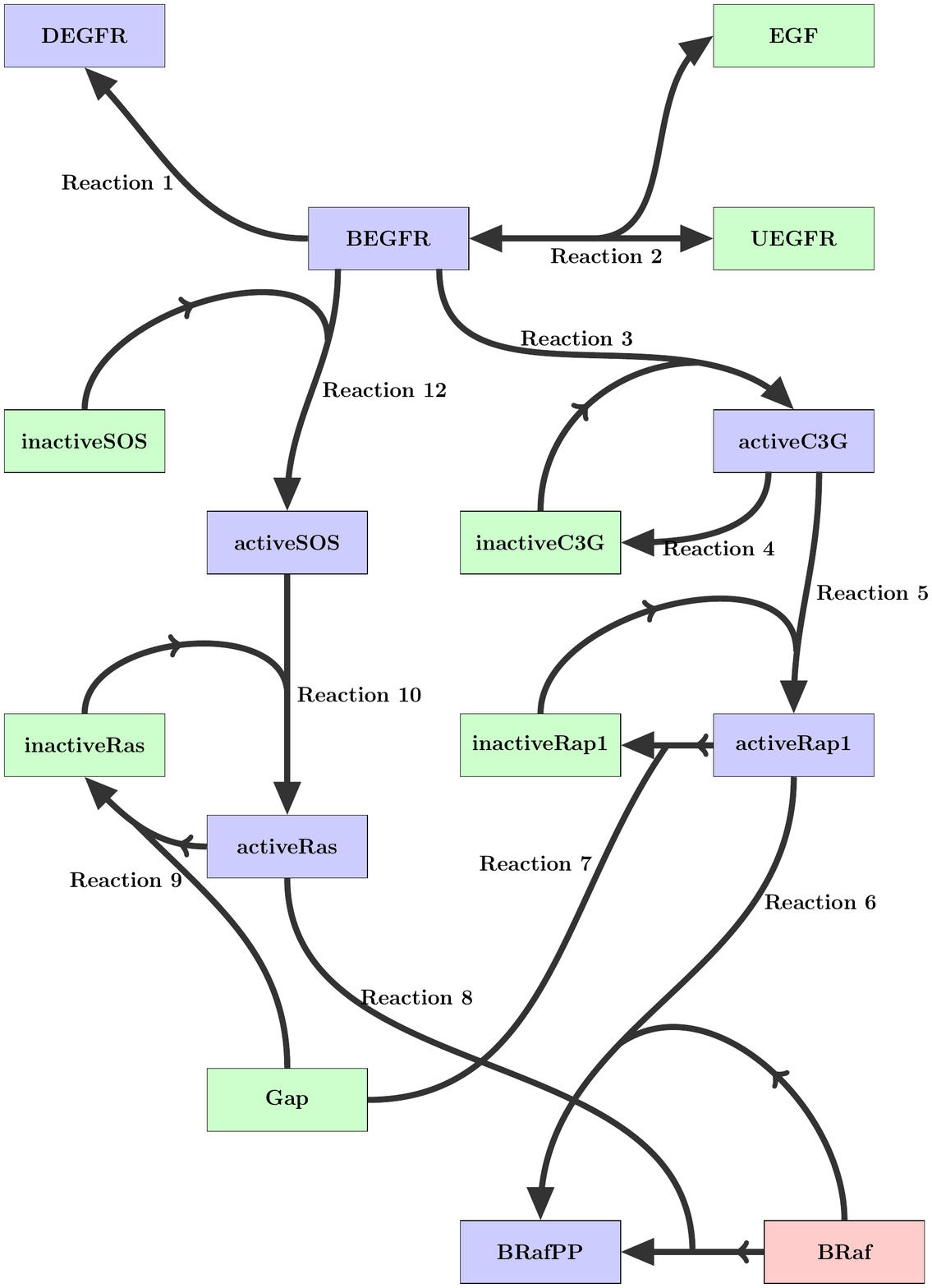}
 \includegraphics[width=0.8\linewidth]{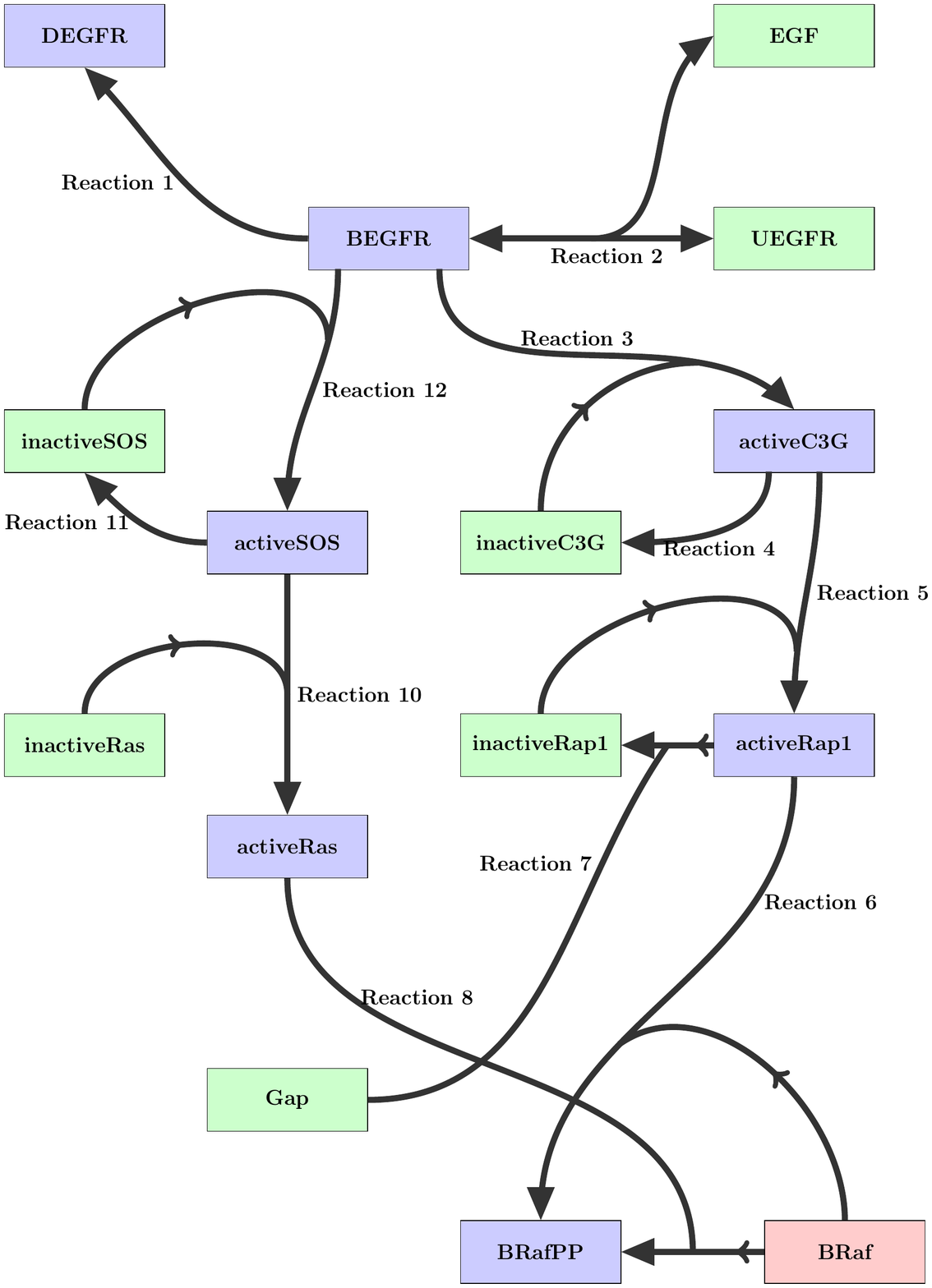}
 \includegraphics[width=0.8\linewidth]{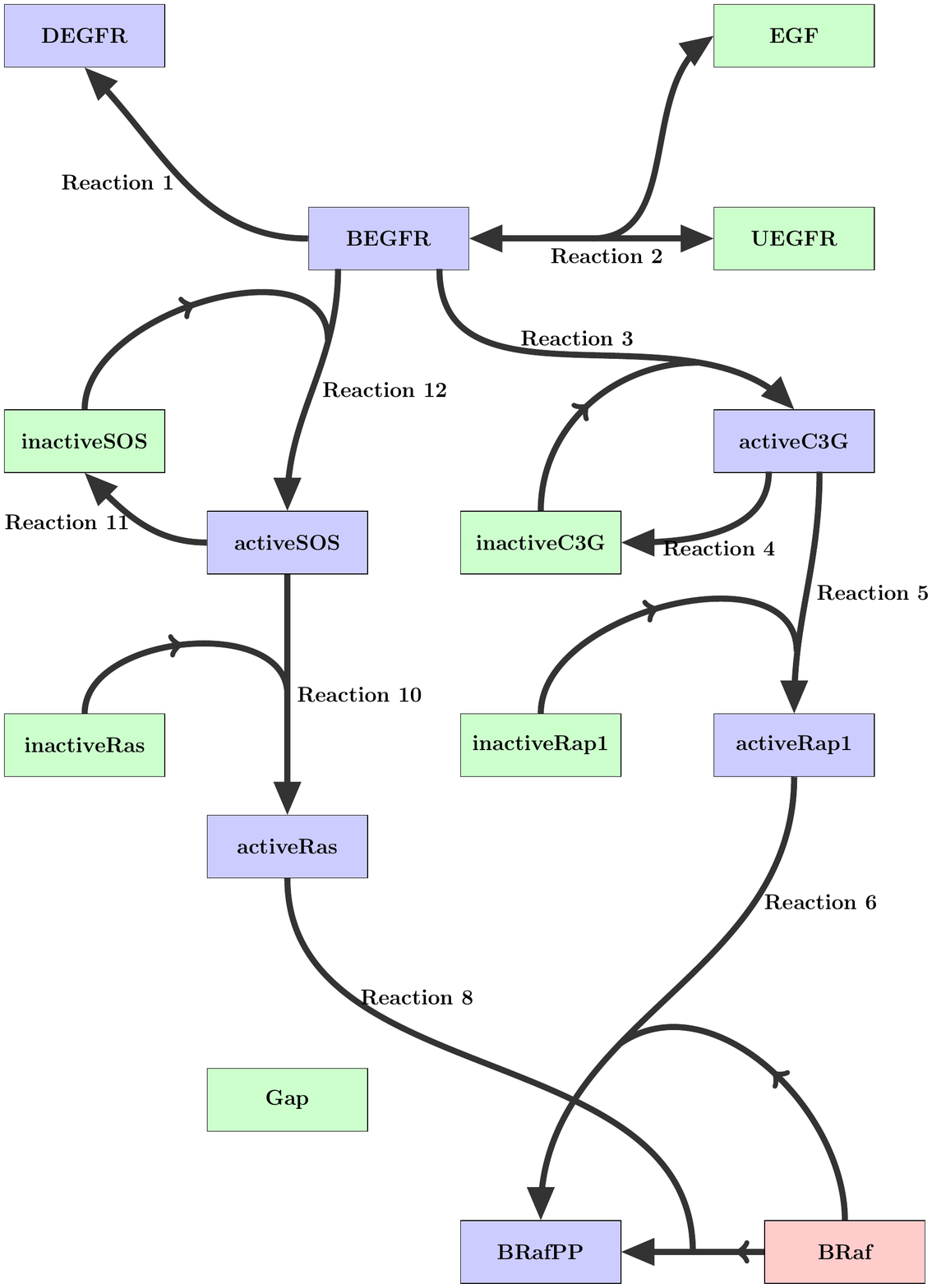}
 \includegraphics[width=0.8\linewidth]{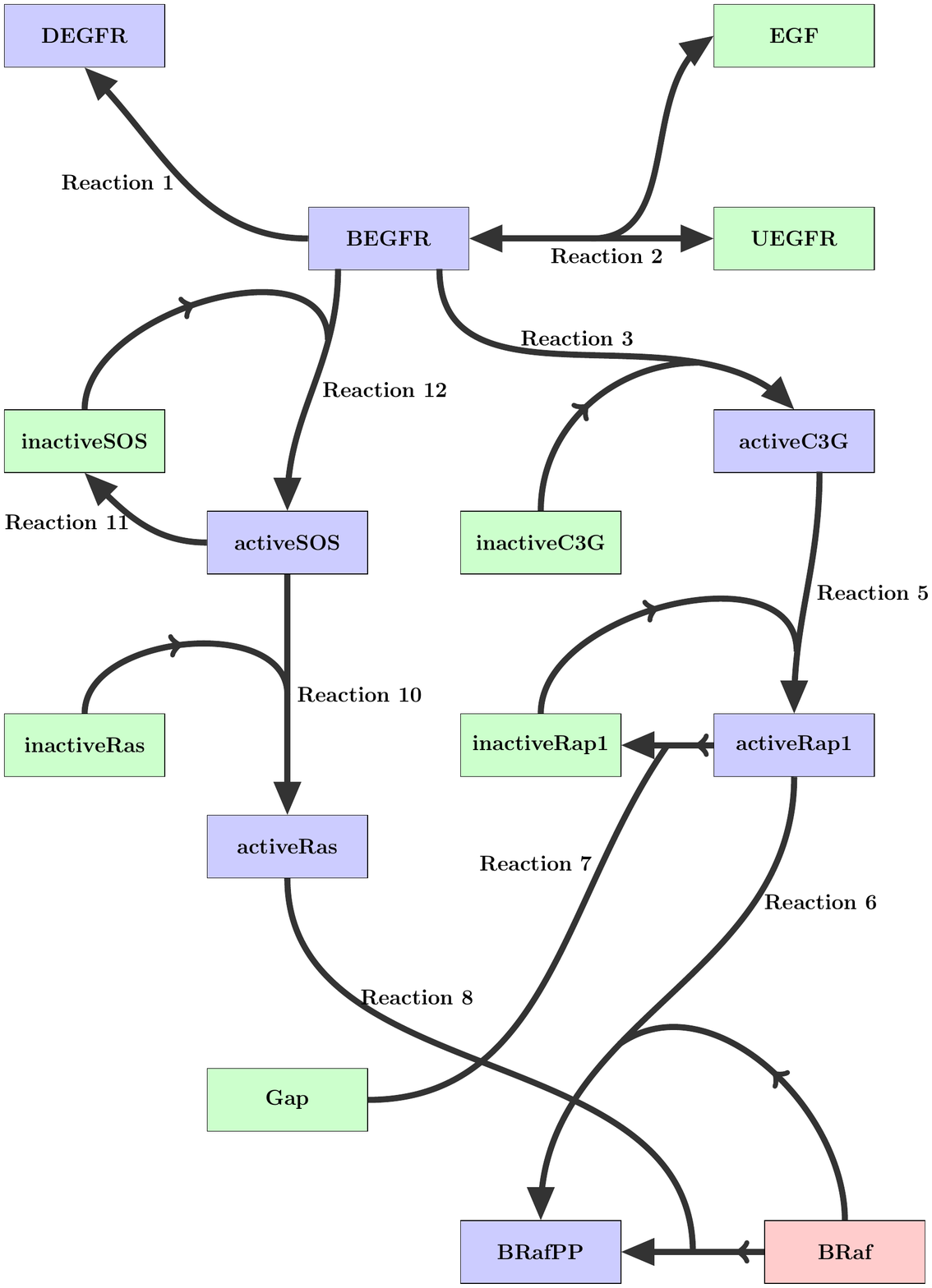}
 \end{center}  
  \label{fig:supp_sub1}
 \end{subfigure} 
 \end{minipage}
 \begin{minipage}[h]{0.3\textwidth} 
\begin{subfigure}[b]{1.0\textwidth}
\begin{center}
\includegraphics[width=0.8\linewidth]{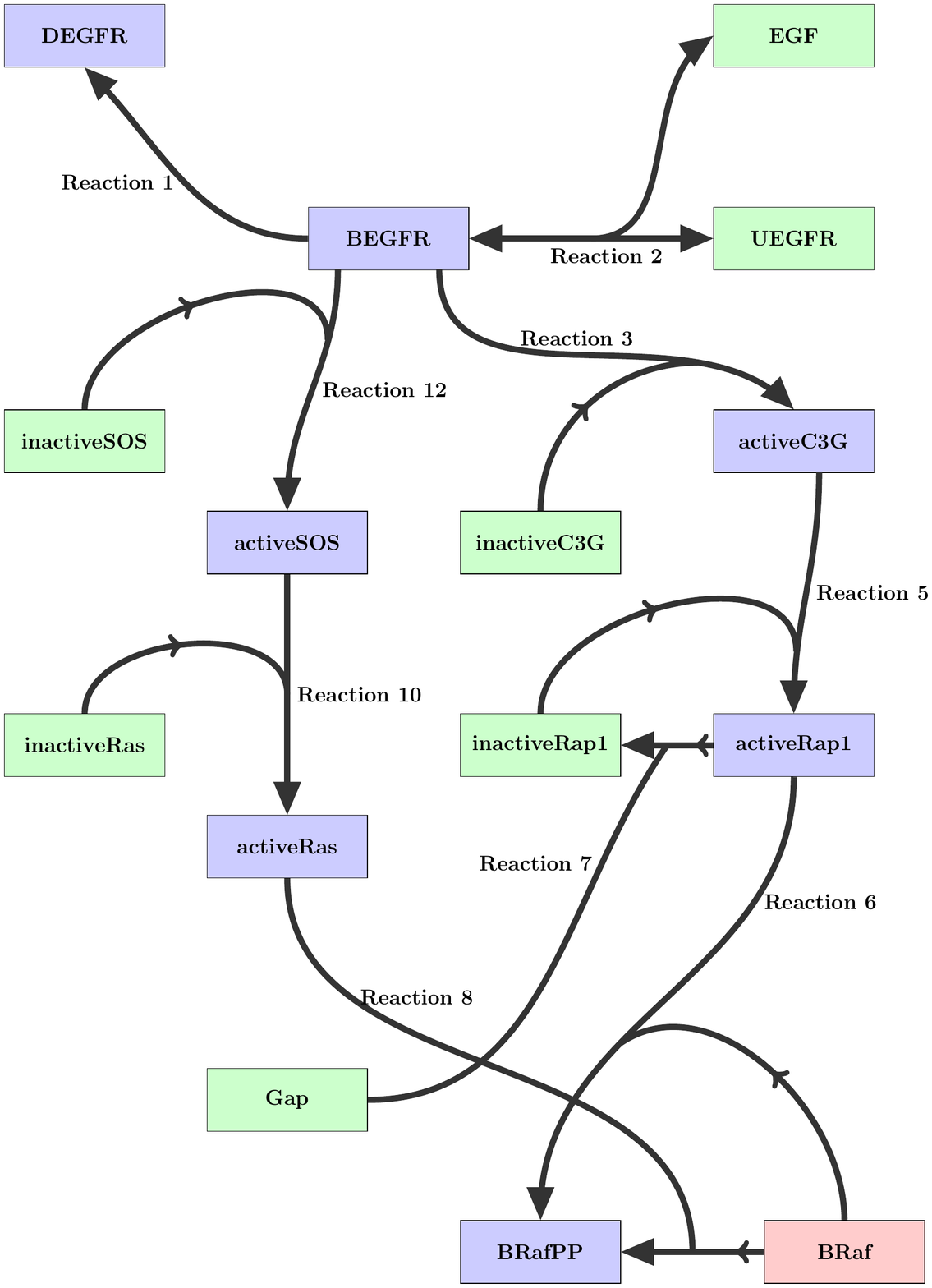}
\includegraphics[width=0.8\linewidth]{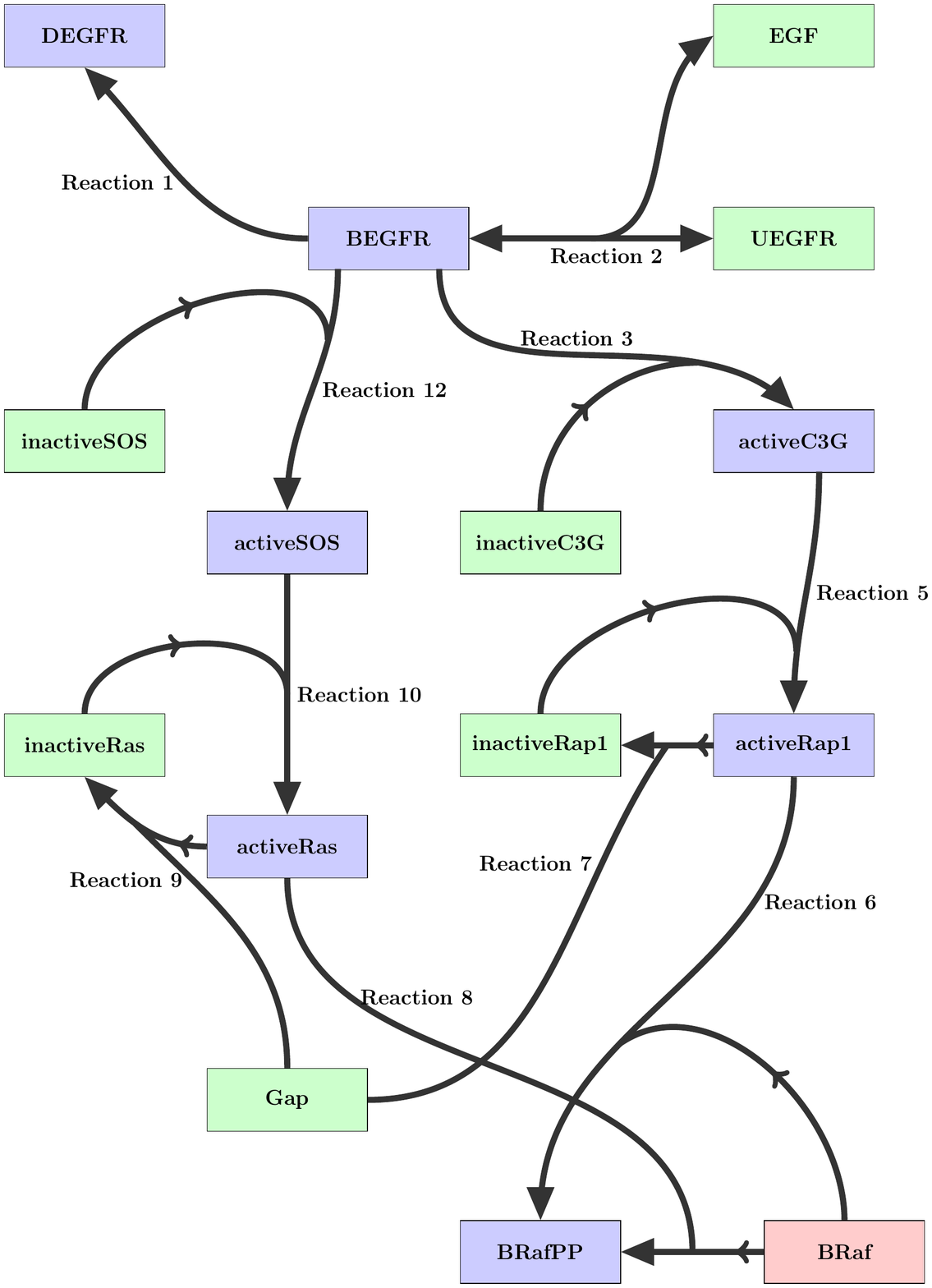}
\includegraphics[width=0.8\linewidth]{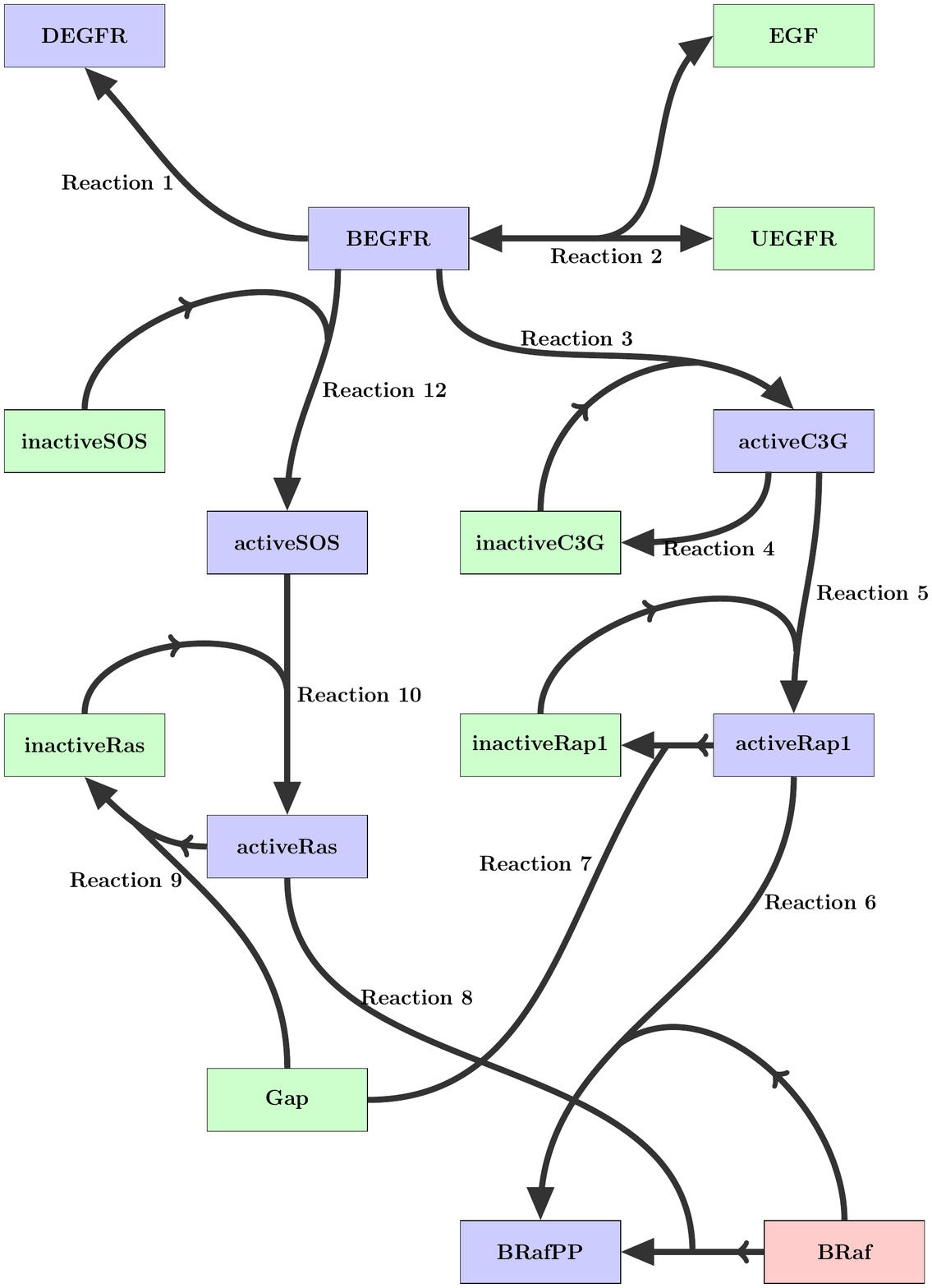}
\includegraphics[width=0.8\linewidth]{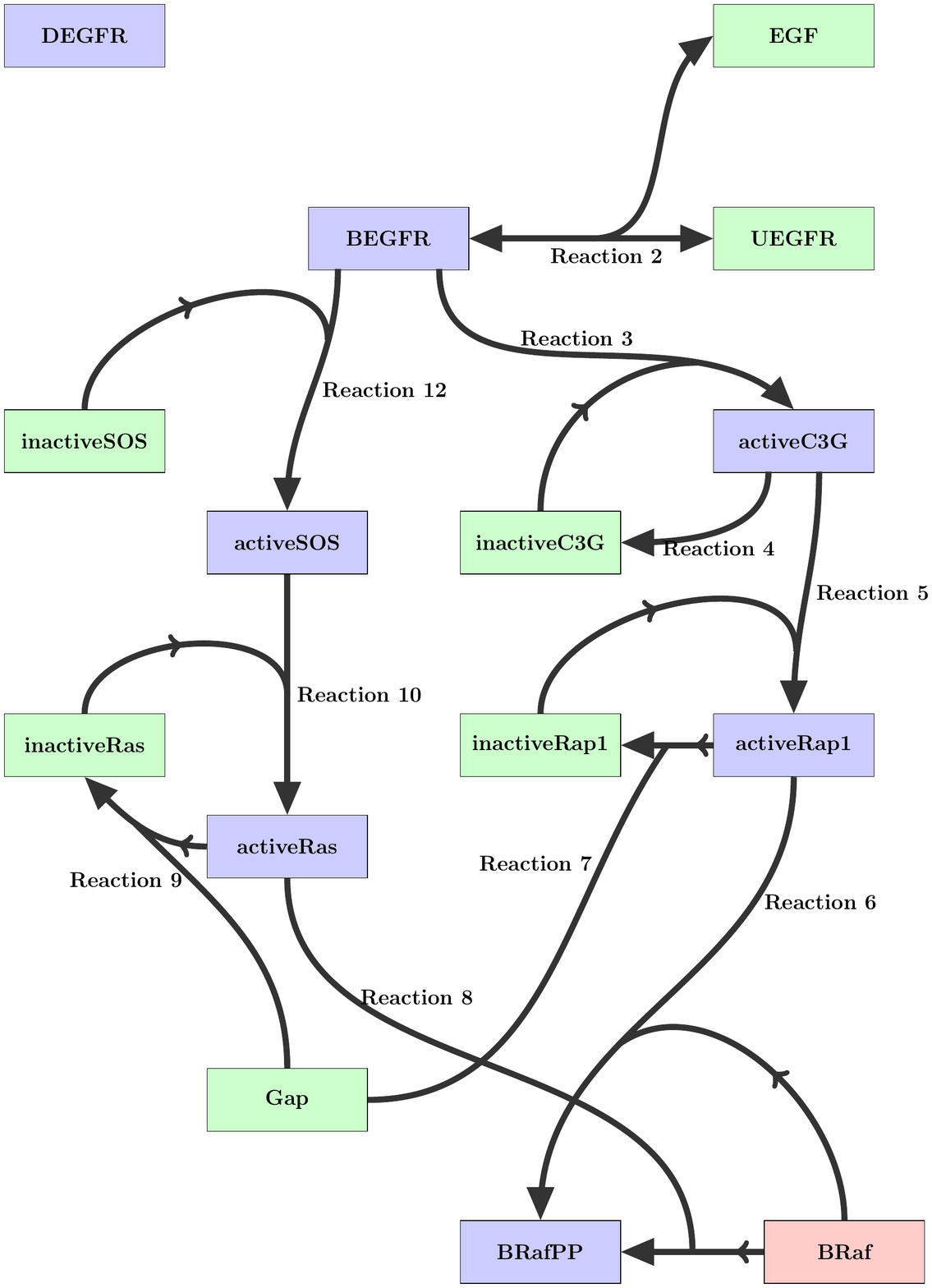}
\end{center}  
\label{fig:supp_sub2}
\end{subfigure}
\end{minipage}
 \begin{minipage}[h]{0.3\textwidth} 
\begin{subfigure}[b]{1.0\textwidth}
\begin{center}
\includegraphics[width=0.8\linewidth]{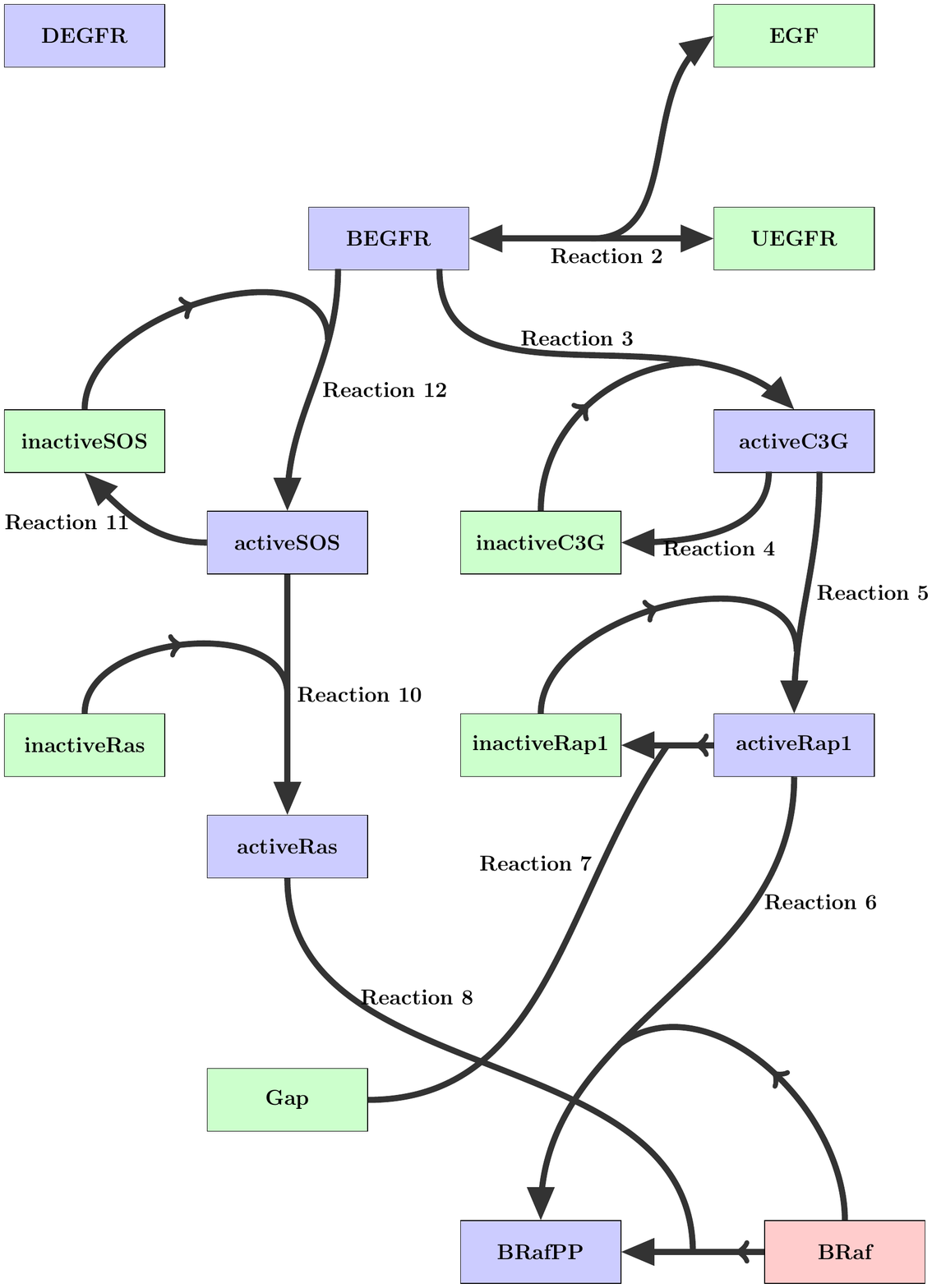}
\includegraphics[width=0.8\linewidth]{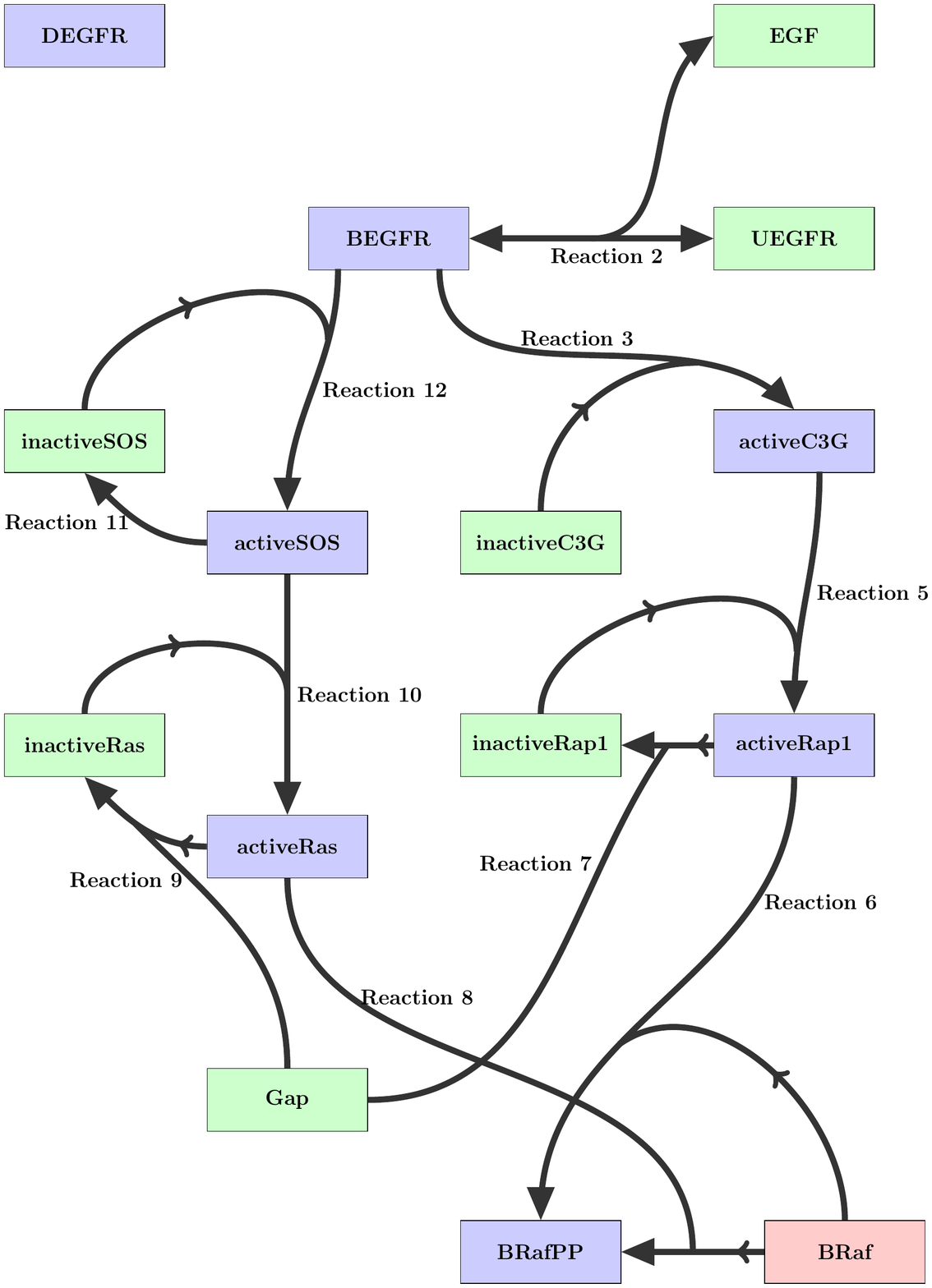}
\includegraphics[width=0.8\linewidth]{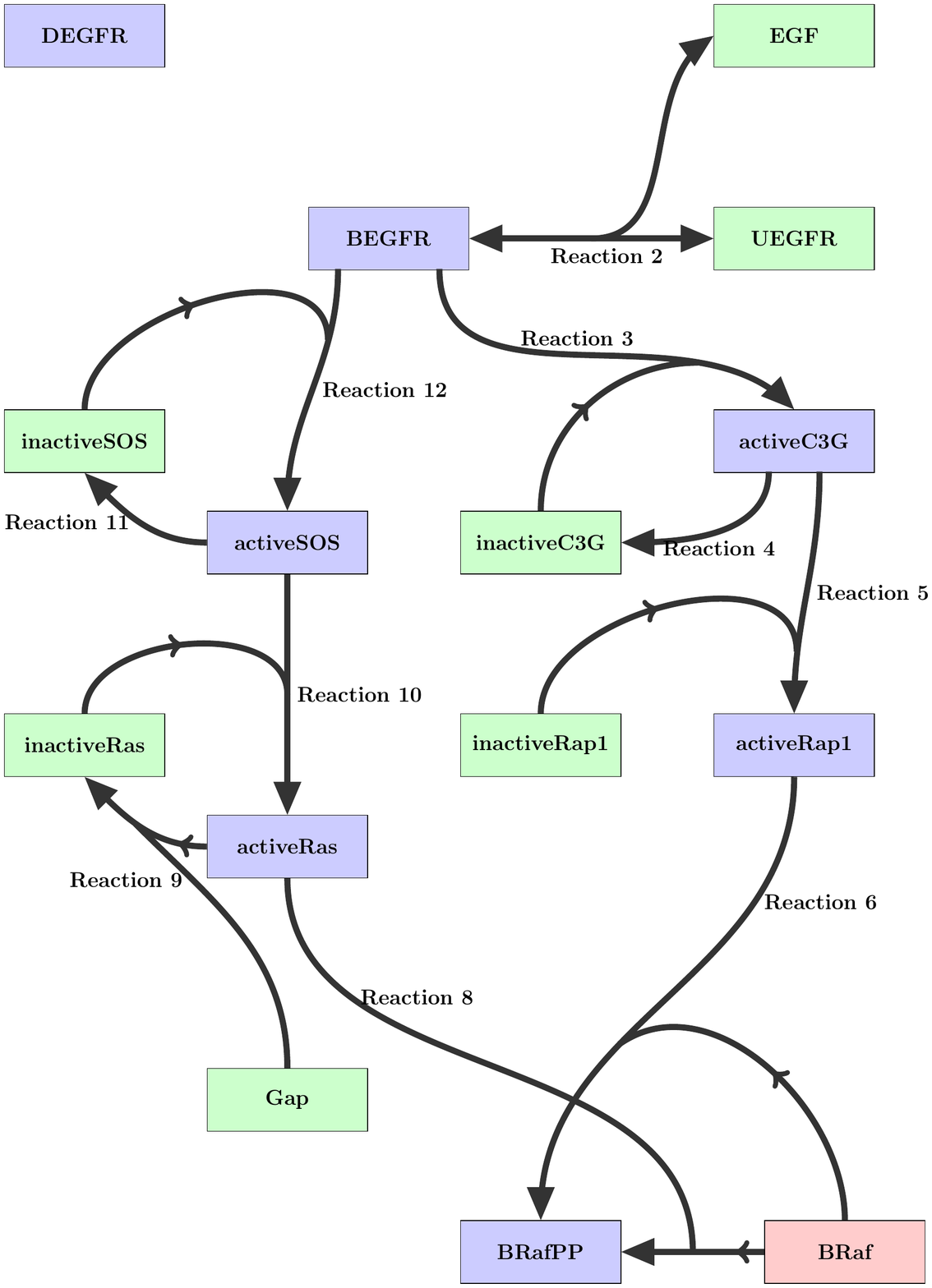}
\includegraphics[width=0.8\linewidth]{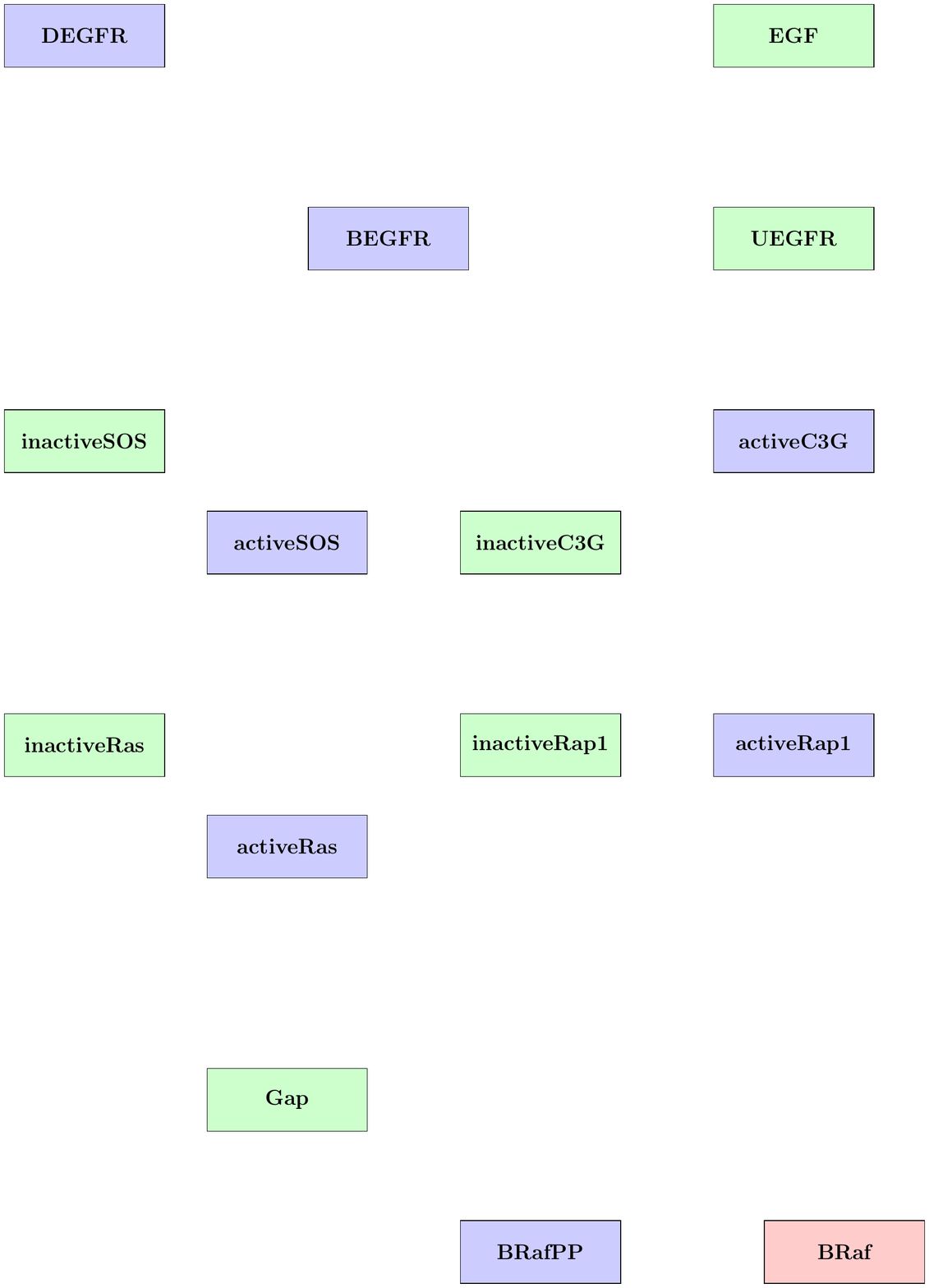}
\end{center}  
\label{fig:supp_sub3}
\end{subfigure}
\end{minipage}
\caption{Effective networks in Example 2 (contd.)}
\label{fig:effective_networks_example2b}
}
\end{figure}

There are twenty-four effective networks in Example 2. Figure \ref{fig:effective_networks_example2} and \ref{fig:effective_networks_example2b} show the network diagrams of all effective networks in Example 2. We again consider a move between the effective network 1 and effective network 2 of Figure \ref{fig:supp_movesinmodelspace} to demonstrate a prototypical between-model move proposal using our sensitivity-based network-aware proposals. Assume that the critical reaction to which the two networks are most sensitive is Reaction 8 and we only include one highly sensitive reaction for proposal construction. Thus in a move from effective network 1 to effective network 2, the rate constant of reaction 8 would not be kept fixed but instead be  proposed according to a distribution. Reaction 4 is the additional reaction in network 2. All other reactions common to the two networks, namely reactions 3, 5, 6, 9, 10, 11, and 12 will have their rate constant values kept fixed during the move. The identity jump function $f$ for the between-model move between the two effective networks is:

\begin{align}
k_{2,\{3,5,6,9,10,11,12\}}=k_{1,\{3,5,6,9,10,11,12\}},\ &k_{2,8}=u_{1,8},\ u_{2,8}=k_{1,8}, k_{2,4}=u_{1,4},
\end{align}

\noindent
where the proposal distributions $q(u_{1,\{4,8\}} \vert k_{1,\{3,5,6,9,10,11,12\}})$ and $q(u_{2,8} \vert k_{1,\{3,5,6,9,10,11,12\}})$ are $\mathcal{N}(u_{1};\mu_1,\Sigma_1)$ and $\mathcal{N}(u_{2};\mu_2,\Sigma_2)$, respectively, with proposal means

\begin{equation}
\mu_{{1}} =\underset{k_{2,\{4,8\}}}{\arg\max}\mbox{  }p(k_{2,\{4,8\}} \vert k_{{1},\{3,5,6,9,10,11,12\}},M_{2},\mathcal{D})
\end{equation}
\noindent
and 
\begin{equation}
\mu_{{2}} =\underset{k_{1,8}}{\arg\max}\mbox{  }p(k_{1,8} \vert k_{{2},\{3,5,6,9,10,11,12\}},M_{1},\mathcal{D})
\end{equation}
\noindent
and proposal covariances

\begin{equation}
\Sigma_{{1}} =-[\nabla^2 \log p(k_{2,\{4,8\}} \vert k_{1,\{3,5,6,9,10,11,12\}},M_{2},\mathcal{D})]^{-1}\big|_{\mu_{{1}}}
\end{equation}
\noindent
and 

\begin{equation}
\Sigma_{{2}} =-[\nabla^2 \log p(k_{1,8} \vert k_{2,\{3,5,6,9,10,11,12\}},M_{1},\mathcal{D})]^{-1}\big|_{\mu_{{2}}}.
\end{equation}

\section{12-dimensional reaction network}
\label{sec:12Dnetwork}
Here we present the details of the set of proposed reactions, the corresponding reaction and species production rate ODE expressions for the 12-reaction network used in Example 1 and 2. The resulting nonlinear system of ordinary differential equations are solved using the multistep BDF integrator available in the SUNDIALS suite \cite{Sundials2005}. 

\subsection{Reactions}

\begin{align*}
&1. \mbox{ } boundEGFR \rightarrow degradedEGFR \\
&2. \mbox{ } EGF+unboundEGFR \leftrightarrow boundEGFR\\
&3. \mbox{ } inactiveC3G+boundEGFR \rightarrow activeC3G + boundEGFR \\
&4. \mbox{ } activeC3G \rightarrow inactiveC3G \\
&5. \mbox{ } inactiveRap1+activeC3G \rightarrow activeRap1+activeC3G \\
&6. \mbox{ } BRaf + activeRap1 \rightarrow BRafPP+activeRap1 \\
&7. \mbox{ } activeRap1 + Gap \rightarrow inactiveRap1 + Gap \\
&8. \mbox{ } BRaf+activeRas \rightarrow BRafPP+activeRas \\
&9. \mbox{ } activeRas+Gap \rightarrow inactiveRas+Gap \\
&10. \mbox{ }inactiveRas+activeSOS \rightarrow activeRas + activeSOS \\
&11. \mbox{ }activeSOS \rightarrow inactiveSOS \\
&12. \mbox{ }inactiveSOS + boundEGFR \rightarrow activeSOS + boundEGFR
\end{align*}

\subsection{Reaction rates}

\begin{align*}
&1.  \mbox{ }k_{1}[boundEGFR]\\
&2.  \mbox{ }k_{2f}[EGF][unboundEGFR]-\mbox{ }k_{2r}[boundEGFR]\\
&3.  \mbox{ }\frac{k_{3}[boundEGFR][inactiveC3G]}{k_{3}^{'}+[inactiveC3G]}\\
&4.  \mbox{ }k_{4}[activeC3G]\\
&5.  \mbox{ }\frac{k_{5}[activeC3G][inactiveRap1]}{k_{5}^{'}+[inactiveRap1]}\\
&6.  \mbox{ }\frac{k_{6}[activeRap1][BRaf]}{k_{6}^{'}+[BRaf]}\\
&7.  \mbox{ }\frac{k_{7}[Gap][activeRap1]}{k_{7}^{'}+[activeRap1]}\\
&8.  \mbox{ }\frac{k_{8}[activeRas][BRaf]}{k_{8}^{'}+[BRaf]}\\
&9.  \mbox{ }\frac{k_{9}[Gap][activeRas]}{k_{9}^{'}+[activeRas]}\\
&10.  \mbox{ }\frac{k_{10}[activeSOS][inactiveRas]}{k_{10}^{'}+inactiveRas}\\
&11.  \mbox{ }\frac{k_{11}[activeSOS]}{k_{11}^{'}+[activeSOS]}\\
&12.  \mbox{ }\frac{k_{12}[boundEGFR][inactiveSOS]}{k_{12}^{'}+[inactiveSOS]}
\end{align*}

\subsection{Species production rates}
\begin{align*}
&1.\mbox{ }\dot{[unboundEGFR]}=-k_{2f}[EGF][unboundEGFR]+k_{2r}[boundEGFR] \\
&2.\mbox{ }\dot{[inactiveSOS]}=-\frac{k_{12}[boundEGFR][inactiveSOS]}{k_{12}^{'}+[inactiveSOS]}+\frac{k_{11}[activeSOS]}{k_{11}^{'}+[activeSOS]}\\
&3.\mbox{ }\dot{[inactiveRas]}=-\frac{k_{10}[activeSOS][inactiveRas]}{k_{10}^{'}+[inactiveRas]}+\frac{k_{9}[Gap][activeRas]}{k_{9}^{'}+[activeRas]}\\
&4.\mbox{ }\dot{[inactiveRap1]}=\frac{k_{7}[Gap][activeRap1]}{k_{7}^{'}+[activeRap1]}-\frac{k_{5}[activeC3G][inactiveRap1]}{k_{5}^{'}+[inactiveRap1]}\\
&5.\mbox{ }\dot{[boundEGFR]}=k_{2f}[EGF][unboundEGFR]-k_{2r}[boundEGFR]-k_{1}[boundEGFR]\\
&6.\mbox{ }\dot{[activeSOS]}=\frac{k_{12}[boundEGFR][inactiveSOS]}{k_{12}^{'}+[inactiveSOS]}-\frac{k_{11}[activeSOS]}{k_{11}^{'}+[activeSOS]}\\
&7.\mbox{ }\dot{[activeRas]}=\frac{k_{10}[activeSOS][inactiveRas]}{k_{10}^{'}+[inactiveRas]}-\frac{k_{9}[Gap][activeRas]}{k_{9}^{'}+[activeRas]}\\
&8.\mbox{ }\dot{[activeRap1]}=\frac{k_{7}[Gap][activeRap1]}{k_{7}^{'}+[activeRap1]}+\frac{k_{5}[activeC3G][inactiveRap1]}{k_{5}^{'}+[inactiveRap1]}\\
&9.\mbox{ }\dot{[EGF]}=-k_{2f}[EGF][unboundEGFR]+k_{2r}[boundEGFR]\\
&10.\mbox{ }\dot{[BRafPP]}=\frac{k_{6}[activeRap1][BRaf]}{k_{6}^{'}+[BRaf]}+\frac{k_{8}[activeRas][BRaf]}{k_{8}^{'}+[BRaf]}\\
&11.\mbox{ }\dot{[BRaf]}=-\frac{k_{6}[activeRap1][BRaf]}{k_{6}^{'}+[BRaf]}-\frac{k_{8}[activeRas][BRaf]}{k_{8}^{'}+[BRaf]}\\
&12.\mbox{ }\dot{[activeC3G]}=\frac{k_{3}[boundEGFR][inactiveC3G]}{k_{3}^{'}+[inactiveC3G]}-k_{4}[activeC3G]\\
&13.\mbox{ }\dot{[inactiveC3G]}=-\frac{k_{3}[boundEGFR][inactiveC3G]}{k_{3}^{'}+[inactiveC3G]}+k_{4}[activeC3G]\\
&14.\mbox{ }\dot{[degradedEGFR]}=k_{1}[boundEGFR]\\
&15.\mbox{ }\dot{[Gap]}=0\\
\end{align*}

\subsection{Initial species concentrations}
\noindent
All simulations using the above reactions are performed with the following initial concentrations:
\begin{align*}
&1.\mbox{ }[unboundEGFR]_{0}=500 \\
&2.\mbox{ }[inactiveSOS]_{0}=1200\\
&3.\mbox{ }[inactiveRas]_{0}=1200\\
&4.\mbox{ }[inactiveRap1]_{0}=1200\\
&5.\mbox{ }[boundEGFR]_{0}=0\\
&6.\mbox{ }[activeSOS]_{0}=0\\
&7.\mbox{ }[activeRas]_{0}=0\\
&8.\mbox{ }[activeRap1]_{0}=0\\
&9.\mbox{ }[EGF]_{0}=1000\\
&10.\mbox{ }[BRafPP]_{0}=0\\
&11.\mbox{ }[BRaf]_{0}=1500\\
&12.\mbox{ }[activeC3G]_{0}=0\\
&13.\mbox{ }[inactiveC3G]_{0}=1200\\
&14.\mbox{ }[degradedEGFR]_{0}=0\\
&15.\mbox{ }[Gap]_{0}=2400\\
\end{align*}

\bibliography{galagali}
\bibliographystyle{plain}

\end{document}